\documentclass[preprint,11pt,nofootinbib]{revtex4-1}
\usepackage{amsmath}
\usepackage{amsfonts}
\usepackage{amssymb}
\usepackage{mathtools}
\usepackage{etoolbox}
\usepackage{tikz}
\usetikzlibrary{shapes.geometric}
\usepackage{slashed}
\usepackage{subcaption}
\usepackage{braket}
\usepackage{listings}
\usepackage{xcolor}
\usepackage{appendix}
\usepackage{hyperref}
\usepackage{enumitem}
\usepackage{diagbox} %for slash in table
\usepackage{listings}
\usepackage{float}
\usepackage{bm}
\usepackage{multirow}
\usepackage{lipsum}
\usepackage{array}
\usepackage{float}
\usepackage{bm}
\usepackage{subcaption}
\usepackage{caption}
\usepackage[left=35mm, top=30mm,bottom=33mm,right=25mm]{geometry}
\usetikzlibrary{calc,matrix,fit}
\DeclareMathOperator{\Tr}{Tr}

\newcommand{\be}{\begin{equation}}
\newcommand{\ee}{\end{equation}}
\newcommand{\bea}{\begin{eqnarray}}
\newcommand{\eea}{\end{eqnarray}}
\newcommand{\non}{\nonumber}
\newcommand{\bi}{\begin{itemize}}
\newcommand{\ei}{\end{itemize}}

\begin{document}
%\vspace*{-100mm}
\begin{flushright}
KANAZAWA-26-02  \\ UTHEP-824 \\ UTCCS-P-180
\end{flushright}
%\vspace*{80mm}
\title{Multi-particle states inverstigation with tensor renormalization group method}

\author{Fathiyya Izzatun Az-zahra$^1$}
\email{Contact author: fathiyya@hep.s.kanazawa-u.ac.jp}
\author{Shinji Takeda$^{1}$}
\email{Contact author: takeda@hep.s.kanazawa-u.ac.jp}
\author{Takeshi Yamazaki$^{2,3}$}
\email{Contact author: yamazaki@het.ph.tsukuba.ac.jp}

\affiliation{${}^1$\footnotesize{Institute for Theoretical Physics, Kanazawa University, Kanazawa 920-1192, Japan}\\
${}^{2}$\footnotesize{Institute of Pure and Applied Sciences, University of Tsukuba, Tsukuba 305-8571, Japan}\\
${}^{3}$\footnotesize {Center for Computational Sciences, University of Tsukuba, Tsukuba 305-8577, Japan}
}
\begin{abstract}
We investigate multi-particle states of the (1+1)d Ising Model using a spectroscopy scheme based on
transfer matrix and tensor renormalization group method.
The scheme begins with computing the energy spectrum of the system from the transfer matrix 
estimated by the coarse-grained tensor network.
The quantum number and momentum of these energy eigenstates are not \textit{a priori} known,
thus we identify them using matrix elements of an interpolating operator that is numerically computed
with an impurity tensor network. 
Furthermore, by observing the dependence of the energy as a function of system size, 
we identify the number of particles of the eigenstates and obtain one-, two-, and three-particle states for a specific quantum number and momentum.
From the two-particle state sector, we compute the scattering phase shift using L\"uscher's formula and
wave function approach, and observe their consistency with theoretical prediction. 
Using the information of the two-particle scattering phase shift, we investigate the degeneracy of the two-particle states, the theoretical prediction of the three-particle finite volume energy and also the degeneracy in the three-particle states.
\end{abstract}
\date{\today}
\maketitle
\tableofcontents
\section{Introduction}
\label{sec:intr}
The study on multi-particle state is important for modern physics, particularly for system that involves strong interaction such as nuclear and high energy physics.  
In nuclear physics, 
the formation of nuclei involves the two- and three-nucleon interactions.
In high energy physics, multi-particle states appear during the decay processes of several hadrons
such as $\rho(770)$, which mainly decays into two pions. 
Other examples are $\omega(782)$ and $\pi(1300)$, which mainly decay into three pions, etc.
Within the framework of lattice quantum chromodynamics (LQCD), these states can be studied
with non-perturbative spectroscopy calculation where Monte Carlo (MC) algorithm is primarily applied for the computation
\cite{PhysRevD.92.014501,PhysRevD.94.034501,PhysRevLett.133.211906,vfr3-5lsb}. 
Despite the success of Monte Carlo algorithms for spectroscopy calculations in LQCD, there are some difficulties that hinder the computation. A large lattice extent in the time direction is required to reduce contamination from higher eigenstates, and very high statistics are needed to suppress noise in the extraction of excited states signals.
Motivated by these issues, tensor network has emerged as an alternative and promising method for spectroscopy calculation.  
The method has two approaches i.e.
the Hamiltonian approach \cite{PhysRevLett.69.2863,PhysRevLett.75.3537,Verstraete:2004cf,Banuls:2016gid,Banuls:2019bmf,Banuls:2019hzc,Schneider:2022lcl}
and the Lagrangian approach
\cite{
Shimizu:2012wfa,
Shimizu:2012zza,
Yu:2013sbi,
Zou:2014rha,
Shimizu:2014uva,
Shimizu:2014fsa,
Takeda:2014vwa,
Yang:2015rra,
Kawauchi:2016xng,
Shimizu:2017onf,
Kadoh:2018hqq,
Kadoh:2018tis,
Kuramashi:2018mmi,
Kuramashi:2019cgs,
Bazavov:2019qih,
Akiyama:2019xzy,
Akiyama:2020ntf,
Akiyama:2020soe,
Akiyama:2021zhf,
Akiyama:2021xxr,
Akiyama:2021glo,
Nakayama:2021iyp,
Akiyama:2022eip,
Akiyama:2023hvt,
Kuwahara:2022ubg,
Hirasawa:2021qvh,
Fukuma:2021cni,
Bloch:2021mjw,
Luo:2022eje,
Bloch:2022vqz,
Jha:2022pgy,
Samberger:2026vpy,
sugimoto2026tensorrenormalizationgroupstudy}.
The spectroscopy scheme using the former approach is presented in Refs.~\cite{Itou:2023img,Itou:2024psm,Matsumoto:2025fjb}, and the latter one is in Refs.~\cite{PhysRevD.110.034514,Az-zahra:2024pqa}.

In the present work, we focus on the Lagrangian approach 
in which the energy spectrum is extracted from the transfer matrix of the system 
and the estimation of this matrix is calculated using one of the tensor renormalization group algorithms 
\cite{PhysRevLett.99.120601,PhysRevB.86.045139,PhysRevLett.103.160601,
evenbly2015tensor,
yang2017loop,
Hauru:2017jbf,
morita2018tensor,
harada2018entanglement,
PhysRevB.99.155101,
PhysRevB.102.054432,
Kadoh:2019kqk,
Kadoh:2021fri,
Arai:2022uee,
Nakayama:2023ytr,homma2023},
for instance, higher order tensor renormalization group algorithm (HOTRG) \cite{PhysRevB.86.045139}.
The quantum number of the energy eigenstate is not a priori known,
so we identify it with a selection rule derived from the symmetry of the system.
The important quantity for the selection rule is the matrix element of an interpolating operator
which can be represented as an impurity tensor network, 
and then coarse-grained by using HOTRG to obtain its estimate.
The momentum can also be identified in a similar manner.

With this method, basically one can extract the energy spectrum of the system using only a single-time slice of the lattice,
and no statistical errors are present because this method is deterministic.
However, the coarse graining with HOTRG introduces systematic errors.
Ref.~\cite{PhysRevD.110.034514} showed that, for two-dimensional system, the transfer matrix estimated from a single-time slice tensor network 
produces eigenvalues that are closely degenerate, which causes large errors during the coarse-graining steps.  To resolve this issue, 
a square tensor network is used
to perform the calculation.
Although this method produces reasonable results, the errors of the extracted energy spectrum increase drastically as the system size becomes larger, particularly for higher excited states.
Consequently, the system sizes that we can explore are limited and the calculation is restricted to low lying-energy eigenstates.

In this paper, we introduce a new coarse-graining strategy to improve the accuracy, and reliably extract the energy of the higher excited states, which are expected to correspond to the multi-particle states.
Additionally, the analysis of the dynamical properties of the two-particle state sector is also performed.
We compute the phase shift of the two-particle scattering state with zero total momentum 
using both the finite volume energy approach based on L\"uscher's formula \cite{Luscher:1985dn,Luscher:1986pf,Luscher:1990ck,LUSCHER1991531}
and the wave function approach \cite{BALOG2001315,Yamazaki:2017gjl,PhysRevD.98.011501,Namekawa:2019xiy}. 
We also compute the phase shift from the states with non-zero total momentum following the procedures given in 
Refs.~\cite{RUMMUKAINEN1995397, PhysRevD.99.014501},
and check the consistency of the phase shift extracted from these different methods.
Lastly, using the information of phase shift, we compute the two- and three-particle state dispersion relation for both zero and non-zero total momentum, and investigate the degeneracy of the energy eigenstates in these sectors.

This paper is organized as follows.
The formulation of the computational scheme is given in Sec.~\ref{Sec:Formulation}.
At the beginning, we briefly review the transfer matrix formalism and the tensor network representation
for the computation of the energy spectrum
and the matrix elements for the quantum number identification in Sec.~\ref{Subsec:TM_TN}.
After that, we explain the new strategy for the coarse graining of the tensor network in Sec.~\ref{subs:coarse-graining}.
We apply the scheme to (1+1)d Ising Model and show the numerical results in Sec.~\ref{Sec:num_res},
where the energy spectrum, quantum number, momentum, number of particles, and wave function
are given in Sec.~\ref{subsec:energy_spectrum}, \ref{subsec:quantum_number}, \ref{subsec:momentum},
\ref{subsec:numb_particles}, \ref{subsec:wf_2p}, respectively. 
In Sec.~\ref{subsec:2p_state}, we present the scattering phase shift of the two-particle state sector.
The phase shifts computed from the energy spectrum with L\"uscher's formula in rest and moving frame are given in Sec.~\ref{ssubec:ps_energy}.
Meanwhile, the phase shifts obtained from the wave function outside and inside interaction range are presented in Sec.~\ref{subsec:ps_2p_wf} and
\ref{subsec:ps_2p_wf_inside}, respectively.
The two-particle states's dispersion relation, and their degeneracy are discussed in \ref{sec:two_p_states_degeneracy}. Moreover, the theoretical prediction of three-particle states from the dispersion relation, and discussion about their degeneracies are presented in \ref{sec:three_p_states_disrel}, and \ref{sec:three_p_states_degeneracy}, respectively. 
The summary is given in the final section. Lastly, we present the derivation of L\"uscher's equation from the wave function inside the interaction range in Appendix~\ref{sec:luschers_from_bswf}

\section{Formulation}
\label{Sec:Formulation}
\subsection{Transfer matrix formulation and its tensor network representation}
\label{Subsec:TM_TN}
Let us start with a brief review of the spectroscopy scheme
using transfer matrix and tensor network that we proposed in \cite{PhysRevD.110.034514}.
The scheme is explained in the framework of the two dimensional scalar field theory 
with nearest-neighbor interactions on lattice. 
However, its extension to higher dimensional systems is straightforward,
and in principle, the scheme is also applicable to fermionic or gauge systems.
The lattice action of the (1+1)d scalar field theory in Euclidean space-time is given by
\be
S[\phi]=
\sum_{{\bm r}\in\Gamma}
\left[
\sum_{\mu=0}^1\frac{1}{2}
\left(
\phi({\bm r}+\hat\mu)-\phi({\bm r})
\right)^2
+
V(\phi({\bm r}))
\right].
\ee
Here, $\phi(\bm r)$ is the scalar field resides on two dimensional square lattice
${\bm r}=(t,x)\in\Gamma$ and $\hat\mu$ is the unit vector for the $\mu=0,1$ direction,
where the $0$ direction ($1$ direction) is considered as the time (space) direction.
The lattice $\Gamma$ has periodic boundary in spatial direction, and can be defined as
\be
\Gamma=\{(t,x)|
t=0,1,2,\hdots,L_{\rm t}-1
\,\,{\rm and}\,\,
x=0,1,2,\hdots, L_{\rm s}-1
\},
\label{eqn:Gamma}
\ee
where $L_{\rm t}$ and $L_{\rm s}$ denote the lattice size in time and space direction, respectively.
Note that the mass term and self-interaction term are already included in the potential $V(\phi({\bm r}))$. 
Accordingly, the partition function of the system is given by
\be
Z=\int \prod_{{\bm r}\in\Gamma} d\phi({\bm r}) e^{-S[\phi]}.
\ee

\begin{figure}[t!]
\centering
\begin{subfigure}[b]{0.25\textwidth}
\centering
\includegraphics[width=3.5cm,height=6cm]{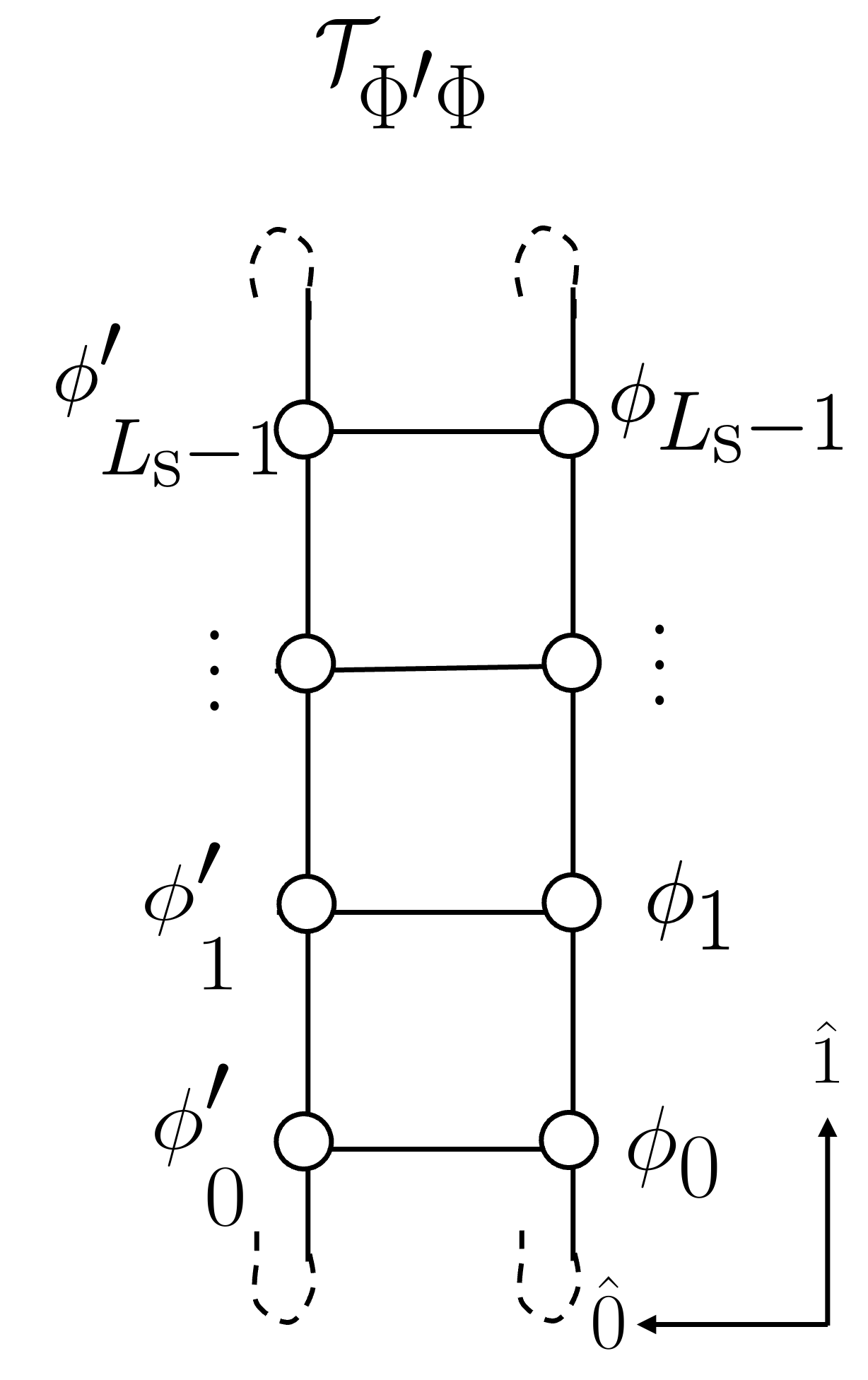}
\caption{}
\label{sfig:transfer_matrix}
\end{subfigure}
\begin{subfigure}[b]{0.25\textwidth}
\centering
\includegraphics[width=3.5cm,height=6cm]{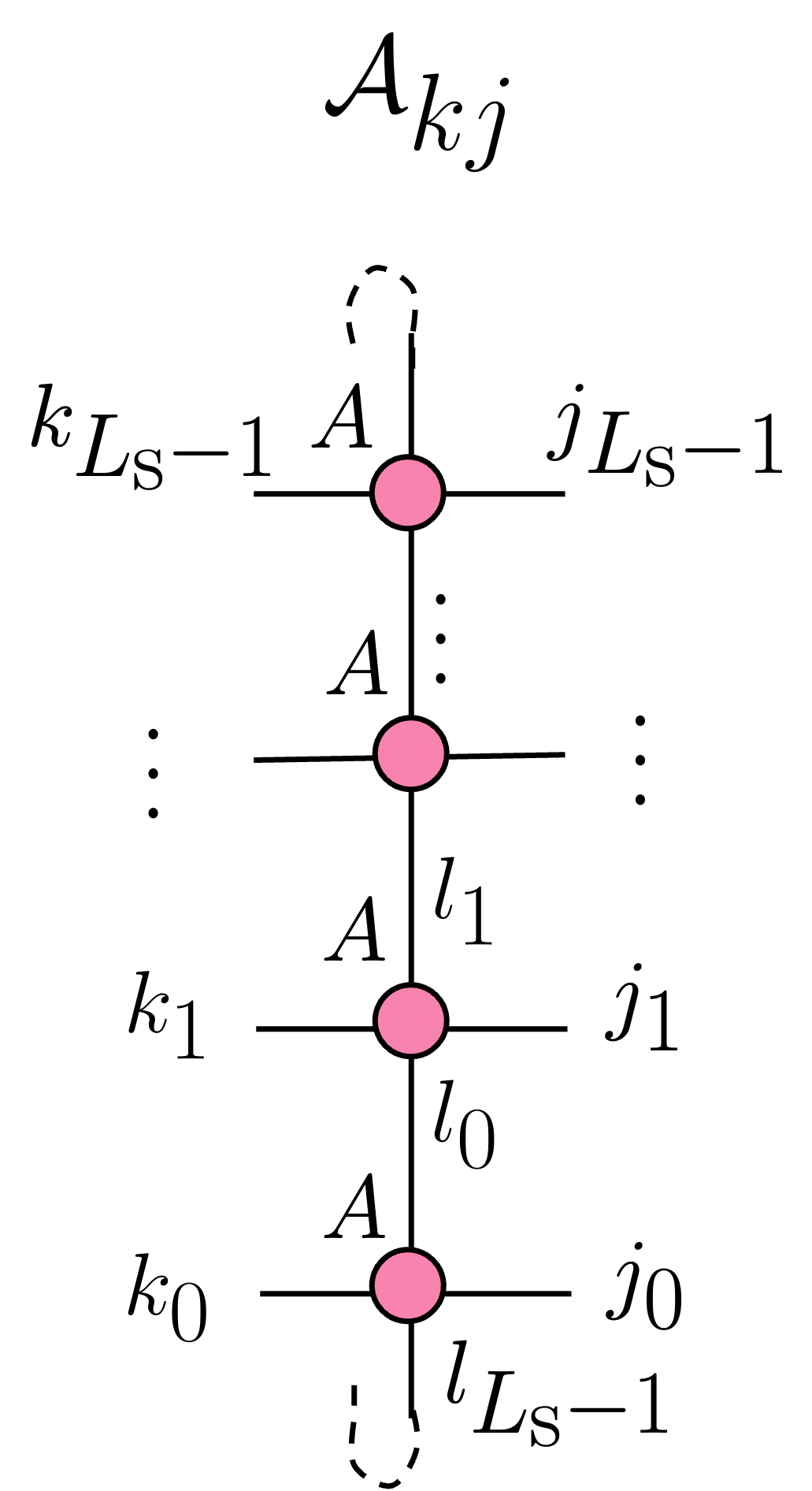}
\caption{}
\label{sfig:tensor_network}
\end{subfigure}
\begin{subfigure}[b]{0.25\textwidth}
\centering
\includegraphics[width=3.5cm,height=6cm]{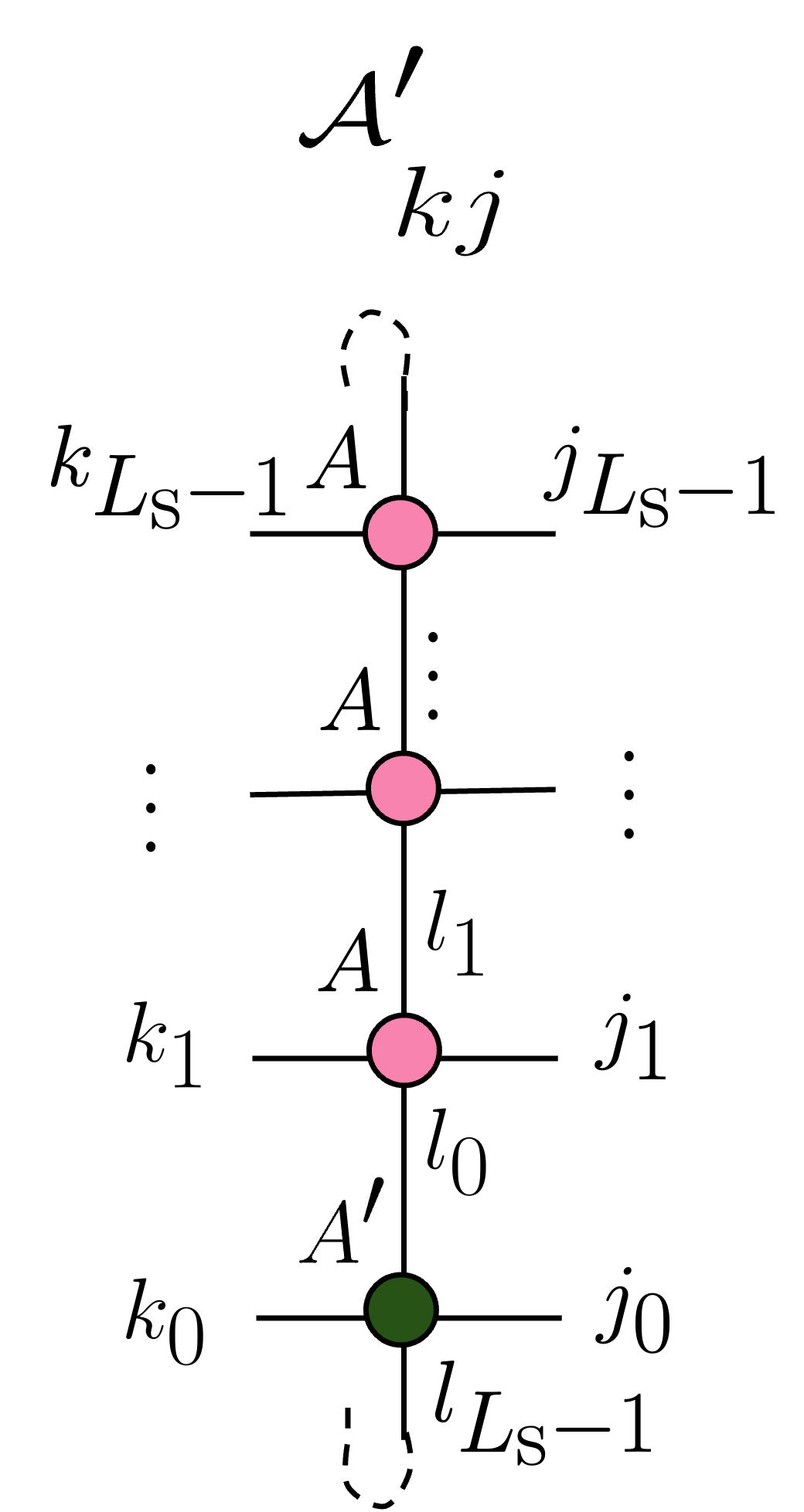}
\caption{}
\label{sfig:impurity_tn}
\end{subfigure}
\caption{
A graphical image of (a) transfer matrix,
 (b) single-time slice tensor network,
and (c) single-time slice impurity tensor network consisted of several pure tensors $A$ and one impurity tensor $A'$.
}
\label{fig:pict_TN}
\end{figure}

This partition function can be reformulated in terms of transfer matrix 
\be\label{eq:partition_function}
Z=
{\rm Tr}
\left[
{\cal T}^{L_{\rm t}}
\right],
\ee
where the transfer matrix ${\cal T}$ (see Fig.~\ref{sfig:transfer_matrix} for the diagram) is given by \cite{montvay1994quantum}
\bea
{\cal T}_{\Phi'\Phi}
&=&
\left(
\prod_{x=0}^{L_{\rm s}-1}
\exp\left[
-\frac{1}{2}(\phi_x^\prime-\phi_x)^2
-\frac{1}{4}V(\phi^\prime_x)
-\frac{1}{4}V(\phi_x)
\right]
\right)
\non\\
&\times&
\left(
\prod_{x=0}^{L_{\rm s}-1}
\exp\left[-\frac{1}{4}(\phi_{x+1}^\prime-\phi_x^\prime)^2
-\frac{1}{8}V(\phi^\prime_{x+1})
-\frac{1}{8}V(\phi^\prime_x)
\right]
\right)
\non\\
&\times&
\left(
\prod_{x=0}^{L_{\rm s}-1}
\exp\left[-\frac{1}{4}(\phi_{x+1}-\phi_x)^2
-\frac{1}{8}V(\phi_{x+1})
-\frac{1}{8}V(\phi_x)
\right]
\right),
\label{eqn:TM_details}
\eea
where 
\begin{alignat}{4}
&\Phi^\prime&=&\{\phi'_x=\phi(t+1&,&x)&|&x=0,1,2,\hdots,L_{\rm s}-1\},
\\
&\Phi&=&\{\phi_x=\phi(t&,&x)&|&x=0,1,2,\hdots,L_{\rm s}-1\}
\label{eqn:integrated_phi}
\end{alignat}
are field configurations on the Euclidean time slice at $t+1$ and $t$.
In this case, ${\cal T}_{\Phi'\Phi}$ is treated as a usual matrix 
where $\Phi$ is treated as integer-valued index for notational convenience.
The diagonalization of ${\cal T}_{\Phi^\prime \Phi}$ is given by
\be
{\cal T}_{\Phi^\prime\Phi}
=
\sum_{a=0}^\infty U_{\Phi^\prime a}\lambda_a (U^\dag)_{a\Phi}
\,\,
\Longleftrightarrow
\,\,
\hat {\cal T}|a\rangle=\lambda_a|a\rangle.
\label{eqn:T_EVD}
\ee
Here $U_{\Phi'a}=\langle\Phi'|a\rangle$ is the field representation of the eigenstate $|a\rangle$.
The eigenvalues $\lambda_a$ give us the information of the energy eigenvalues $E_a$ of the system
\be\label{eq:lambda_to_energy}
\lambda_a=e^{-E_a}
\hspace{10mm}
\mbox{ for } a=0,1,2,3,\ldots.
\ee 
where $\lambda_0\geq \lambda_1\geq \ldots$. 
Instead of energy $E_a$, the energy gaps $\omega_a$ 
\be\label{eq:egap}
\omega_a=E_a-E_0
\hspace{10mm}
\mbox{ for } a=1,2,3,\ldots
\ee
where $E_0$ is the ground state energy, are more useful.
So that, hereafter, the energy gap spectrum $\omega_a$ will be mentioned as energy spectrum for simplicity.

The quantum number of each eigenstate $|a\rangle$
is not \textit{a priori} known.
Therefore, we identify it using matrix elements 
\be\label{eq:matrix_elements_def}
\langle b|\hat{\mathcal{O}}_q|a\rangle= \left( U^{\dagger}{\mathcal{O}_q}U\right)_{ba},
\ee
where $U$ is the unitary matrix from Eq.~(\ref{eqn:T_EVD}),
$\hat{\mathcal{O}}_q$ is an interpolating operator with quantum number $q$ 
and $\mathcal{O}_q$  is the field representation of the interpolating operator
\be\label{eq:int_operator}
(\mathcal{O}_q)_{\Phi'\Phi}=\langle \Phi'|\hat{\mathcal{O}}_q|\Phi\rangle.
\ee 
A selection rule derived from the symmetry
determines the quantum number of the eigenstate $|a\rangle$.
As derived in \cite{PhysRevD.110.034514},
the selection rule for the system with discrete symmetry is given by
\be\label{eq:selection_rule}
\langle b| \hat{\mathcal{O}}_q|a\rangle \neq 0 \rightarrow q_bqq_a=1,
\ee
where $q_a,q_b$ are the quantum numbers for eigenstates $|a\rangle$ and $|b\rangle$, respectively.
Solving Eq.~(\ref{eq:selection_rule}) for known $q_b$ and $q$, yields the quantum number of the state $|a\rangle$.
The quantum number in systems with continuous symmetries can also be classified using a similar approach, see \cite{PhysRevD.110.034514}.
%%%%%%%%%
%%%%%%%%%

After presenting the formalism for computing the energy spectrum
and identifying the corresponding quantum numbers,
we now discuss how to compute them numerically.
Because the dimensionality of transfer matrix is extremely large,
we apply an approximation method to compute its eigenvalues and extract the energy spectrum,
which in our case is based on tensor network techniques.
To construct the tensor network representation of the transfer matrix in Eq.~(\ref{eqn:TM_details}),
we first decompose the transfer matrix into
\be
{\cal T}_{\Phi^\prime\Phi}
=
\sum_k Y_{\Phi^\prime k} Y^\dag_{k\Phi},
\label{eqn:TM_YY}
\ee
where $k=\{k_x|x=0,1,2,\ldots,L_{\rm s}-1\}$ is a newly defined integrated index 
and the matrix $Y$ is defined as
\bea\label{eq:Y}
Y_{\Phi k}
&=&
\left(
\prod_{x=0}^{L_{\rm s}-1}
\sum_{k_x=0}^\infty
u_{\phi_x{k_x}}
\sqrt{\sigma_{k_x}}
\right)
\non\\&\times&
\left(
\prod_{x=0}^{L_{\rm s}-1}
\exp\left[-\frac{1}{4}(\phi_{x+1}-\phi_x)^2
-\frac{1}{8}V(\phi_{x+1})
-\frac{1}{8}V(\phi_x)
\right]
\right).
\eea
The matrix $u_{\phi_xk_x}$ and $\sigma_{k_x}$ 
are obtained from the eigenvalue decomposition (EVD) of the Boltzmann weight in Eq.~(\ref{eqn:TM_details}) 
for the fields $(\phi,\phi'\in\mathbb{R})$, specifically from the temporal hopping term, namely
\be
\exp\left[
-\frac{1}{2}(\phi^\prime-\phi)^2
-\frac{1}{4}V(\phi^\prime)
-\frac{1}{4}V(\phi)
\right]
=
\sum_{k=0}^\infty
u_{\phi^\prime{k}}
\sigma_{k}
(u^\dag)_{k\phi}.
\label{eqn:hopping1}
\ee
Using the matrix $Y$ in Eq.~(\ref{eq:Y}), the partition function in Eq.~(\ref{eq:partition_function})
can be rewritten as
\be\label{eq:partition_func_Y}
Z=\Tr_{\Phi}{\left[{\cal T}^{L_{\rm t}}\right]}=\Tr_{\Phi}{\left[(YY^{\dagger})^{L_{\rm t}}\right]}
=\Tr_k{\left[(Y^{\dagger}Y)^{L_{\rm t}}\right]}=\Tr_k{\left[{\cal A}^{L_{\rm t}}\right]}.
\ee
Here, we have defined
\be\label{one_time_slice_tensor_network}
{\cal A}\coloneqq Y^{\dagger}Y.
\ee
As seen in Fig.~\ref{sfig:tensor_network}, 
$\mathcal{A}$ can be written as a product of rank-four tensors 
\be\label{eq:pure_Tensornetwork}
{\cal A}_{kj}=\sum_{\{l\}} \prod_{x=0}^{L_{\rm s}-1} A_{k_xl_xj_xl_{x-1}}
\ee
where the tensor $A_{k_xl_xj_xl_{x-1}}$ is computed by 
\be\label{eq:A}
A_{k_xl_xj_xl_{x-1}}\coloneqq\sqrt{\sigma_{k_x}\sigma_{l_x}\sigma_{j_x}\sigma_{l_{x-1}}}\sum_{\phi_x}
\left(u^{\dagger}\right)_{k_x\phi_x}\left(u^{\dagger}\right)_{l_x\phi_x}u_{\phi_xj_x}u_{\phi_xl_{x-1}}.
\ee
The EVD of this single-time slice tensor network $\mathcal{A}_{kj}$ is given by
\be\label{eq:EVD_A}
{\cal A}_{kj}= W_{ka}\lambda_a W^{\dagger}_{aj},
\ee
where $\lambda_a$ are the same eigenvalues as in Eq.~(\ref{eqn:T_EVD}) and $W$ is a diagonalization matrix.

Next, we consider the matrix elements $\langle b|\hat{\mathcal{O}}_q|a\rangle$.
This matrix also has very large dimensionality, so we apply the tensor network methods to approximate it as well.
For this purpose, we first rewrite $\langle b|\hat{\mathcal{O}}_q|a\rangle$ in the tensor network language.
The complete procedure for formulating matrix elements in terms of tensor network is given in \cite{PhysRevD.110.034514},
where the final expression is given by
\be\label{eq:matrix_elements}
\langle b|\hat{\mathcal{O}}_q|a\rangle=\left(\lambda^{-m+1/2}W^{\dagger}\mathcal{A}^{m-1}\mathcal{A}'\mathcal{A}^mW\lambda^{-m-1/2}\right)_{ba}.
\ee
Here, ${\cal A}^{m-1}{\cal A}'{\cal A}^m$ represents an impurity tensor network,
and $1\leq m\leq L_{\rm t}/2$ specifies the position of $\cal A'$ within the network (see Fig.~\ref{fig:coarse-graining_imp}).
Note that, ${\cal A}'$ is the impurity version of $\mathcal{A}$, that is 
$\mathcal{A}'\coloneqq Y^{\dagger}\mathcal{O}_qY$,
where $Y$ is the matrix given in Eq.~(\ref{eq:Y}).
For a single field $\mathcal{O}_q=\phi_x$ at lattice site $x$, an impurity tensor $A'$ is defined as
\be
A_{k_xl_xj_xl_{x-1}}^\prime
\coloneqq
\sqrt{\sigma_{k_x}\sigma_{l_x}\sigma_{j_x}\sigma_{l_{x-1}}}
\sum_{\phi_x}
\phi_x
(u^\dag)_{k_x\phi_x}
(u^\dag)_{l_x\phi_x}
u_{\phi_xj_x}
u_{\phi_xl_{x-1}}.
\label{eq:impurity_tensor}
\ee
In Fig.~\ref{sfig:impurity_tn} we show an image of $\mathcal{A}'$, which is
composed of several pure tensors $A$ and the single impurity tensor $A'$ located at $x=0$.
The mathematical expression of the single-time slice impurity tensor network $\mathcal{A}'_{kj}$ 
shown in terms of $A$ and $A'$ is given by
\be
\mathcal{A}'_{kj}=\sum_{\{l\}}A'_{k_0l_0j_0l_{L_{\rm s}-1}}\prod_{x=1}^{L_{\rm s}-1}A_{k_xl_xj_xl_{x-1}}.
\ee

\subsection{Coarse-graining of tensor network}
\label{subs:coarse-graining}
Here, we explain how to coarse grain the tensor network representation
of transfer matrix and matrix elements derived in the previous section
using higher order tensor renormalization group (HOTRG) algorithm \cite{PhysRevB.86.045139}.
In the following, the bond dimension of the initial tensor is denoted by $\chi$.
%%%%%%%%%
\begin{figure}[t!]
\centering
\begin{subfigure}[b]{1\textwidth}
\centering
\includegraphics[width=12cm,height=4cm]{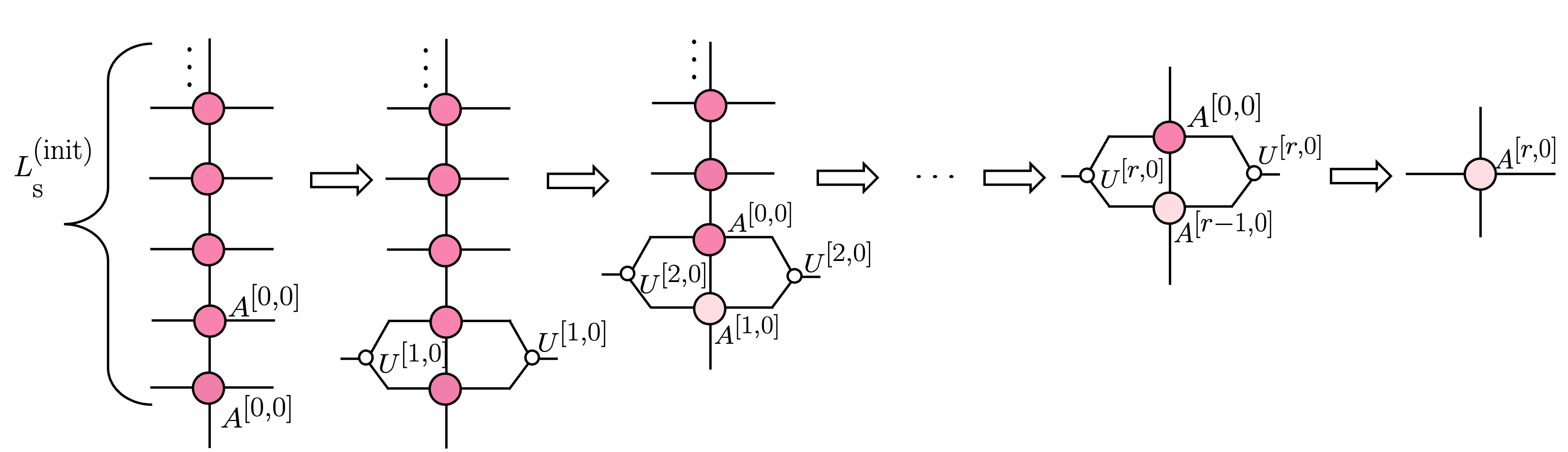}
\caption{}
\label{sfig:init_tn_pure}
\end{subfigure}
\begin{subfigure}[b]{1\textwidth}
\centering
\includegraphics[width=12cm,height=4cm]{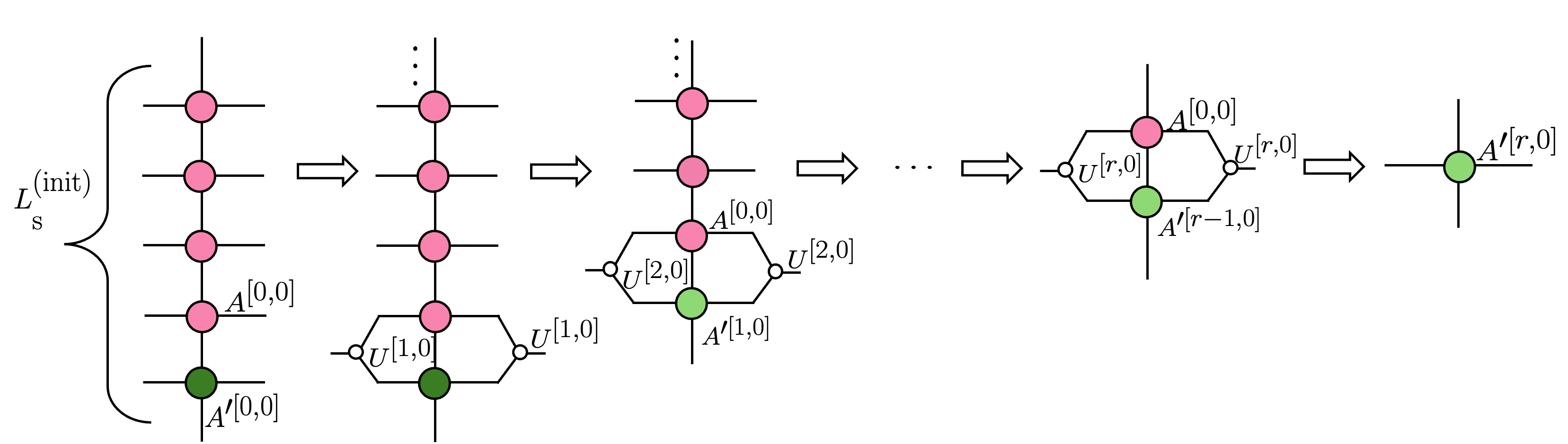}
\caption{}
\label{sfig:init_tn_impure}
\end{subfigure}
\caption{(a-b) The coarse-graining procedure 
for constructing the initial pure and impurity tensor networks with size $L_{\rm s}^{(\rm init)}=r+1$ for $r\in \{0,2,4,6\}$.
The dark pink (dark green) circle denotes the bare pure (impurity) tensor $A^{[0,0]} (A'^{[0,0]})$,
and the light pink (light green) circle denotes the pure (impurity) tensor network after $r$-th contraction, $A^{[r,0]} (A'^{[r,0]})$.}
\label{fig:coarse-graining_init}
\end{figure}
%%%%%%%%%

It is well-known that single coarse-graining iteration of HOTRG naturally reduces the tensor network size
by a factor of two, thereby
restricting the accessible system sizes to $L_{\rm s}=2^n$,
where $n=0,1,2,\ldots$ is the number of coarse-graining step. 
To access a wider range of spatial sizes, 
we prepare an initial single-time slice tensor network
\footnote{The idea of constructing the initial tensor network for transfer matrix was first proposed in Ref.~\cite{PhysRevB.107.205123}.} with spatial size $L_{\rm s}^{(\rm init)}$
and then embed it into the main tensor network such that the total spatial size becomes $L_{\rm s}=L_{\rm s}^{(\rm init)}\times 2^n$.
Note that, throughout this work, we use initial tensor sizes $L_{\rm s}^{(\rm init)}\in \{1,3,5,7\}$.

For this purpose, we denote the tensor $A$ in Eq.~(\ref{eq:A}) as a bare tensor $A^{[0,0]}$.
The size of initial tensor network is set as $L_{\rm s}^{(\rm init)}=r+1$ for $r\in\{0,2,4,6\}$.
As shown in Fig.~\ref{sfig:init_tn_pure},
contractions are applied to the bare tensors $A^{[0,0]}$ 
using HOTRG-like algorithm repeatedly until only a single tensor $A^{[r,0]}$ remains.
We refer to this as HOTRG-like algorithm because the coarse-graining is applied to two different tensors
(see Fig.~\ref{fig:coarse-graining_init}), although the procedure is essentially the same as in the standard HOTRG algorithm.
To be concrete, a single coarse-graining step that produces the new tensor $A^{[s,0]}$
from the previously coarse-grained tensor $A^{[s-1,0]}$ is given by 
\be\label{eq:tensor_network_init_cont}
\begin{aligned}
A^{[s,0]}_{c_0a_1b_0a_0} &= 
\vcenter{\hbox{\includegraphics[height=3cm]{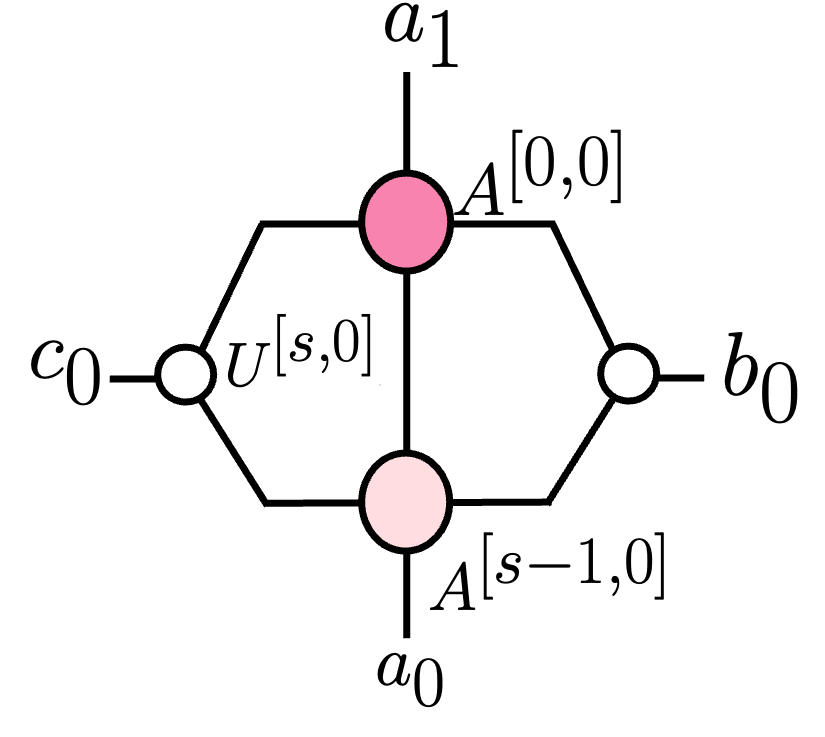}}}
\hspace{10mm}
(s=1,2,\ldots,r).
\end{aligned}
\ee
For $s=1$, $A^{[s-1,0]}$ corresponds to the bare tensor.
The tensor $U^{[s,0]}$ in Eq.~(\ref{eq:tensor_network_init_cont}) is an isometry 
that is computed from the EVD of the following self-adjoint matrix 
\be\label{eq:mat_for_isometry}
M_{(k_0k_1,k_0'k_1')}^{[s,0]}=
\vcenter{\hbox{\includegraphics[height=3cm]{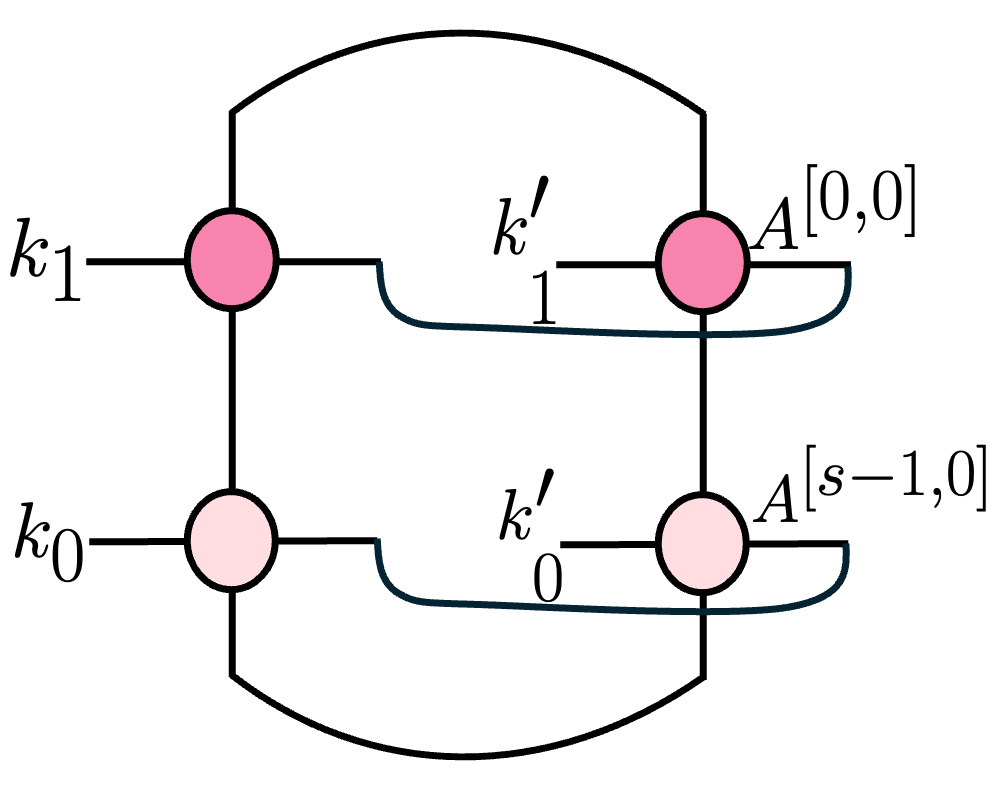}}}.
\ee
Applying EVD and truncating the eigenvalues of $M_{(k_0k_1,k_0'k_1')}^{[s,0]}$ yield
\be\label{eq:isometry}
M_{(k_0k_1,k_0'k_1')}^{[s,0]}\approx\sum_{c_0=0}^{\chi} \vcenter{\hbox{\includegraphics[height=2cm]{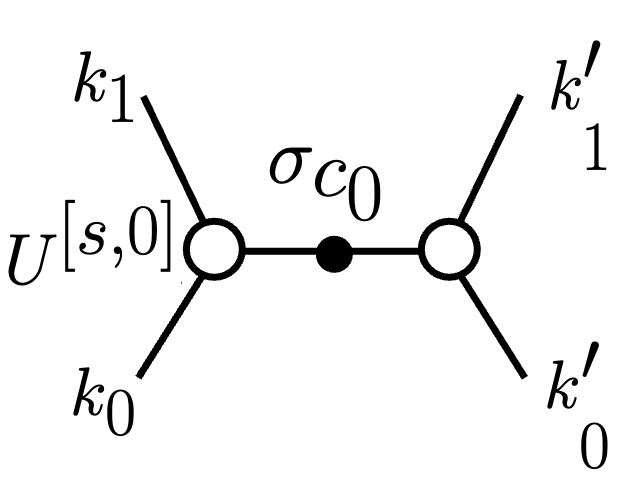}}}
\ee
where $U^{[s,0]}$ is the isometry and $\sigma_{c_0}$ are the eigenvalues of $M$
that are truncated up to the cut-off bond dimension $\chi$.
See \cite{PhysRevB.86.045139} for a complete description of the isometry.

%%%%%%%%%
\begin{figure}[t!]
\begin{center}
\begin{tabular}{c}
\includegraphics[width=12cm,pagebox=cropbox,clip]{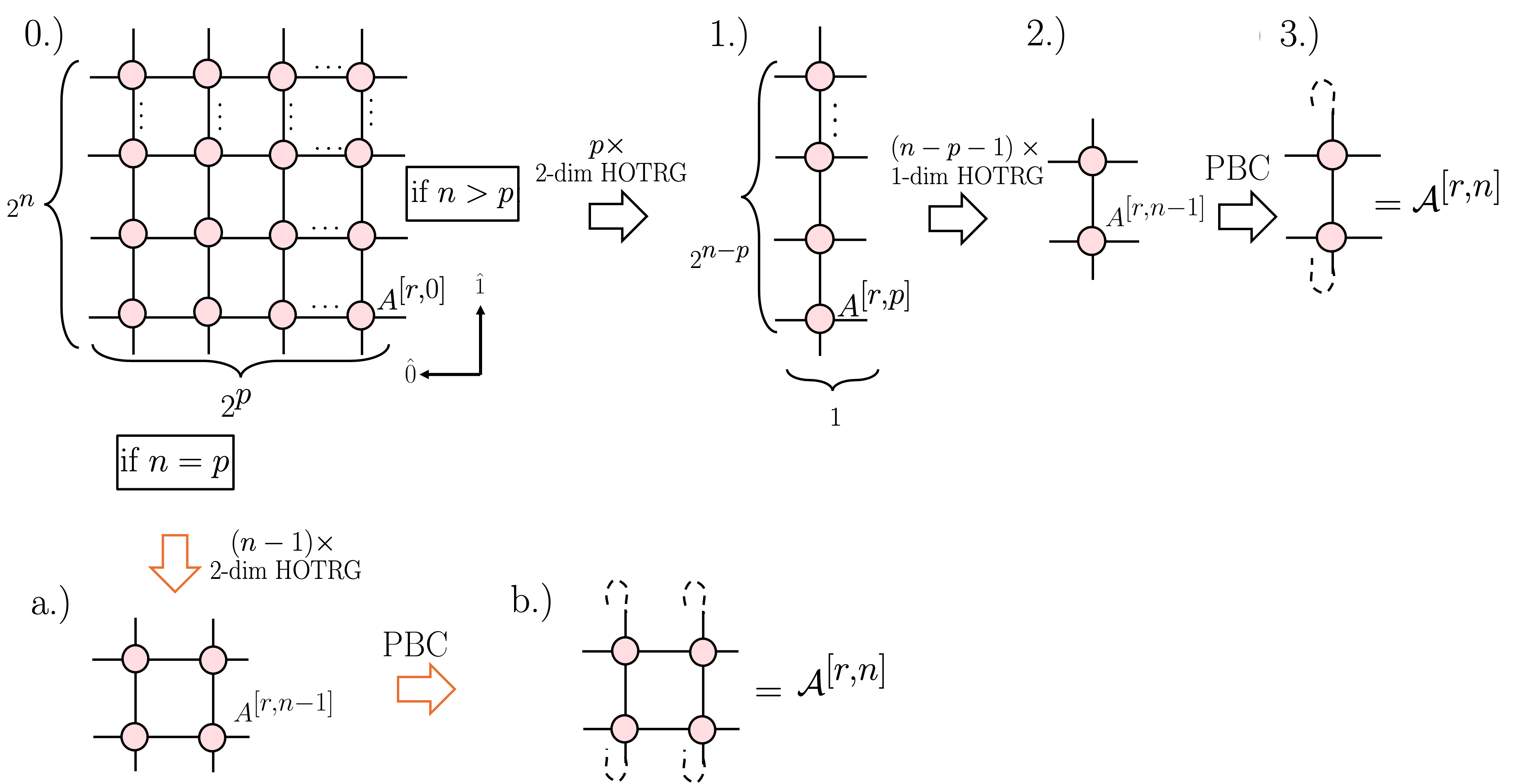}
\end{tabular}
\end{center}
\caption{
The coarse-graining procedure of the main tensor network using HOTRG. 
The procedure 0.)-3.) correspond to the case $n>p$, while the procedures 0.)-b.) apply to the case $n=p$.
See the main text for detail descriptions.
}
\label{fig:coarse-graining}
\end{figure}

After completing the coarse-graining of the initial tensor network with size $L_{\rm s}^{(\rm init)}$, we obtain the tensor $A^{[r,0]}$.
As mentioned earlier, we embed this tensor into the main tensor network, which then consists of $2^p\times 2^n$ tensors $A^{[r,0]}$, as illustrated in Fig.~\ref{fig:coarse-graining}.
Accordingly, the total size of the main tensor network is $L_{\rm t}\times L_{\rm s}=2^p\times (L_{\rm s}^{(\rm init)}\times 2^n)$.
The coarse-graining of this network is performed using the standard HOTRG algorithm \cite{PhysRevB.86.045139}, where the new tensor $A^{[r,i]}$ for $i=1,2,\ldots,n$ is obtained from previously coarse-grained tensors $A^{[r,i-1]}$ by coarse-graining for the time and spaces direction.

An important point to note is that, $p$ and $n$
do not only determine the size of tensor network, but
also represent the number of coarse graining steps in the time and spatial directions, respectively. These values are chosen such that
\be
n\geq p. 
\ee
As illustrated in Fig.~\ref{fig:coarse-graining}, 
when $n>p$, the coarse graining procedure is as follows: 1.) Apply $p$ coarse-graining iterations in both directions
using two-dimensional HOTRG. 2.) Continue with $(n-p-1)$ coarse-graining iterations in the spatial direction using the one-dimensional HOTRG. 3.) Perform an exact contraction to the last two remained tensors $A^{[r,n-1]}$ with the periodic boundary condition (PBC), obtaining
\be\label{eq:exact_cont}
\left({\cal A}^{L_{\rm t}}\right)_{kj}\approx
\sum_{l_0,l_1}{A}^{[r,n-1]}_{k_0l_0j_0l_1}{A}^{[r,n-1]}_{k_1l_1j_1l_0}\eqqcolon{\cal A}^{[r,n]}_{k_0k_1,j_0j_1}.
\ee
Meanwhile, the coarse graining procedure when $n=p$ is as follows: a.) Apply $(n-1)$ coarse graining iterations in both directions using the two-dimensional HOTRG. b.) Perform an exact contraction
to the last four remained tensors with PBC, resulting in
\be\label{eq:exact_cont_2}
\left({\cal A}^{L_{\rm t}}\right)_{kj}\approx
\sum_{l_0,l_1,m_0,m_1,n_0,n_1}{A}^{[r,n-1]}_{k_0l_0n_0l_1}{A}^{[r,n-1]}_{k_1l_1n_1l_0}{A}^{[r,n-1]}_{n_0m_0j_0m_1}{A}^{[r,n-1]}_{n_1m_1j_1m_0}\eqqcolon{\cal A}^{[r,n]}_{k_0k_1,j_0j_1}.
\ee
Subsequently, we apply the EVD to ${\cal A}^{[r,n]}$ as
\be\label{eq:pure_tensor_coarsegrained}
{\cal A}^{[r,n]}=W^{[r,n]}\lambda^{[r,n]}W^{[r,n]\dagger},
\ee
where $W^{[r,n]}$ is a unitary matrix that approximates $W$ in Eq.~(\ref{eq:EVD_A}) and $\lambda^{[r,n]}$ denotes the numerical eigenvalues.
These are related to the eigenvalues of $\cal A$ in Eq.~(\ref{eq:EVD_A}) by
\be\label{eq:energy_hotrg_eigenval}
\lambda_a\approx \left(\lambda_a^{[r,n]}\right)^{1/L_{\rm t}}.
\ee
Finally, the estimated energy spectrum is given by
\begin{equation}\label{egaptn}
\omega_a\approx\frac{1}{L_{\rm t}}\log\left(\frac{\lambda_0^{[r,n]}}{\lambda_a^{[r,n]}}\right)
\eqqcolon
\omega_a^{[{\rm hotrg}]}
\,\,\,\,
\mbox{ for } a=1,2,3,\hdots.
\end{equation}

%%%%%%%%%
\begin{figure}[t!]
\begin{center}
\begin{tabular}{c}
\includegraphics[width=12cm,pagebox=cropbox,clip]{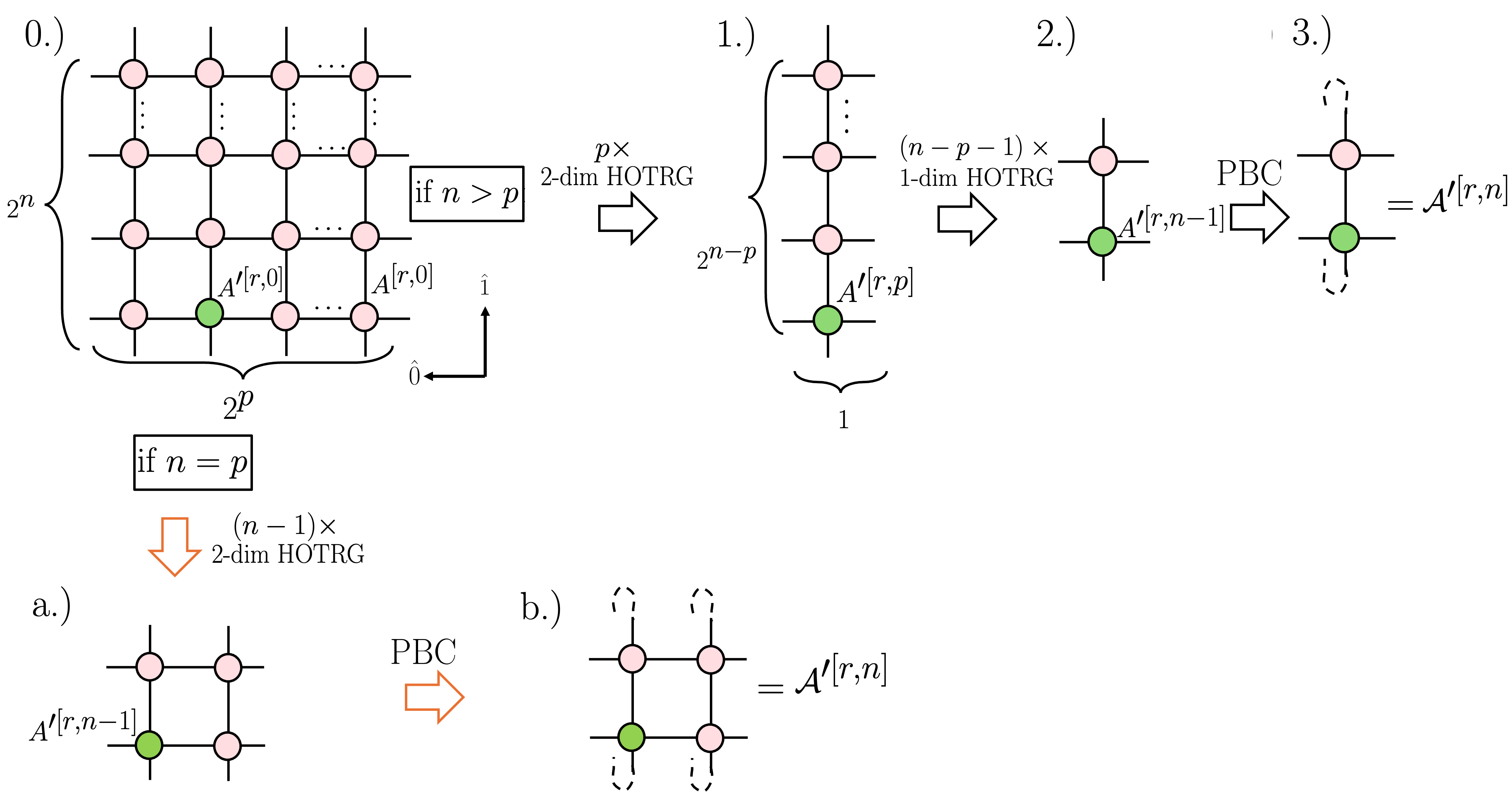}
\end{tabular}
\end{center}
\caption{
The coarse-graining procedure for the impurity tensor network ${\cal A}^{m-1}{\cal A}'{\cal A}^{m}$ with $m=2^{p-1}$. 
Procedure 0.)-3.) is for $n>p$ while procedure 0.)-b.) is for $n=p$.
}
\label{fig:coarse-graining_imp}
\end{figure}
%%%%%%%%%%%%%

Next, we explain the coarse graining procedure for the impurity tensor network 
${\cal A}^{m-1}{\cal A}'{\cal A}^{m}$ in Eq.~(\ref{eq:matrix_elements}).
Similar to the pure tensor case, 
we begin by preparing the initial impurity tensor network.
The bare impurity tensor is denoted by $A^{'[0,0]}=A'$.
To obtain the $s$-th coarse-grained impurity tensor $A^{'[s,0]}$, we perform 
\be\label{eq:tensor_network_init_cont_2}
A'^{[s,0]}_{c_0a_1b_0a_0}=
\vcenter{\hbox{\includegraphics[height=3cm]{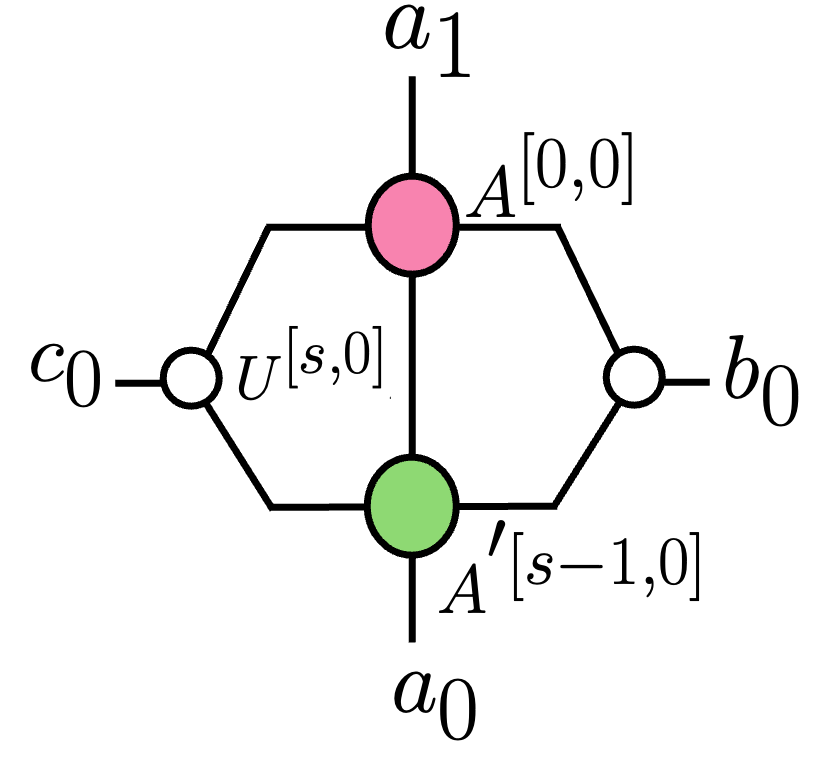}}},
\hspace{10mm}
(s=1,2,\ldots, r)
\ee
where $A^{'[s-1,0]}$ is the previously coarse grained impurity tensor,
and $U^{[s,0]}$ is the same isometry with the pure tensor case obtained from Eqs.~(\ref{eq:mat_for_isometry}-\ref{eq:isometry}).
After $r$ coarse-graining iterations, we obtain $A'^{[r,0]}$,
which is then inserted to the main impurity tensor network,
as shown in Fig.~\ref{fig:coarse-graining_imp}. 
The new tensor $A'^{[r,i]}$ in this main impurity tensor network is obtained
from the previous tensors $A'^{[r,i-1]}$ and $A^{[r,i-1]}$.
As in the pure tensor case,
when $n>p$, the coarse graining of the impurity tensor network is carried out according to the procedure 0.)-3.) shown in Fig.~\ref{fig:coarse-graining_imp}. In the last step, we perform an exact contraction for the remaining one pure tensor and one impurity tensor along with PBC,
and obtain the estimate of ${\cal A}^{m-1}{\cal A}'{\cal A}^{m}$, that is
\be
({\cal A}^{m-1}{\cal A}'{\cal A}^{m})_{kj}\approx 
\sum_{l_0,l_1}{A}'^{[r,n-1]}_{k_0l_0j_0l_1}{A}^{[r,n-1]}_{k_1l_1j_1l_0}\eqqcolon{\cal A}'^{[r,n]}_{k_0k_1,j_0j_1}.
\ee
On the other hand, when $n=p$, we follow the procedure 0.)-b.).
In the final step, an exact contraction of one impurity tensor and three pure tensors is performed to obtain
\be\label{eq:imp_exact_cont_2}
({\cal A}^{m-1}{\cal A}'{\cal A}^{m})_{kj}\approx 
\sum_{l_0,l_1,m_0,m_1,n_0,n_1}{A}^{'[r,n-1]}_{k_0l_0n_0l_1}{A}^{[r,n-1]}_{k_1l_1n_1l_0}{A}^{[r,n-1]}_{n_0m_0j_0m_1}{A}^{[r,n-1]}_{n_1m_1j_1m_0}\eqqcolon{\cal A}'^{[r,n]}_{k_0k_1,j_0j_1}.
\ee
After the building blocks of matrix elements have been estimated, we compute the approximate matrix elements given in Eq.~(\ref{eq:matrix_elements}), using the following expression
\be\label{eq:approx_Bba}
\langle b|\hat{\mathcal{O}}_q|a\rangle\approx\left(\left(\lambda^{[r,n]}\right)^{-(m-1/2)/L_{\rm t}}W^{[r,n]\dagger}{\cal A}'^{[r,n]}
W^{[r,n]}\left(\lambda^{[r,n]}\right)^{-(m+1/2)/L_{\rm t}}\right)_{ba}
\eqqcolon\langle b|\hat{\mathcal{O}}_q|a\rangle^{[\rm hotrg]}.
\ee
%%%%%%%%%%%%%%%%%%%%%%%%%%%%%%%%%%%%%%%%%
\section{Numerical results}
\label{Sec:num_res}
In this section, we demonstrate the spectroscopy technique to investigate the multi-particles states of the (1+1)d Ising model under PBC with no external magnetic field.
The partition function of the model is given by
\be
Z=\sum_{\{s\}}e^{-\frac{1}{T} H}
\ee
where $T$ is the temperature, $H$ is the Hamiltonian of the system that is given by
\be
H[s]=-J\sum_{\bm{r}\in\Gamma}\sum_{\mu=0}^{1}s(\bm{r}+\hat{\mu})s(\bm{r}),
\ee
where $s=\pm 1$ and $J$ is set to be unity.
The initial bare tensor $A^{[0,0]}$ of this model can be written as \cite{PhysRevD.88.056005}
\be
A^{[0,0]}_{abcd}=\sqrt{\sigma_a\sigma_b\sigma_c\sigma_d}\sum_{s=\pm 1} (u^{\dagger})_{as}
(u^{\dagger})_{bs}u_{sc}u_{sd}
\ee
where $u$ and $\sigma$ are obtained from the SVD of the Boltzmann factor in the partition function, that is
\be
e^{\frac{1}{T} s's}=\sum_{k=1}^2u_{s'k}\sigma_k(u^{\dagger})_{ks}.
\ee
The details can be found in Ref.~\cite{PhysRevD.110.034514}.
In this work, the temperature of the system is fixed at $T=2.44$, which lies in the symmetric phase.
\subsection{Energy spectrum}
\label{subsec:energy_spectrum}
%%%%%%%%%
\begin{figure}[t!]
\centering
\begin{subfigure}[b]{0.4\textwidth}
\centering
\includegraphics[width=8cm,height=6cm]{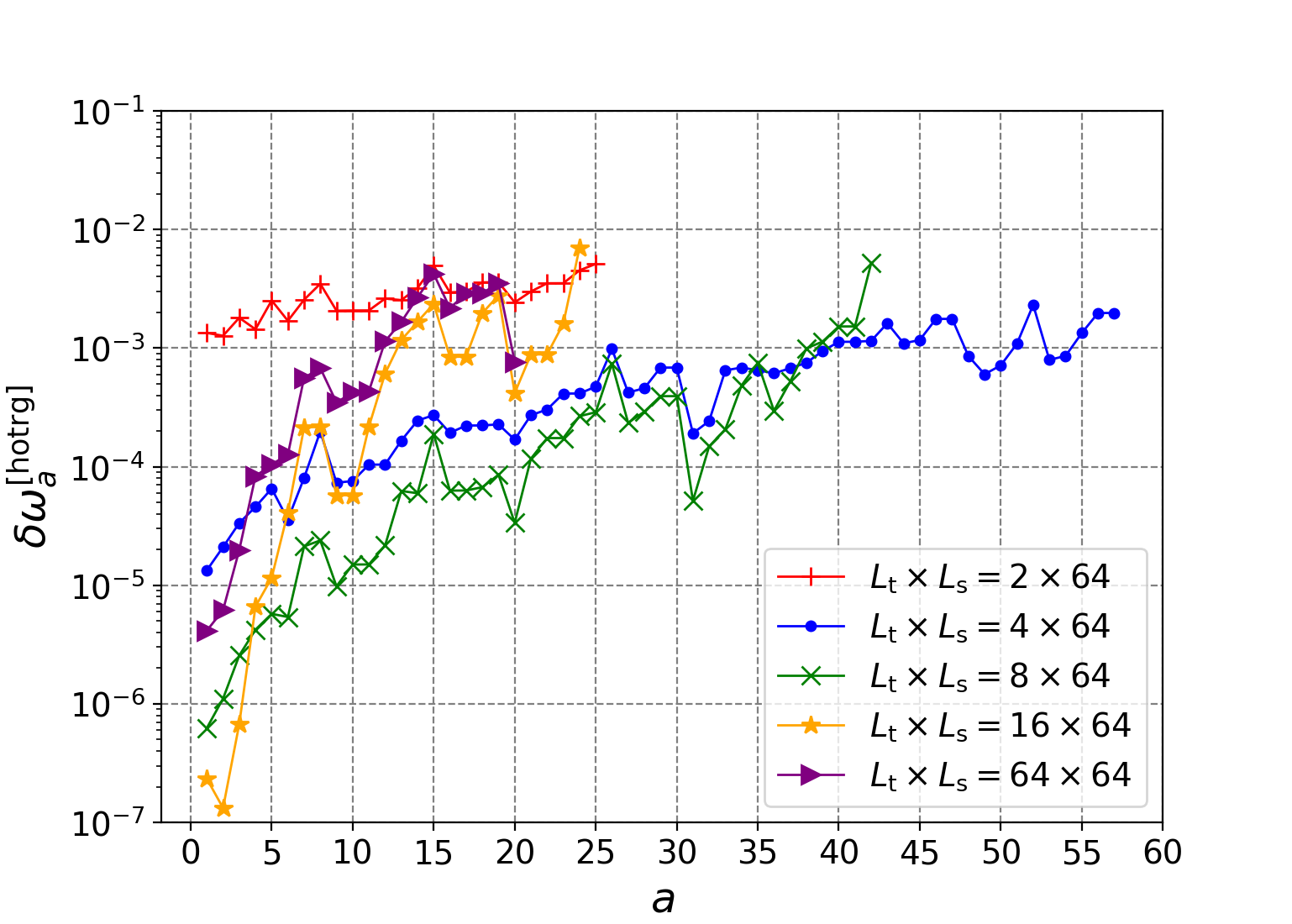}
\caption{}
\label{sfig:rel-error}
\end{subfigure}\hspace{10mm}
\begin{subfigure}[b]{0.4\textwidth}
\centering
\includegraphics[width=8cm,height=6cm]{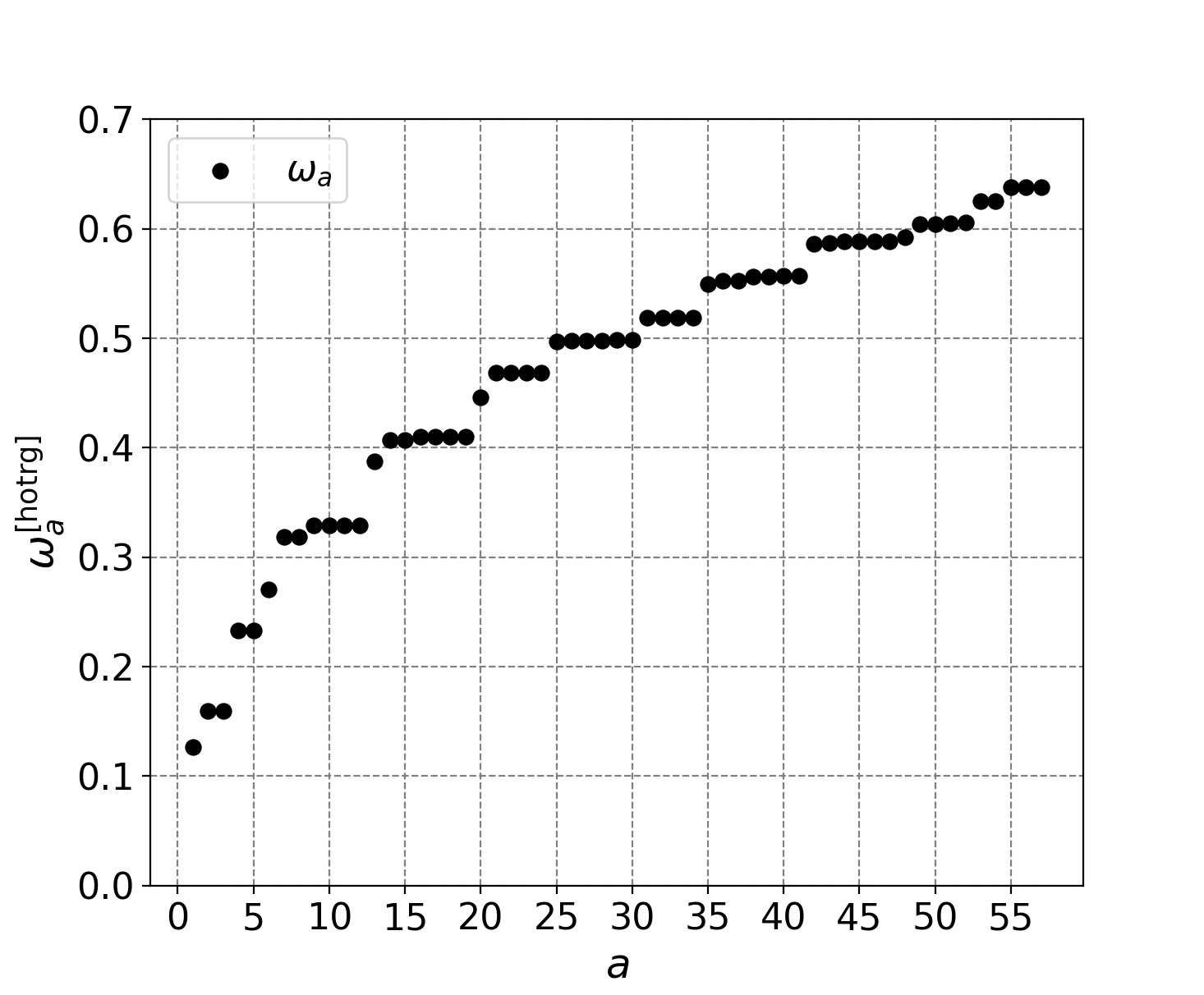}
\caption{}
\label{sfig:energy}
\end{subfigure}
\caption{
(a) The relative error of the energy spectrum $\omega_a^{[\rm hotrg]}$ 
for system size $L_{\rm s}=64$ computed with $\chi=80$ and $L_{\rm t}=2,4,8,16,64$.
(b) The energy spectrum $\omega_a^{[\rm hotrg]}$ for system size $L_{\rm s}=64$,
computed using HOTRG with $\chi=80$ and $L_{\rm t}=4$.
}
\label{fig:err_energy_spectrum_l64}
\end{figure}
%%%%%%%%%

%%%%%%%%%
In \cite{PhysRevD.110.034514},
a square tensor network was employed
for all calculations.
Using this setup, we were able to investigate the spectrum up to system size $L_{\rm s} = 64$, where the highest excited state for which the quantum numbers are correctly identified corresponds to eigenstate number $a = 20$, and the error in the energy spectrum is of order $O(10^{-3})$. 
In the present work, we slightly modify the coarse-graining strategy to improve the accuracy of the calculations, thereby enabling access to higher eigenstates and larger system sizes.
Specifically, we compute the spectrum at fixed $L_{\rm s}$  while varying system size in time direction $L_{\rm t}=2,4,\ldots,L_{\rm s}$,
where $L_{\rm t}=L_{\rm s}$ correspond to a square tensor network.
The case $L_{\rm t}=1$ is excluded, as it was demonstrated in Ref.~\cite{PhysRevD.110.034514}
that this choice leads to large numerical error.

%%%%%%%%%%%%%
\begin{figure}[t!]
\centering
\includegraphics[width=8cm,height=6cm]{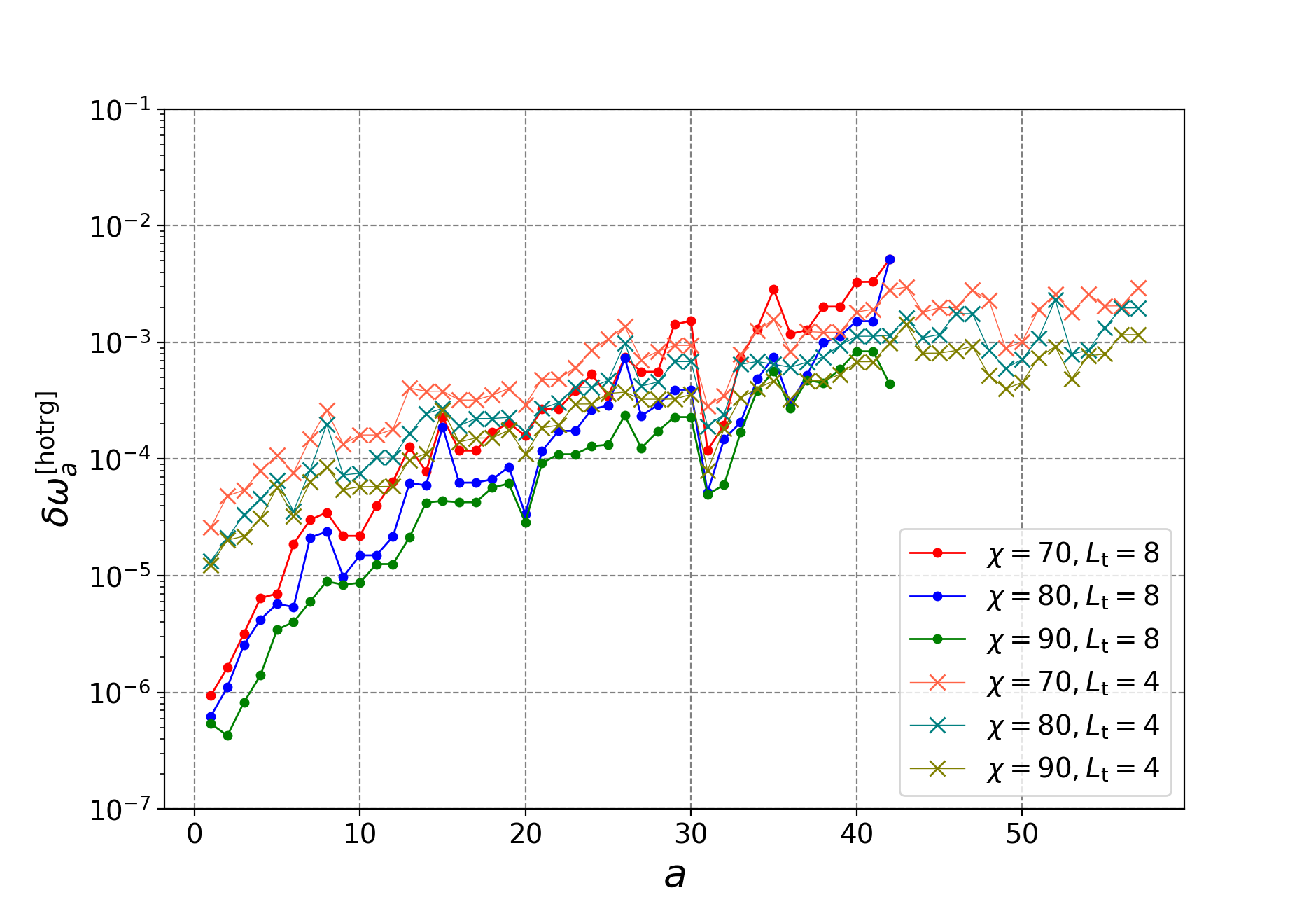}
\caption{Relative errors of energy spectrum for $L_{\rm s}=64$, computed with $\chi=70,80,90$
and $L_{\rm t}=4,8$.}
\label{fig:rel_err_dcut}
\end{figure}
%%%%%%%%%%%%%%%

We compute $\omega_a^{[\rm hotrg]}$ using Eq.~(\ref{egaptn}) for $L_{\rm s}=64$
and $L_{\rm t}=2,4,8,16,64$, following the procedure introduced in Sec.~\ref{subs:coarse-graining}.
As shown in Fig.~\ref{sfig:rel-error},
the tensor network with $L_{\rm t}=4,8$ 
yields smaller relative error, 
$\delta\omega_a= \left| \frac{\omega_a^{[\rm hotrg]}-\omega_a^{[\rm exact]}}{\omega_a^{[\rm exact]}}\right |$,  
compared to other choices $L_{\rm t}=2,16,64$.
Here, the exact spectrum $\omega_a^{[\rm exact]}$ is computed
using Kaufman's formula \cite{ PhysRev.76.1232},
see the appendix of Ref.~\cite{PhysRevD.110.034514} for the concise summary.
There are two main reasons why $L_{\rm t}=4,8$ produce relatively smaller errors compared to the other choices $L_{\rm t}=2,16,64$. 
First, the number of the coarse-graining steps required 
for $L_{\rm t}=4,8$ is smaller than for $L_{\rm t}=16,64$
resulting in smaller accumulated coarse-graining error.
Second, the eigenvalues of the transfer matrix tend to be more degenerate for smaller $L_{\rm t}$, particularly when $L_{\rm s}$ is large,
which leads to large errors after each coarse-graining step.
This explains why $L_{\rm t}=4,8$ yield smaller errors compared to $L_{\rm t}=2$.
Furthermore, when comparing the results for $L_{\rm t}=4$ and $L_{\rm t}=8$,
we observe that the low-lying eigenstates are determined more accurately for $L_{\rm t}=8$,
whereas the higher eigenstates exhibit smaller error for $L_{\rm t}=4$.
Since each choice, $L_{\rm t}=4,8$ has its own advantages and disadvantages,
we employ both in the computations presented in this paper.
In Fig.~\ref{sfig:energy}, we present the energy spectrum $\omega_a^{[\rm hotrg]}$ of the
system for $L_{\rm s}=64$ computed using $\chi=80$
and $L_{\rm t}=4$.
In principle, our scheme allows the extraction of $\omega_a^{[\rm hotrg]}$ up to $a=\chi^2$. However, we only present the result for which the numerical errors are in order $O(10^{-3})$.
For the case $L_{\rm s}=64$, $L_{\rm t}=4$ and $\chi=80$, these correspond to the eigenstates $a=0-57$.

Next, we examine how the relative error 
of the energy spectrum changes over different cut-off bond dimension.
We compute the energy spectrum for $L_{\rm s}=64$ and $L_{\rm t}=4,8$,
using $\chi=70,80,90$, 
and present the results in Fig.~\ref{fig:rel_err_dcut}.
From the figure, we clearly observe that the relative error decreases as $\chi$ increases for both $L_{\rm t}=4,8$, as expected.

\subsection{Quantum number}
\label{subsec:quantum_number}
One of the quantum numbers for (1+1)d Ising model
is $q=\{+1,-1\}$ which arises from the internal $\mathbb{Z}_2$ symmetry of the spin field.
To classify the eigenstates according to this quantum number,
we compute matrix elements of a spin field operator $\hat{s}_x$ (we use $\hat{s}_0$ at $x=0$ in practice), 
namely $\langle \Omega|\hat{s}_x|a\rangle$ where 
$\langle\Omega|$ is the ground state
whose quantum number is $q_{\Omega}=+1$.
From the selection rule of the discrete symmetry given in Eq.~(\ref{eq:selection_rule}),
 if $\langle \Omega|\hat{s}_0|a\rangle\neq 0$,
then the eigenstate $|a\rangle$ has a quantum number $q_a=-1$.
To estimate the matrix elements,
we coarse grain the corresponding impurity tensor network for $L_{\rm s}=64$ and $L_{\rm t}=4,8$ (with the impurity placed at $m=2,4$, respectively), using $\chi=80$, and subtitute the result into Eq.~(\ref{eq:approx_Bba}).
%%%%%%%%%
\begin{figure}[t!]
\centering
\begin{subfigure}[b]{0.4\textwidth}
\centering
\includegraphics[width=7cm,height=8.5cm]{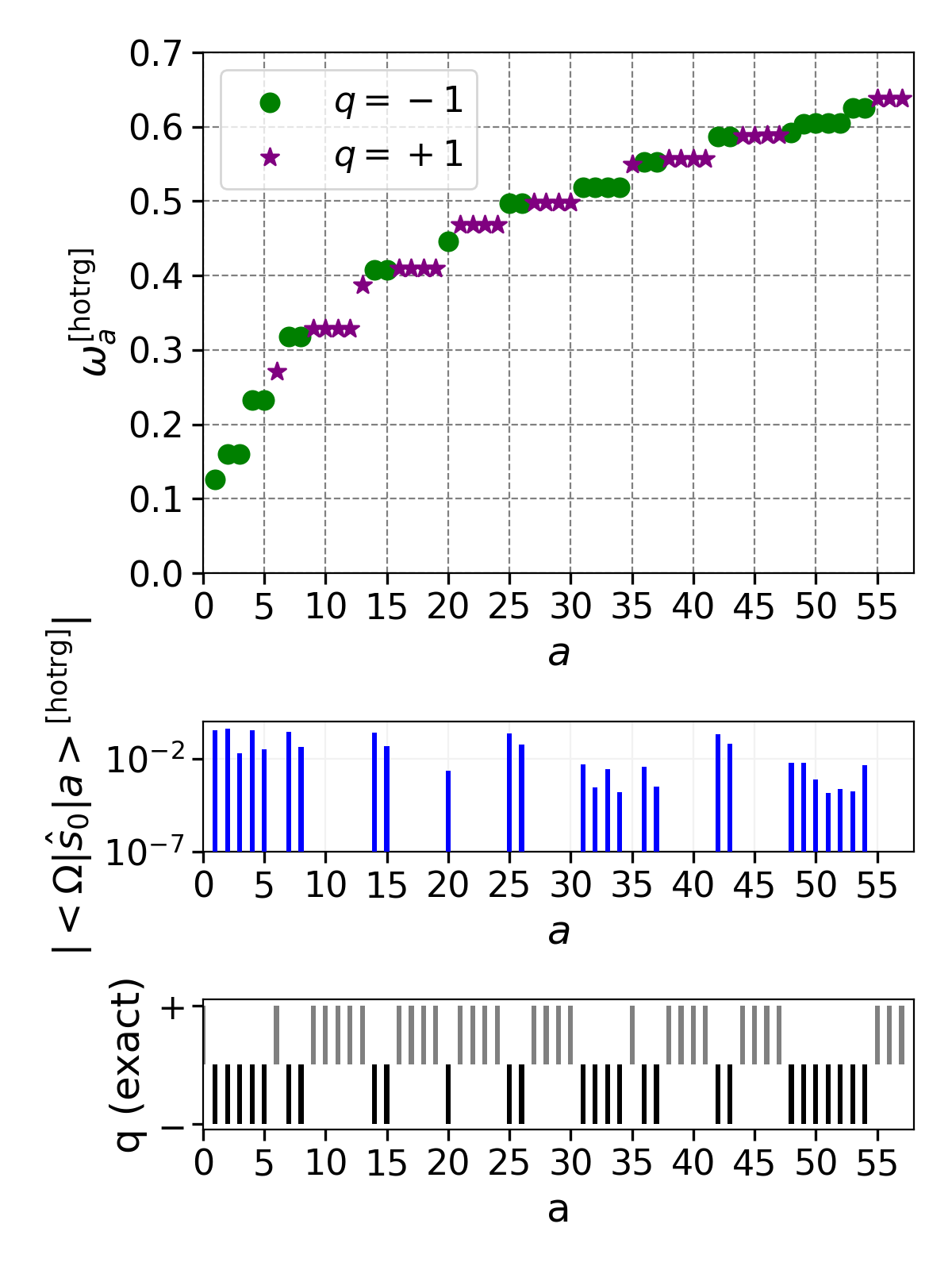}
\caption{}
\label{sfig:spec_l64_lt4}
\end{subfigure}\hspace{20mm}
\begin{subfigure}[b]{0.4\textwidth}
\centering
\includegraphics[width=7cm,height=8.5cm]{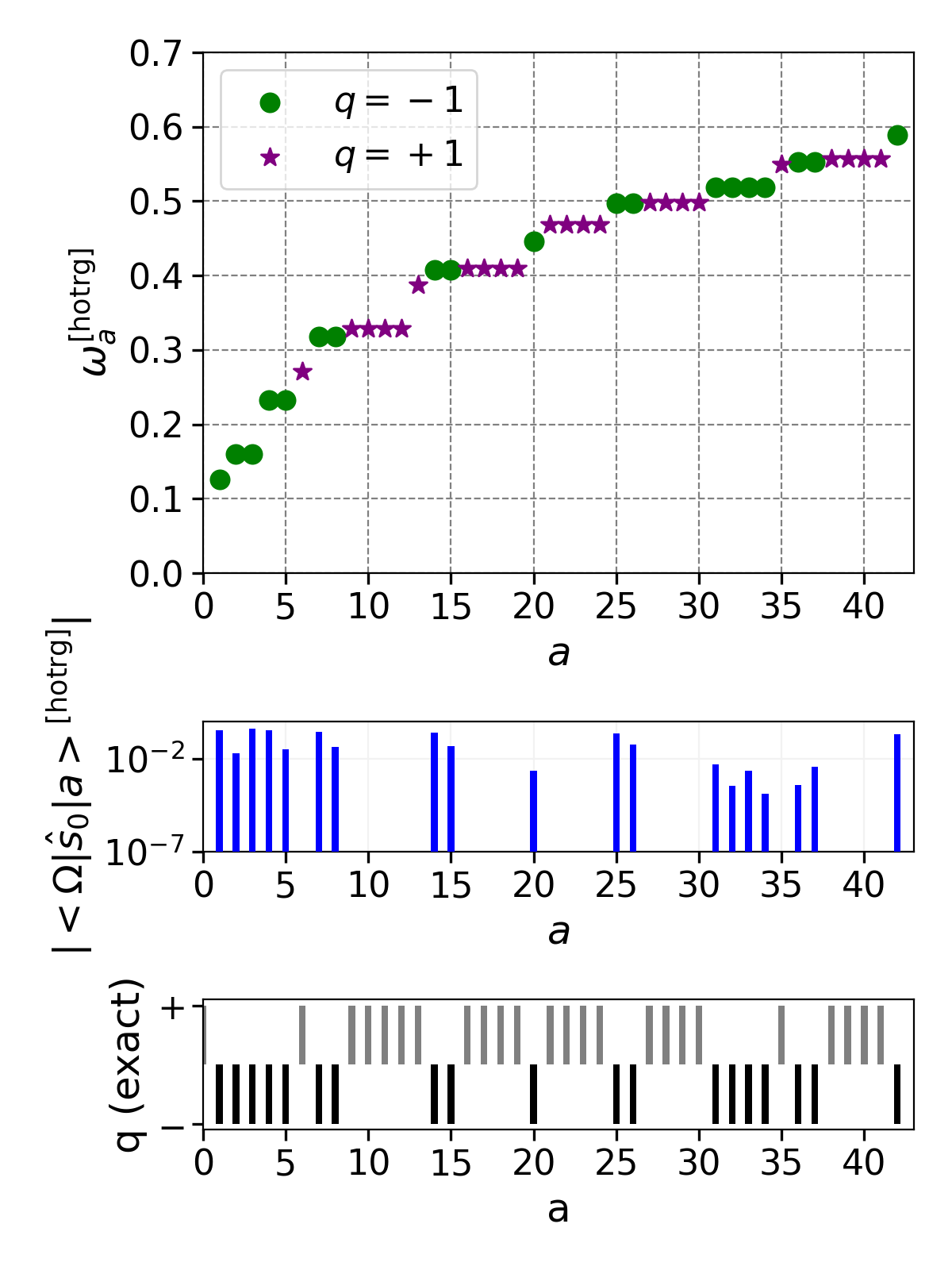}
\caption{}
\label{sfig:spec_l64_lt8}
\end{subfigure}
\caption{
Energy spectrum $\omega_a^{[\rm hotrg]}$ and absolute value of matrix elements $|\langle\Omega|\hat{s}_0|a\rangle^{[\rm hotrg]}|$
for quantum number classification, together with the exact quantum number as comparison for $L_{\rm s}=64$
computed with $\chi=80$ and using (a) $L_{\rm t}=4$ and (b) $L_{\rm t}=8$.}
\label{fig:energy_spectrum_l64}
\end{figure}
We present the energy spectrum $\omega_a^{[\rm hotrg]}$
previously shown in Fig.~\ref{sfig:energy} with the additional corresponding quantum number in the upper panel of Fig.~\ref{fig:energy_spectrum_l64}.
The quantum number classification is based on $|\langle\Omega|\hat{s}_0|a\rangle^{[\rm hotrg]}|$ 
shown in the middle panel.
Meanwhile, the bottom panel displays the exact quantum number for comparison, 
computed following Ref.~\cite{PhysRev.76.1232}.
From the figures, using the matrix elements obtained with  $L_{\rm t}=4$, we can correctly identify the quantum number for $L_{\rm s}=64$ up to $a= 57$,
whereas those obtained with $L_{\rm t}=8$ 
are accurate only up to $a=42$. 
In general, these results indicate that, by using tensor network with $L_{\rm t}=4,8$, 
we are able to identify more higher-energy eigenstates correctly compared to the previous results reported in Refs.~\cite{PhysRevD.110.034514,Az-zahra:2024pqa} using square lattice.

%%%%%%%%%%%
\subsection{Momentum}
\label{subsec:momentum}
\begin{figure}[t!]
\centering
\begin{subfigure}[b]{0.4\textwidth}
\centering
\includegraphics[width=6cm,height=2cm]{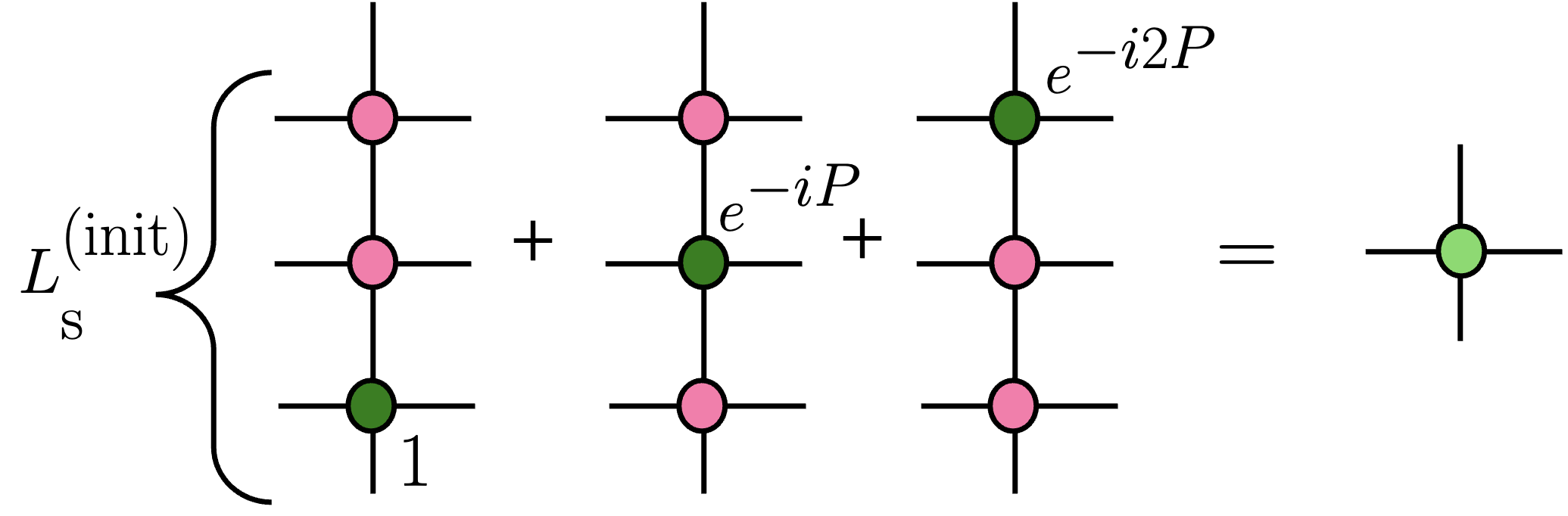}
\caption{}
\label{sfig:momentum_TN_init}
\end{subfigure}
\begin{subfigure}[b]{0.4\textwidth}
\centering
\includegraphics[width=3.5cm,height=2cm]{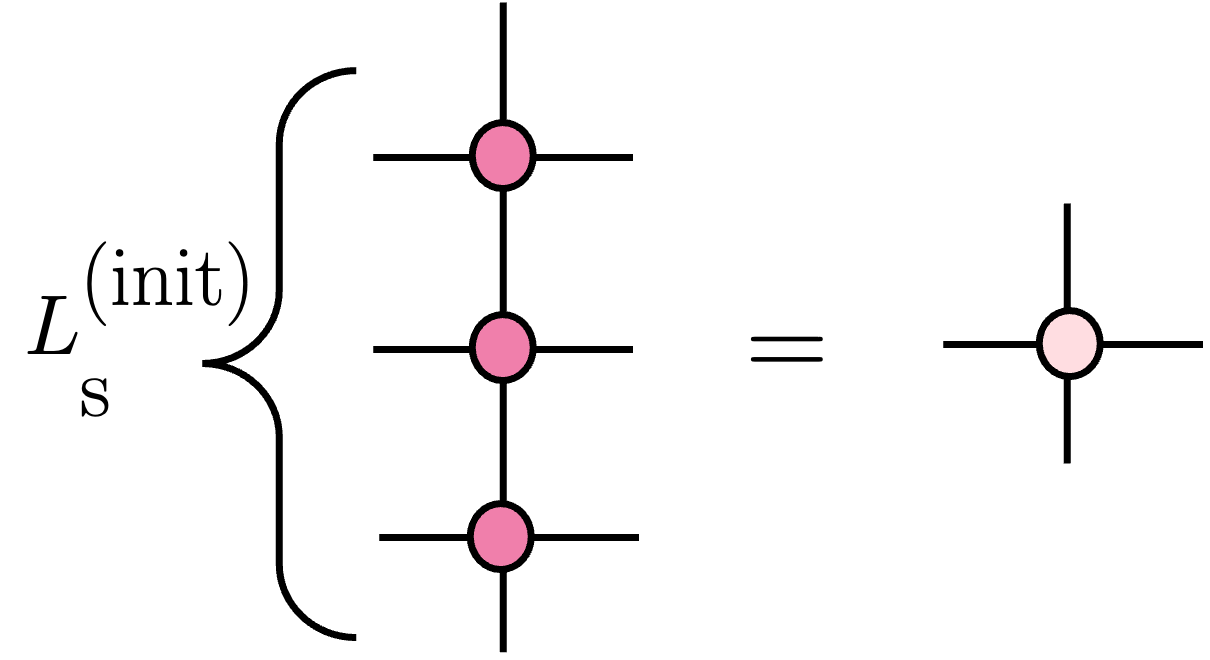}
\caption{}
\label{sfig:pure_TN_init}
\end{subfigure}
\begin{subfigure}[b]{1\textwidth}
\centering
\includegraphics[width=12cm,height=3cm]{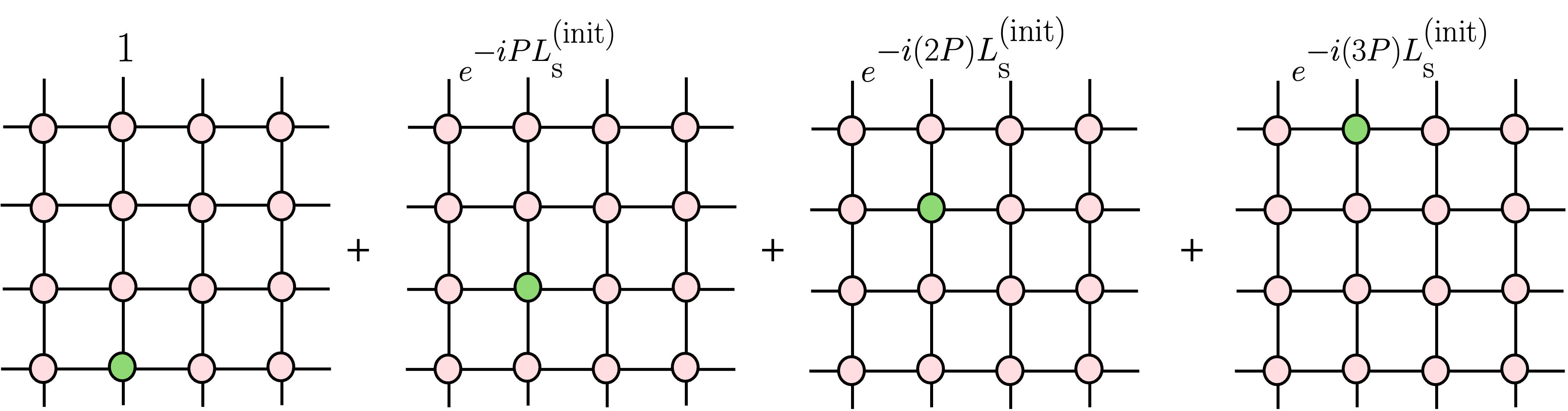}
\caption{}
\label{sfig:momentum_TN}
\end{subfigure}
\caption{(a) The initial tensor network representation with size $L_{\rm s}^{(\rm init)}=3$ for $\hat{\mathcal{O}}_1(P)$ denoted by $A'^{[2,0]}$ (light green circle) which is
constructed from one bare impurity tensor $A'^{[0,0]}$ (dark green circle) with a proper momentum factor, and two bare pure tensors (dark pink circle).
(b) Initial pure tensor network (light pink circle) constructed from three bare pure tensors.  
(c) The main impurity tensor network with total size $L_{\rm s}=L_{\rm s}^{(\rm init)}\times 4$ for 
computing $\langle \Omega|\hat{\mathcal{O}}_1(P)|a\rangle$.}
\label{fig:momentum_TN}
\end{figure}

The momentum of the eigenstate $|a\rangle$ in $q=-1$ sector can be identified
by computing the matrix elements of an operator with momentum $P$ 
\be
\langle \Omega|\hat{\mathcal{O}}_1(P)|a\rangle,
\ee
where
\be\label{eq:momentum_op_1p}
\hat{\mathcal{O}}_1(P)=\frac{1}{L_{\rm s}}\sum_{x=0}^{L_{\rm s}-1}\hat{s}_x e^{-iPx}
\ee
 and $P=2\pi n/L_{\rm s}$ is the discrete momentum with $n=0,1,2,\ldots,L_{\rm s}-1$.
The selection rule for momentum states that
for a fixed $P$, if the following condition 
\be\label{eq:selection_rule_qmin}
\langle \Omega|\hat{\mathcal{O}}_1(P)|a\rangle \approx  \langle \Omega|\hat{\mathcal{O}}_1(P)|a\rangle^{[\rm hotrg]}\neq 0
\ee
is satisfied, then the eigenstate $|a\rangle$ carries the momentum $P$.
In Fig.~\ref{fig:momentum_TN}, we show the impurity tensor network diagram corresponding to the operator $\hat{\mathcal{O}}_1(P)$.
These networks are then coarse-grained to obtain $\langle \Omega|\hat{\mathcal{O}}_1(P)|a\rangle^{[\rm hotrg]}$.
%%%%%%%%%
%%%%%%%%%

\begin{table}[t!]
\begin{center}
\caption{The absolute value of matrix elements 
$| \langle\Omega|\hat{\mathcal{O}}_1(P)|a\rangle^{[\rm hotrg]}|$
with $|P|=0\mbox{--}12\pi/L_{\rm s}$ and $L_{\rm s}=64$ computed using $\chi=80$ and $L_{\rm t}=4$ for eigenstate $|a\rangle$ in $q=-1$ sector.}
\label{Tab:spec_momB_1pstate}
\begin{tabular}{|c|c|c|c|c|c|c|c|c|c|}
\hline
\diagbox{$a$}{$P$}&$0$ & $2\pi/L_{\rm s}$  & $4\pi/L_{\rm s}$ & $6\pi/L_{\rm s}$ & $8\pi/L_{\rm s}$ & $10\pi/L_{\rm s}$ & $12\pi/L_{\rm s}$ & $|P|$&$\omega_a^{[\rm hotrg]}$\\
\hline\hline
%%%%%%%
1&0.316100&$<10^{-6}$&0.000001&$<10^{-6}$&$<10^{-6}$&$<10^{-6}$&0.000001&0&0.126232\\
2&$<10^{-6}$&0.198509&$<10^{-6}$&0.000001&$<10^{-6}$&$<10^{-6}$&$<10^{-6}$&$2\pi/L_{\rm s}$&0.159791\\
3&$<10^{-6}$&0.198503&$<10^{-6}$&0.000009&$<10^{-6}$&0.000009&$<10^{-6}$&$2\pi/L_{\rm s}$&0.159793\\
4&0.000006&$<10^{-6}$&0.164112&$<10^{-6}$&0.000002&$<10^{-6}$&0.000004&$4\pi/L_{\rm s}$&0.232696\\
5&$<10^{-6}$&$<10^{-6}$&0.164106&$<10^{-6}$&0.000015&$<10^{-6}$&0.000012&$4\pi/L_{\rm s}$&0.2327\\
7&$<10^{-6}$&0.000006&$<10^{-6}$&0.139802&$<10^{-6}$&0.000011&$<10^{-6}$&$6\pi/L_{\rm s}$&0.31818\\
8&$<10^{-6}$&0.000010&$<10^{-6}$&0.139800&$<10^{-6}$&0.000056&$<10^{-6}$&$6\pi/L_{\rm s}$&0.318218\\
14&0.000002&$<10^{-6}$&0.000009&$<10^{-6}$&0.122905&$<10^{-6}$&0.000023&$8\pi/L_{\rm s}$&0.407404\\
15&$<10^{-6}$&$<10^{-6}$&0.000017&$<10^{-6}$&0.122902&$<10^{-6}$&0.000067&$8\pi/L_{\rm s}$&0.407415\\
20&0.002199&$<10^{-6}$&0.000005&$<10^{-6}$&0.000019&$<10^{-6}$&0.000008&0&0.445858\\
25&$<10^{-6}$&0.000003&$<10^{-6}$&0.000017&$<10^{-6}$&0.110530&$<10^{-6}$&$10\pi/L_{\rm s}$&0.497221\\
26&$<10^{-6}$&0.000002&$<10^{-6}$&0.000068&$<10^{-6}$&0.110537&$<10^{-6}$&$10\pi/L_{\rm s}$&0.497475\\
31&$<10^{-6}$&0.001747&$<10^{-6}$&0.000590&$<10^{-6}$&0.000003&$<10^{-6}$&$2\pi/L_{\rm s},6\pi/L_{\rm s}$&0.51878\\
32&$<10^{-6}$&0.000840&$<10^{-6}$&0.000777&$<10^{-6}$&0.000059&$<10^{-6}$&$2\pi/L_{\rm s},6\pi/L_{\rm s}$&0.518806\\
33&$<10^{-6}$&0.001749&$<10^{-6}$&0.000576&$<10^{-6}$&0.000194&$<10^{-6}$&$2\pi/L_{\rm s},6\pi/L_{\rm s}$&0.519017\\
34&$<10^{-6}$&0.002347&$<10^{-6}$&0.000303&$<10^{-6}$&0.000002&$<10^{-6}$&$2\pi/L_{\rm s},6\pi/L_{\rm s}$&0.519035\\
36&0.000015&$<10^{-6}$&0.001822&$<10^{-6}$&0.000002&$<10^{-6}$&0.000006&$4\pi/L_{\rm s}$&0.552576\\
37&$<10^{-6}$&$<10^{-6}$&0.001776&$<10^{-6}$&0.000001&$<10^{-6}$&0.000090&$4\pi/L_{\rm s}$&0.552611\\
42&0.000018&$<10^{-6}$&0.000016&$<10^{-6}$&0.000017&$<10^{-6}$&0.101022&$12\pi/L_{\rm s}$&0.586424\\
43&$<10^{-6}$&$<10^{-6}$&0.000004&$<10^{-6}$&0.000083&$<10^{-6}$&0.101045&$12\pi/L_{\rm s}$&0.586699\\
48&0.005158&$<10^{-6}$&0.000009&$<10^{-6}$&0.000055&$<10^{-6}$&0.000240&0&0.592084\\
49&0.000065&$<10^{-6}$&0.002497&$<10^{-6}$&0.000585&$<10^{-6}$&0.000219&$4\pi/L_{\rm s},8\pi/L_{\rm s}$&0.604513\\
50&$<10^{-6}$&$<10^{-6}$&0.001916&$<10^{-6}$&0.000979&$<10^{-6}$&0.000018&$4\pi/L_{\rm s},8\pi/L_{\rm s}$&0.604581\\
51&0.000014&$<10^{-6}$&0.001231&$<10^{-6}$&0.001279&$<10^{-6}$&0.000082&$4\pi/L_{\rm s},8\pi/L_{\rm s}$&0.604808\\
52&$<10^{-6}$&$<10^{-6}$&0.001949&$<10^{-6}$&0.000984&$<10^{-6}$&0.000440&$4\pi/L_{\rm s},8\pi/L_{\rm s}$&0.605546\\
53&$<10^{-6}$&0.002032&$<10^{-6}$&0.000050&$<10^{-6}$&0.000014&$<10^{-6}$&$2\pi/L_{\rm s}$&0.625633\\
54&$<10^{-6}$&0.002014&$<10^{-6}$&0.000051&$<10^{-6}$&0.000014&$<10^{-6}$&$2\pi/L_{\rm s}$&0.625672\\
\hline
\end{tabular}
\end{center}
\end{table}
Table~\ref{Tab:spec_momB_1pstate} lists 
the absolute values of the matrix elements $| \langle \Omega|\hat{\mathcal{O}}_1(P)|a\rangle^{[\rm hotrg]}|$ for the
system size $L_{\rm s}=64$ computed with $L_{\rm t}=4$ and $\chi=80$, 
which are used to identify momentum in $q=-1$ sector together with the selection rule in Eq.~\ref{eq:selection_rule_qmin}.
For this identification, we treat $| \langle \Omega|\hat{\mathcal{O}}_1(P)|a\rangle^{[\rm hotrg]}|<10^{-6}$ as zero.
However, some matrix elements take values of order $O(10^{-6})$ and $O(10^{-5})$.
To clarify the behavior of these values, we compute the matrix elements using a larger bond dimension, $\chi = 120$, and observe that these values decrease, indicating that they are not signals but instead arise from coarse-graining errors.

On the other hand, when the matrix elements are of order $O(10^{-4})$, we regard them as a signal if they remain consistent or increase for the larger bond dimension $\chi = 120$, and we regard them as an error if their values decrease.
As one example, eigenstates $a = 31$--$34$ have nonzero values of order $O(10^{-4})$ at three different absolute value total momenta, $|P| = 2\pi/L_{\rm s}$, $6\pi/L_{\rm s}$, and $10\pi/L_{\rm s}$. In this case, we conclude that the nonzero values appearing at $|P| = 10\pi/L_{\rm s}$ are fake signals, because the values decrease when $\chi = 120$.
With this selection rule,
we can clearly classify the momentum $|P|=0-12\pi/L_{\rm s}$ of the energy eigenstates when the states are non-degenerate and two-fold degenerate.
For four-fold degenerate energy, for instance, the corresponding eigenstates $a=31-34$ are associated with two different absolute total momenta, $|P|=2\pi/L_{\rm s}$ and $6\pi/L_{\rm s}$.
In this case, the eigenstates may be described by linear combination of these corresponding momenta. Accordingly, the operator associated with each momentum can have non-zero overlap with the four states, leading to non-zero matrix elements at both absolute momenta $|P|=2\pi/L_{\rm s},6\pi/L_{\rm s}$.

\begin{figure}[t!]
\centering
\includegraphics[width=10cm,height=8cm]{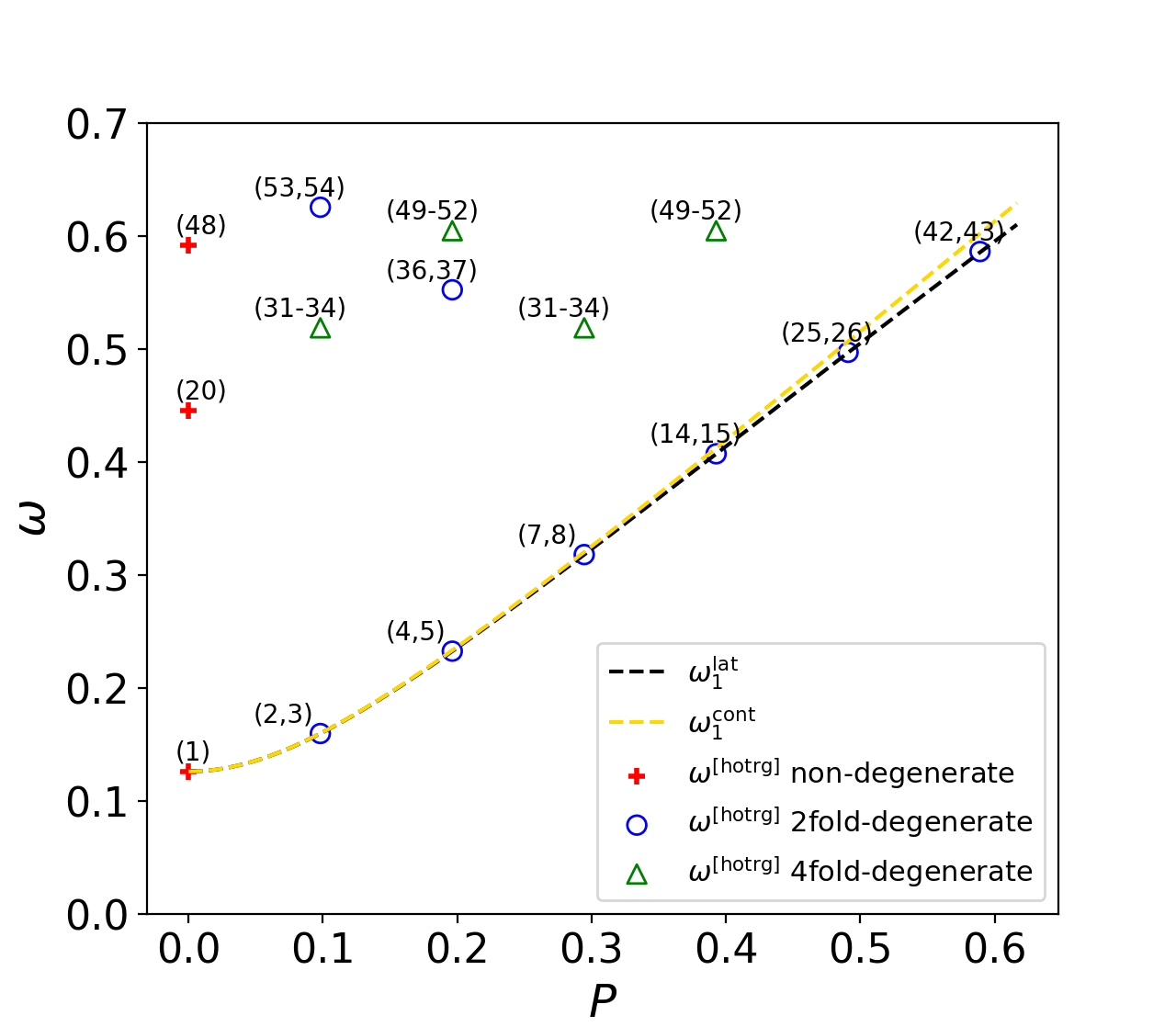}
\caption{The energy spectrum $\omega$ for $q=-1$ sector as a function of $P$ for $L_{\rm s}=64$
computed with $L_{\rm t}=4$ and $\chi=80$. The numbers in parentheses are the energy-level ordering, while the dashed lines are one-particle dispersion relation in Eq.~(\ref{eq:cont_disrel_1p}) and (\ref{eq:lat_disrel_1p}).}
\label{fig:momentum_disrel_p_l64}
\end{figure}

Next, 
we examine the relation between the numerical energy and momentum
and compare it with both continuum dispersion relation of one-particle state
\be\label{eq:cont_disrel_1p}
\omega_1^{\rm cont}=\sqrt{m^2+P^2}
\ee
and lattice dispersion relation
\be\label{eq:lat_disrel_1p}
\omega_1^{\rm lat}=\cosh^{-1}(1-\cos P+\cosh m).
\ee
Here, $m$ denotes the exact rest mass at $T=2.44$ in large volume limit, 
\be\label{eq:rest_mass}
m=0.12621870.
\ee
The number of particles of the eigenstates in $q=-1$ sector will be investigated in Sec.~\ref{subsec:numb_particles}. But here, by comparing with tensor network results and the one-particle dispersion relation, we roughly try to identify the one-particle state.
From such identification, among all eigenstates listed in Table~\ref{Tab:spec_momB_1pstate} for $L_{\rm s}=64$, state numbers $a=1,2,3,4,5,7,8,14,15,25,26,42,43$ are categorized as the one-particle state (see Fig.~\ref{fig:momentum_disrel_p_l64}).
On the other hand, it can be checked that the remaining states do not fit the one-particle dispersion relation (see Fig.~\ref{fig:momentum_disrel_p_l64}); in fact, they follow three-particle state dispersion relation, as will be explained in Sec.~\ref{sec:3p-states}.

\begin{figure}[t!]
\centering
\includegraphics[width=14cm,height=10cm]{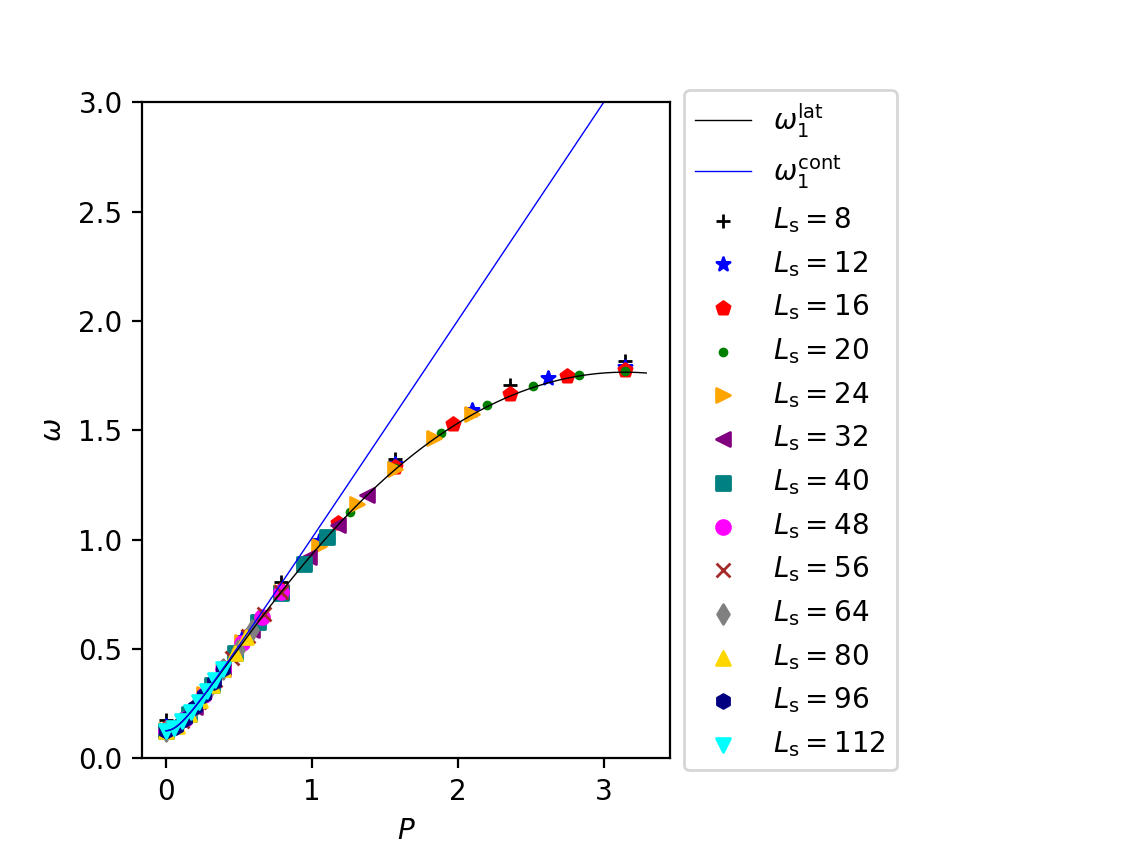}
\caption{One-particle state energy spectrum for $L_{\rm s}=8$–$112$ computed with $L_{\rm t}=4$ and $\chi=80$, together with the one-particle dispersion relation.}
\label{fig:momentum_disrel_p}
\end{figure}

In addition, we perform the same calculation of energy spectrum and the momentum identification for other system sizes $L_{\rm s}=8-112$, and present only data which follow the one-particle dispersion relation in Fig.~\ref{fig:momentum_disrel_p}. 
Note that, the one-particle energy eigenstates with non-zero momentum are two-fold degenerate, sharing the same absolute momentum.
For example, eigenstate number $a=2,3$ of $L_{\rm s}=64$ both have $|P|=2\pi/L_{\rm s}$; 
see Fig.~\ref{fig:energy_spectrum_l64} and Table~\ref{Tab:spec_momB_1pstate}. 
However, this degeneracy is slightly broken
because of the truncation effect, which can be reduced by increasing cut-off bond dimension.
We take the average of the corresponding energy $\omega$ 
and plot these average values in Fig.~\ref{fig:momentum_disrel_p}.
From the figure, we observe that the one-particle state data agree well with both dispersion relations at low-momentum region, while at higher momentum, the data are well described by the lattice dispersion relation as expected.

%%%%%%%%%
%%%%%%%%%

Next, we move to the identification of the momentum for the $q=+1$ sector.
In this case, we compute matrix elements
\be\label{eq:mat_elements_mom_2p}
\langle \Omega|\hat{\mathcal{O}}_2(P,p)|a\rangle
\ee
where
\be\label{eq:operator_mom_2p}
\hat{\mathcal{O}}_2(P,p)=\frac{1}{L_{\rm s}^2}\sum_{x_1,x_2}\hat{s}_{x_1}\hat{s}_{x_2}e^{-ip_1x_1}e^{-ip_2x_2}
\ee
and $p_j$ $(j=1,2)$ are discrete momentum $p_j=\frac{2\pi n_j}{L_{\rm s}}$
with $n_j=0,1,2,\ldots,L_{\rm s}-1$. 
The total momentum is given by $P=p_1+p_2$,
whose values are discrete 
\be\label{eq:total_mom}
P=\frac{2\pi d}{L_{\rm s}}
\ee
with $d=0,1,2,\ldots,L_{\rm s}-1$,
and 
\be\label{eq:relative_momentum}
p=(p_1-p_2)/2
\ee
is the relative momentum\footnote{In our calculation, we use the relative momentum $p = P/2$.}. 
For a fixed total momentum $P$,
if the matrix element is nonzero
\be
\langle \Omega|\hat{\mathcal{O}}_2(P,p)|a\rangle\neq 0
\ee
then the eigenstate $|a\rangle$ has total momentum $P$ irrespective of $p$.
%%%%%%%%%
\begin{figure}[t!]
\centering
\begin{subfigure}[b]{1\textwidth}
\centering
\includegraphics[width=12.2cm,height=2cm]{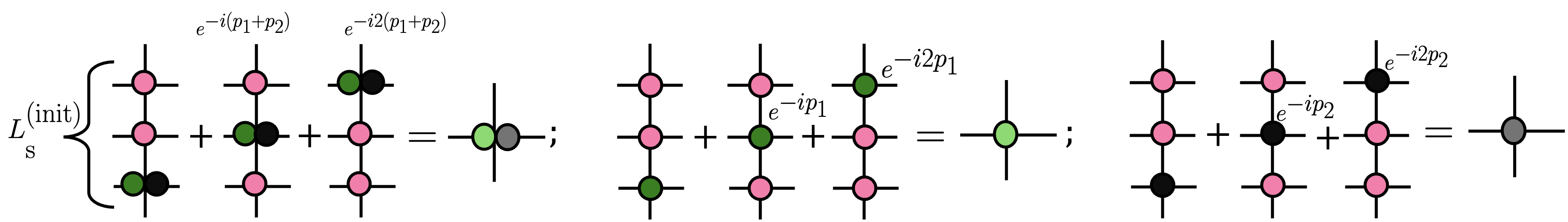}
\caption{}
\label{sfig:momentum_TN_init_2p}
\end{subfigure}
\begin{subfigure}[b]{1\textwidth}
\centering
\includegraphics[width=12.2cm,height=11.7cm]{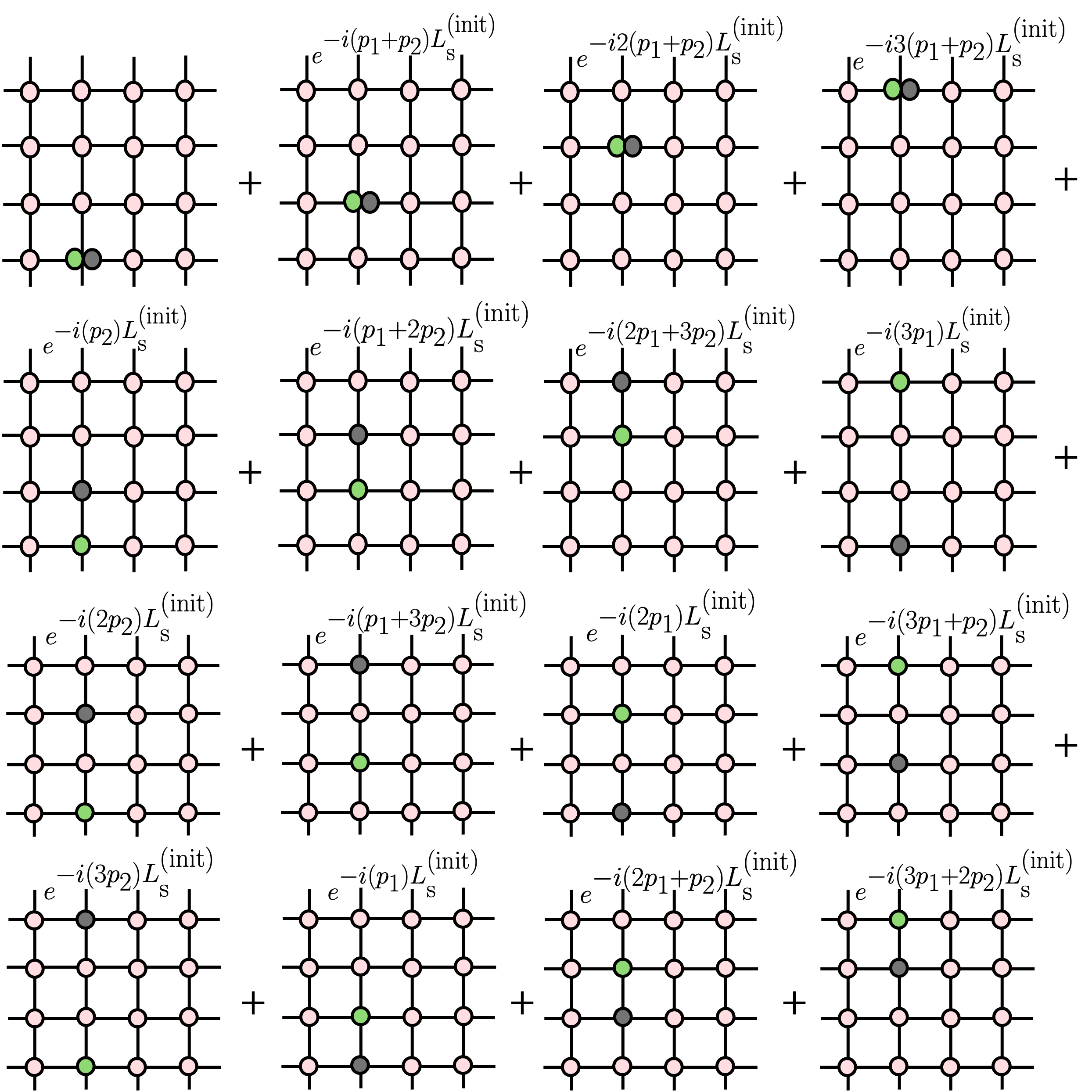}
\caption{}
\label{sfig:momentum_TN_2p}
\end{subfigure}
\caption{(a) The pair of light green and gray circle represents the contracted initial impurity tensor network for a two fields in momentum space. Meanwhile, the single light green (gray) circle denotes a single field in momentum space. These networks are constructed from bare pure tensors (dark pink circle) and some bare impurity tensors with proper momentum factors, namely: pair of black and green circle, and single dark green (black) circle. Here, initial network with $L_{\rm s}^{(\rm init)}=3$ is presented for readability.
(b) The main impurity tensor network with total size $L_{\rm t}\times L_{\rm s}=4\times (L_{\rm s}^{(\rm init)}\times 4)$ for the estimation of $\langle \Omega|\hat{\mathcal{O}}_2(P,p)|a\rangle$.}
\label{fig:momentum_TN_2p}
\end{figure}
%%%%%%%%%
To estimate $\langle \Omega|\hat{\mathcal{O}}_2(P,p)|a\rangle$,
we coarse grain the impurity tensor network shown in Fig.~\ref{fig:momentum_TN_2p}.
The numerical results of the matrix elements
for $L_{\rm s}=64$
in total momentum $|d|=0$--$5$ sectors are listed
in Table~\ref{Tab:spec_momB_2pstate}.
Similar to the $q = -1$ case, matrix elements with value $|\langle \Omega|\hat{\mathcal{O}}_2(P,p)|a\rangle| < 10^{-6}$ are treated as zero. Some elements of order $O(10^{-6})$ and $O(10^{-5})$ are considered as noise,  as we observe that their values decrease for larger bond dimension $\chi = 90$. For elements of order $O(10^{-4})$, some are treated as a signal when their values increase for larger $\chi$, and as a noise when they decrease.

\begin{table}[t!]
\begin{center}
\caption{The absolute value of matrix elements $| \langle\Omega|\hat{\mathcal{O}}_2(P,p)|a\rangle^{[\rm hotrg]}|$
with $|P|=0\mbox{--}10\pi/L_{\rm s}$, for $L_{\rm s}=64$ computed using $\chi=80$ and $L_{\rm t}=4$ for $|a\rangle$ in $q=+1$ sector.}
\label{Tab:spec_momB_2pstate}
\begin{tabular}{|c|c|c|c|c|c|c|c|c|}
\hline
\diagbox{$a$}{$P$}&$0$ & $2\pi/L_{\rm s}$  & $4\pi/L_{\rm s}$ & $6\pi/L_{\rm s}$ & $8\pi/L_{\rm s}$  &$10\pi/L_{\rm s}$  & $|P|$ & $\omega_a^{[\rm hotrg]}$\\
\hline\hline
6&0.206871&$<10^{-6}$&0.000008&$<10^{-6}$&0.000010&$<10^{-6}$&0&0.270811\\
9&$<10^{-6}$&0.092225&$<10^{-6}$&0.000012&$<10^{-6}$&0.000011&$2\pi/L_x$&0.329028\\
10&0.000019&$<10^{-6}$&0.045078&$<10^{-6}$&0.000001&$<10^{-6}$&$4\pi/L_x$&0.329029\\
11&$<10^{-6}$&$<10^{-6}$&0.045050&$<10^{-6}$&0.000010&$<10^{-6}$&$4\pi/L_x$&0.329038\\
12&$<10^{-6}$&0.092206&$<10^{-6}$&0.000028&$<10^{-6}$&0.000032&$2\pi/L_x$&0.329038\\
13&0.067467&$<10^{-6}$&0.000007&$<10^{-6}$&0.000005&$<10^{-6}$&0&0.38727\\
16&$<10^{-6}$&0.000002&$<10^{-6}$&0.053592&$<10^{-6}$&0.000027&$6\pi/L_x$&0.410097\\
17&$<10^{-6}$&0.000052&$<10^{-6}$&0.053624&$<10^{-6}$&0.000012&$6\pi/L_x$&0.410109\\
18&$<10^{-6}$&$<10^{-6}$&0.081047&$<10^{-6}$&0.000024&$<10^{-6}$&$4\pi/L_x$&0.410109\\
19&0.000069&$<10^{-6}$&0.081006&$<10^{-6}$&0.000058&$<10^{-6}$&$4\pi/L_x$&0.410111\\
21&0.000049&$<10^{-6}$&0.000040&$<10^{-6}$&0.008696&$<10^{-6}$&$8\pi/L_x$&0.468347\\
22&$<10^{-6}$&0.037856&$<10^{-6}$&0.000022&$<10^{-6}$&0.000002&$2\pi/L_x$&0.468362\\
23&$<10^{-6}$&$<10^{-6}$&0.000120&$<10^{-6}$&0.008682&$<10^{-6}$&$8\pi/L_x$&0.468414\\
24&$<10^{-6}$&0.037783&$<10^{-6}$&0.000105&$<10^{-6}$&0.000057&$2\pi/L_x$&0.468414\\
27&$<10^{-6}$&0.000014&$<10^{-6}$&0.079396&$<10^{-6}$&0.000171&$6\pi/L_x$&0.498128\\
28&0.000009&$<10^{-6}$&0.000075&$<10^{-6}$&0.059665&$<10^{-6}$&$8\pi/L_x$&0.498146\\
29&$<10^{-6}$&$<10^{-6}$&0.000077&$<10^{-6}$&0.059764&$<10^{-6}$&$8\pi/L_x$&0.498258\\
30&$<10^{-6}$&0.000172&$<10^{-6}$&0.079394&$<10^{-6}$&0.000118&$6\pi/L_x$&0.498259\\
35&0.045862&$<10^{-6}$&0.000492&$<10^{-6}$&0.000071&$<10^{-6}$&0&0.549592\\
38&0.000788&$<10^{-6}$&0.033879&$<10^{-6}$&0.000083&$<10^{-6}$&$4\pi/L_x$&0.556537\\
39&$<10^{-6}$&0.000050&$<10^{-6}$&0.000143&$<10^{-6}$&0.013012&$10\pi/L_x$&0.556644\\
40&$<10^{-6}$&0.000184&$<10^{-6}$&0.000152&$<10^{-6}$&0.012959&$10\pi/L_x$&0.556748\\
41&$<10^{-6}$&$<10^{-6}$&0.033987&$<10^{-6}$&0.000209&$<10^{-6}$&$4\pi/L_x$&0.556748\\
44&0.000192&$<10^{-6}$&0.000104&$<10^{-6}$&0.082079&$<10^{-6}$&$8\pi/L_x$&0.588224\\
45&$<10^{-6}$&0.000030&$<10^{-6}$&0.000231&$<10^{-6}$&0.066092&$10\pi/L_x$&0.588267\\
46&$<10^{-6}$&0.000053&$<10^{-6}$&0.000320&$<10^{-6}$&0.066321&$10\pi/L_x$&0.588615\\
47&$<10^{-6}$&$<10^{-6}$&0.000421&$<10^{-6}$&0.081789&$<10^{-6}$&$8\pi/L_x$&0.588615\\
\hline
\end{tabular}
\end{center}
\end{table}

We observe that the energy in $q=+1, P\neq 0$ sector exhibits a four-fold degeneracy\footnote{For $L_{\rm s}=64$, the four-fold degenerate states are: $a=9$--$12,16$--$19,21$--$24,27$--$30,38$--$41,44$--$47$.}, as also seen in Fig.~\ref{fig:energy_spectrum_l64}.
After identifying the momentum,
it turns out that the four-fold degenerate states belong to two different absolute values of total momentum. This occurrence is related to the phase shift of the (1+1)d Ising model,
which will be explained Sec.~\ref{ssubec:ps_energy}.
Furthermore, all states in Table~\ref{Tab:spec_momB_2pstate} are two-particle states, identified by looking at their behavior over system size, and also by dispersion relation, as will be explained in Sec.~\ref{subsec:numb_particles} and Sec.~\ref{sec:two_p_states_degeneracy}, respectively.

%%%%%%%%%%%%
\subsection{Number of particles}
\label{subsec:numb_particles}
%%%%%%%%%
\begin{figure}[t!]
\centering
\begin{subfigure}[b]{0.4\textwidth}
\centering
\includegraphics[width=6cm,height=6.5cm]{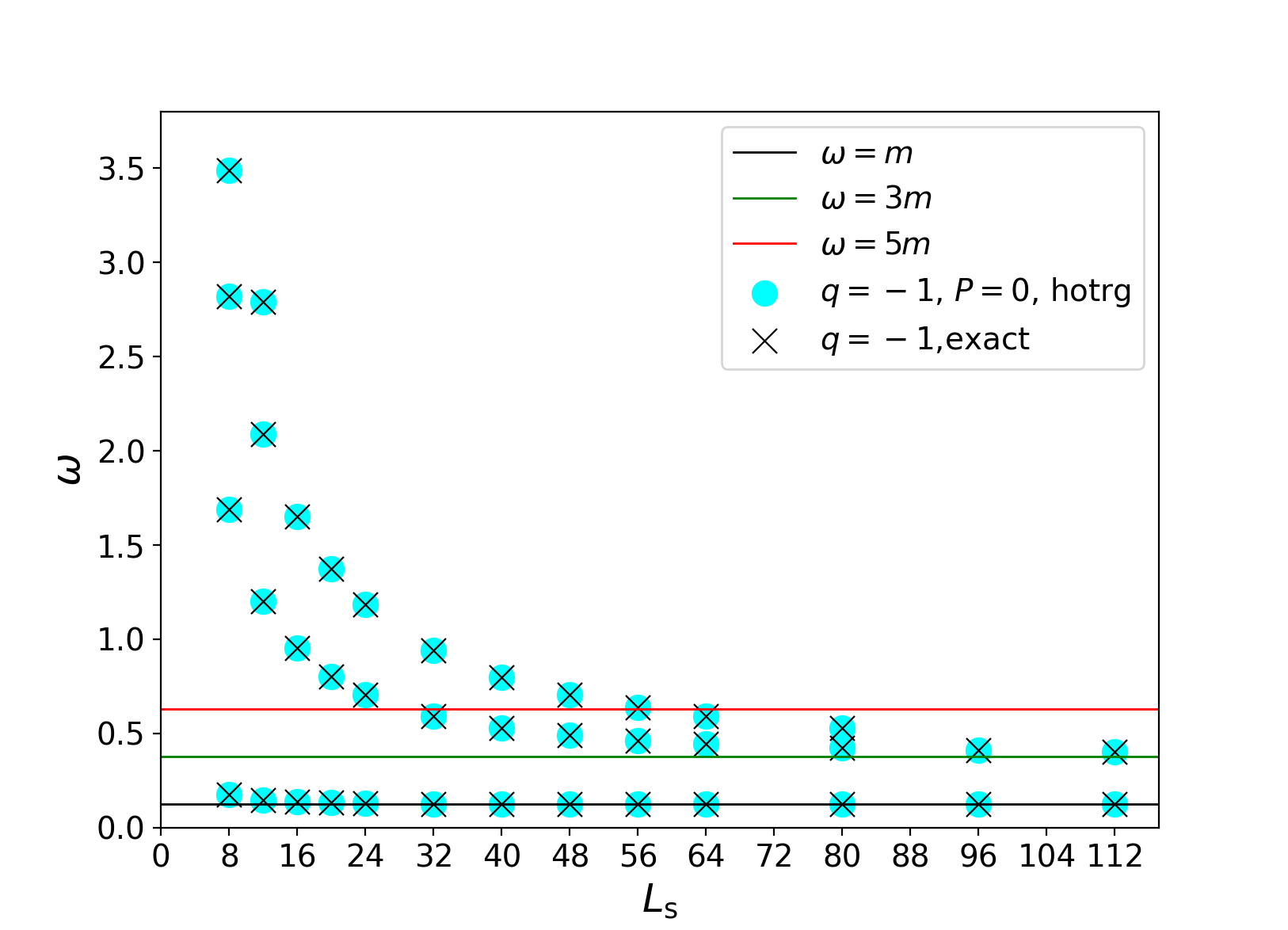}
\caption{}
\label{sfig:spec_qmin_lt4}
\end{subfigure}\hspace{10mm}
\begin{subfigure}[b]{0.4\textwidth}
\centering
\includegraphics[width=6cm,height=6.5cm]{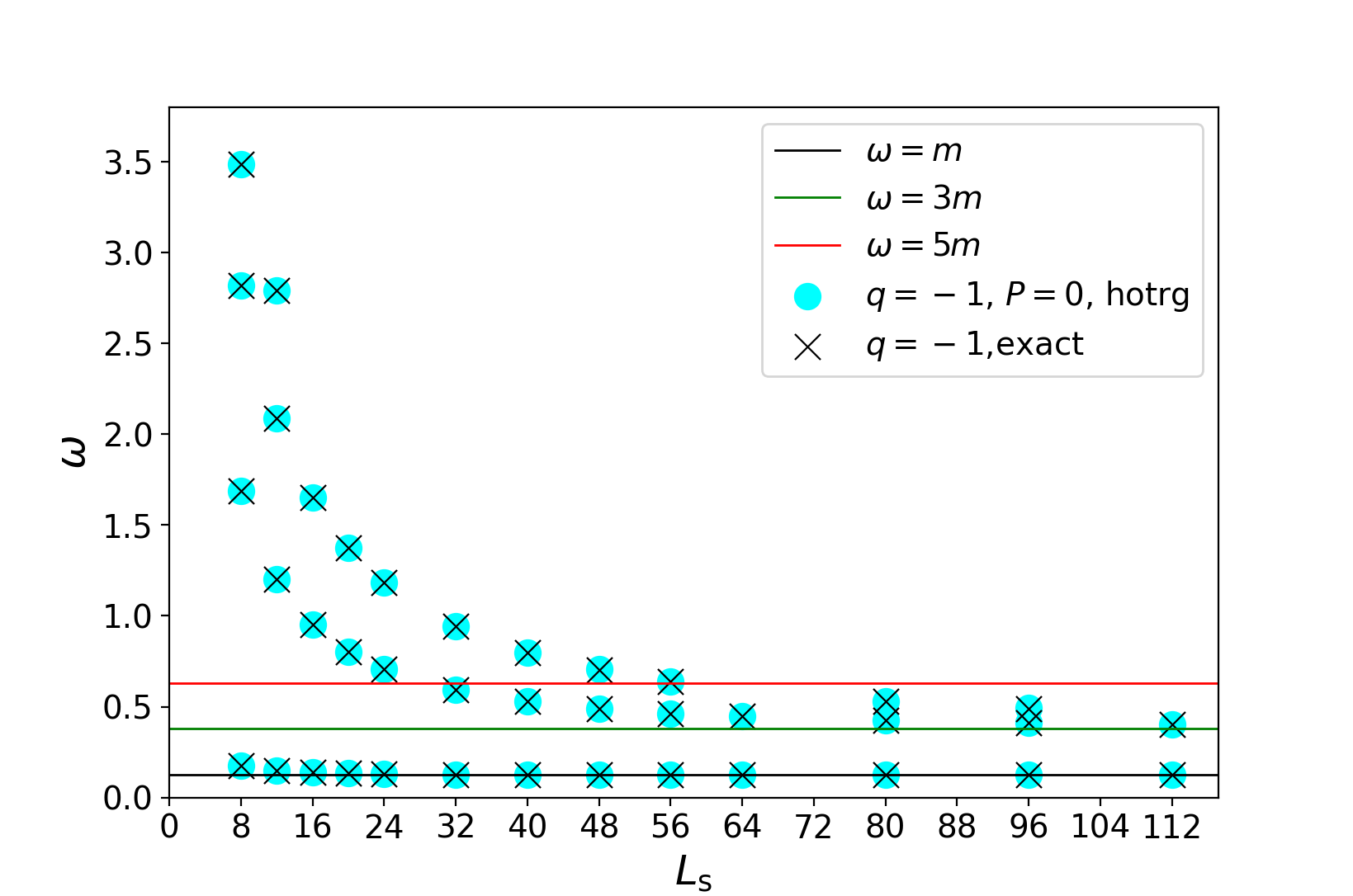}
\caption{}
\label{sfig:spec_qmin_lt8}
\end{subfigure}
\caption{Energy spectrum $\omega_a^{[\rm hotrg]}$ from $q=-1$ and $P=0$ sector 
over system size $L_{\rm s}=8-112$ computed using $\chi=80$ with (a) $L_{\rm t}=4$, and (b) $L_{\rm t}=8$.}
\label{fig:energy_oversize_qmin}
\end{figure}
%%%%%%%%%

The simplest way to identify the number of particles contained in the eigenstates of a given sector is to examine the energy dependence on the system size.
First, we analyze the $q=-1, P=0$ sector. 
From Fig.~\ref{fig:energy_oversize_qmin},
three distinct energy levels are observed.
The lowest level 
corresponds to the one-particle state energy with $P=0$, since it approaches $m$ in large system size,
where $m$ is the rest mass given in Eq.~(\ref{eq:rest_mass}).
Meanwhile, the second and third levels are the three-particle states as they approach $3m$
as the system size increases.

\begin{figure}[t!]
\centering
\begin{subfigure}[b]{0.4\textwidth}
\centering
\includegraphics[width=6cm,height=6.5cm]{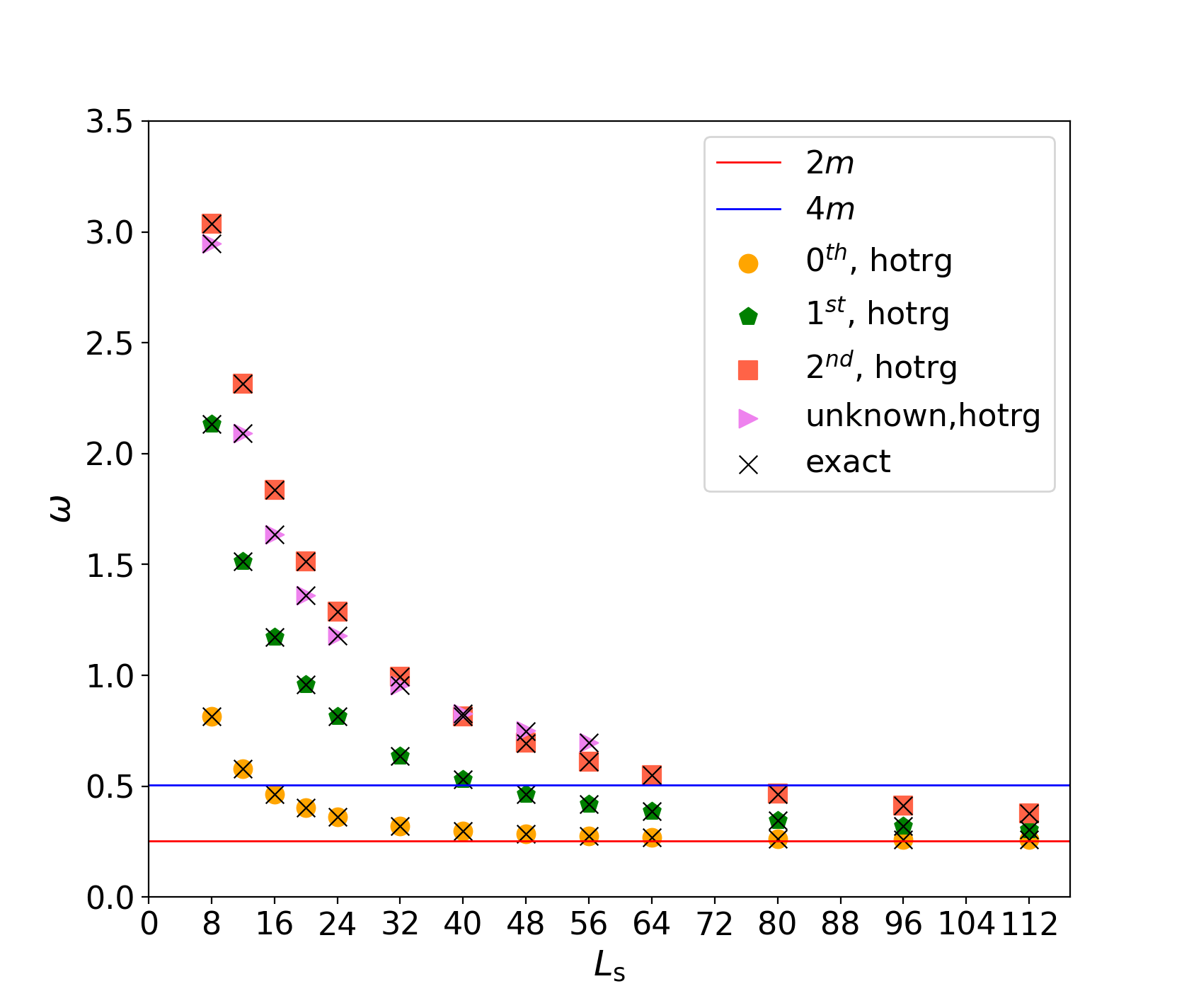}
\caption{}
\label{sfig:w_vs_l_qposlt4}
\end{subfigure}
\begin{subfigure}[b]{0.4\textwidth}
\centering
\includegraphics[width=6cm,height=6.5cm]{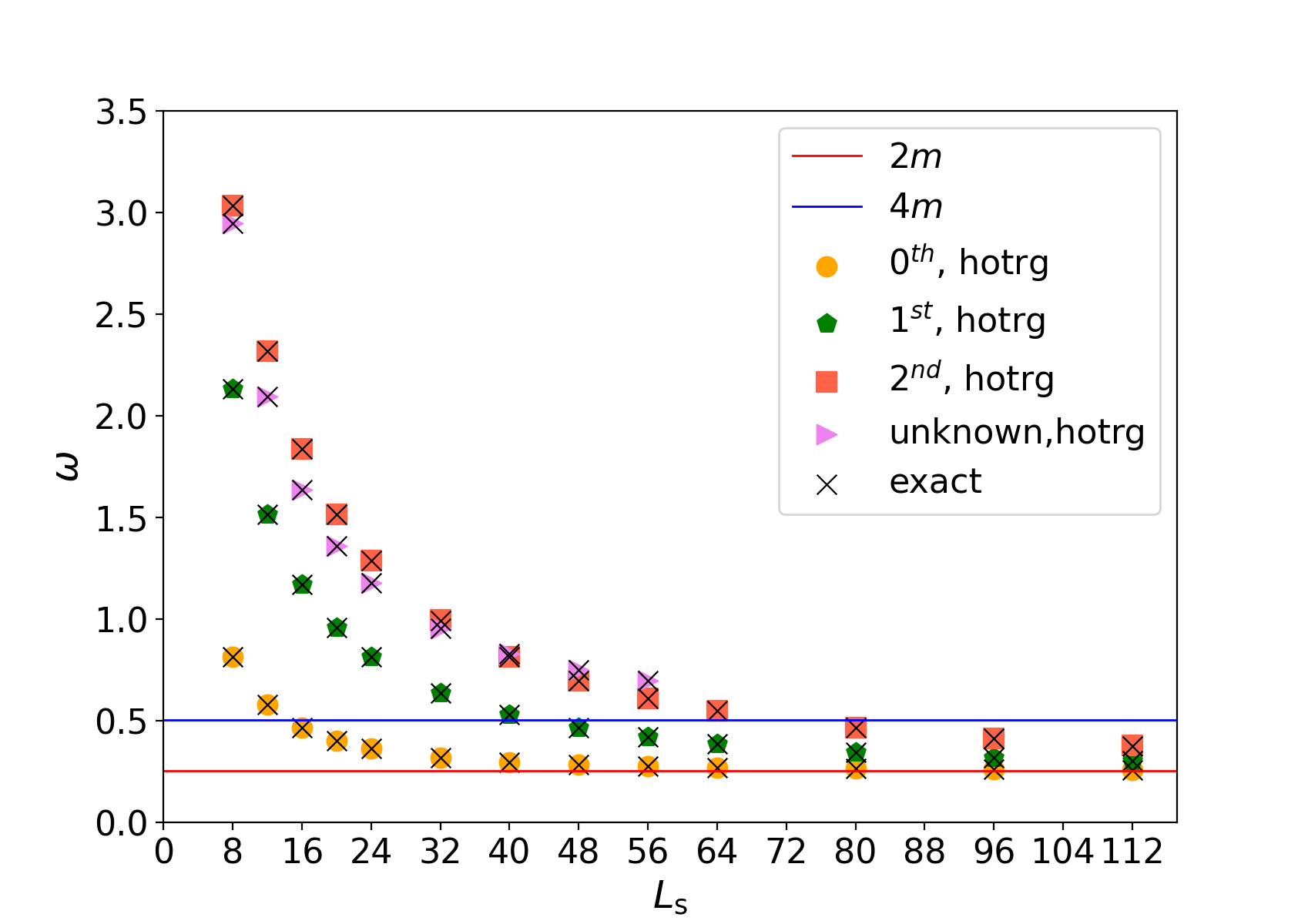}
\caption{}
\label{sfig:w_vs_l_qposlt8}
\end{subfigure}
\caption{Energy spectrum $\omega_a^{[\rm hotrg]}$ from $q=+1,P=0$ sector 
over system size $L_{\rm s}=8-112$ computed using $\chi=80$ with (a) $L_{\rm t}=4$ 
and (b) $L_{\rm t}=8$.}
\label{fig:energy_oversize_qpos}
\end{figure}

Next, we analyze energy spectrum in the $q=+1, P=0$ sector across system size $L_{\rm s}$.
From Fig.~\ref{fig:energy_oversize_qpos}, the networks with $L_{\rm t}=4$ and $L_{\rm t}=8$ yield almost the same  energy level in the sector for each system size.
The three energy levels in this sector shown by orange, green, and red markers, which are classified by looking at the shape of the corresponding wave function, see Sec.~\ref{subsec:wf_2p}, are the two-particle state energy as they approach $\omega=2m$ in the large system size.

%%%%%%%%%
%%%%%%%%%

On the other hand, 
the number of particles corresponding to the energy level shown by violet marker in Fig.~\ref{fig:energy_oversize_qpos} cannot be clearly identified.
Such states, in which the number of particles is not well identified, are found only in relatively small volumes (up to $L_{\rm s}=56)$ in our calculations,
therefore we cannot observe its behavior in the large volume. 
Furthermore, in Sec.~\ref{subsec:wf_2p}, it will be shown that the wave function of this unidentified state is also different from the two-particle case with smaller amplitude. 
We expect that, this is four-particle state but we need further investigation.

\begin{figure}[t!]
\centering
\includegraphics[width=16cm,height=7cm]{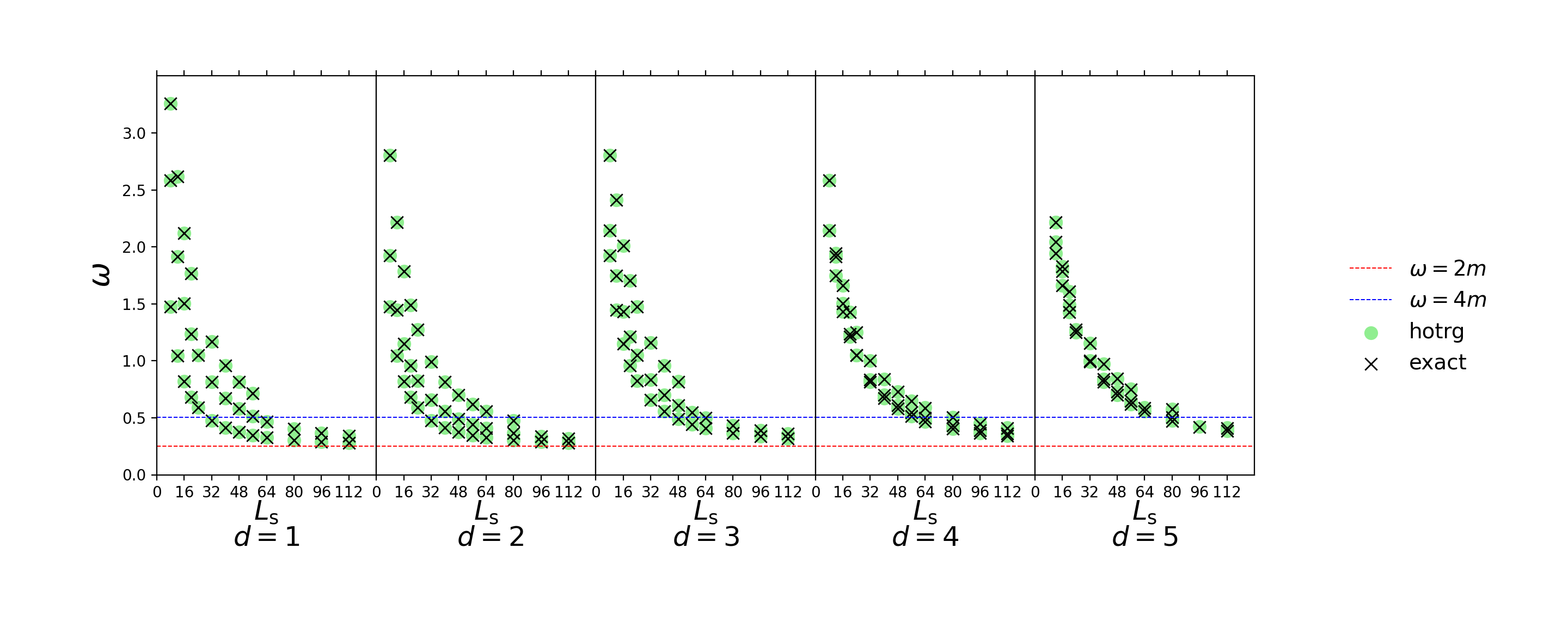}
\caption{Energy spectrum in the $q=+1$ and $d=1$--$5$ sectors over system size $L_{\rm s}$, computed with $L_{\rm t}=4$ and $\chi=80$. The HOTRG data (green circles), represent the average energy of the degenerate states.}
\label{fig:2p_energy_oversize_nonzeromom}
\end{figure}

In addition, the energy spectrum in $q=+1$ for non-zero momentum sectors
as function of system size
is given in Fig.~\ref{fig:2p_energy_oversize_nonzeromom}.
In the figure, for this $d=1$--$5$ sectors, 
one can clearly see that several energy levels approach $2m$ at large volumes, while this behavior is not evident for some other levels. Therefore, we cannot conclude from this figure alone that all the states shown here are two-particle states. However, as we will see later in Sec.~\ref{sec:two_p_states_degeneracy}, by comparing them with the dispersion relation for two-particle states, we find that all these states are indeed two-particle states.

%%%%%%%%%
%%%%%%%%%%%%%%%%%%%%

%%%%%%%%%%%%
\subsection{Wave function}
\label{subsec:wf_2p}
%%%%%%%%%
Wave function is also a useful quantity, 
as it allows direct extraction of important dynamical observables
such as scattering phase shift.
The procedure for computing the wave function with tensor network for $q=-1$ sector 
has already been presented in \cite{PhysRevD.110.034514},
thus we focus here on the $q=+1$ sector.
%%%%%%%%%
\begin{figure}[t!]
\centering
\includegraphics[width=12cm,height=7cm]{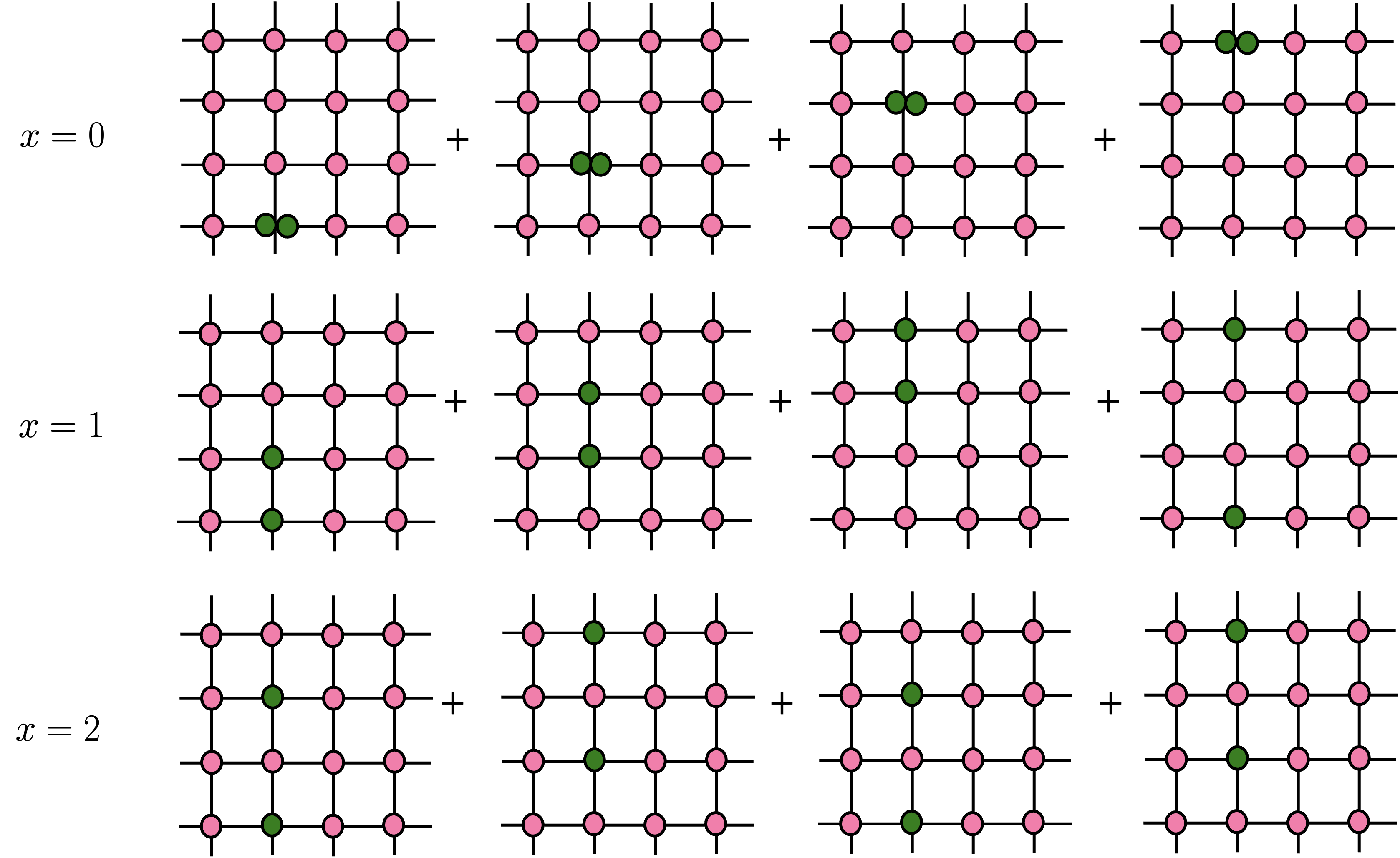}
\caption{The tensor network representation 
for the computation of
two-particle state wave function $\psi_a(x)$ with system size $L_{\rm s}=4$
and $L_{\rm t}=4$. Due to PBC, it is sufficient to consider $x=0,1,2$.
Note that the green circles represent the impurity tensors.}
\label{fig:TN_wf}
\end{figure}
%%%%%%%%%

%%%%%%%%%
\begin{figure}[t!]
\centering
\begin{subfigure}[b]{0.4\textwidth}
\centering
\includegraphics[width=7cm,height=6cm]{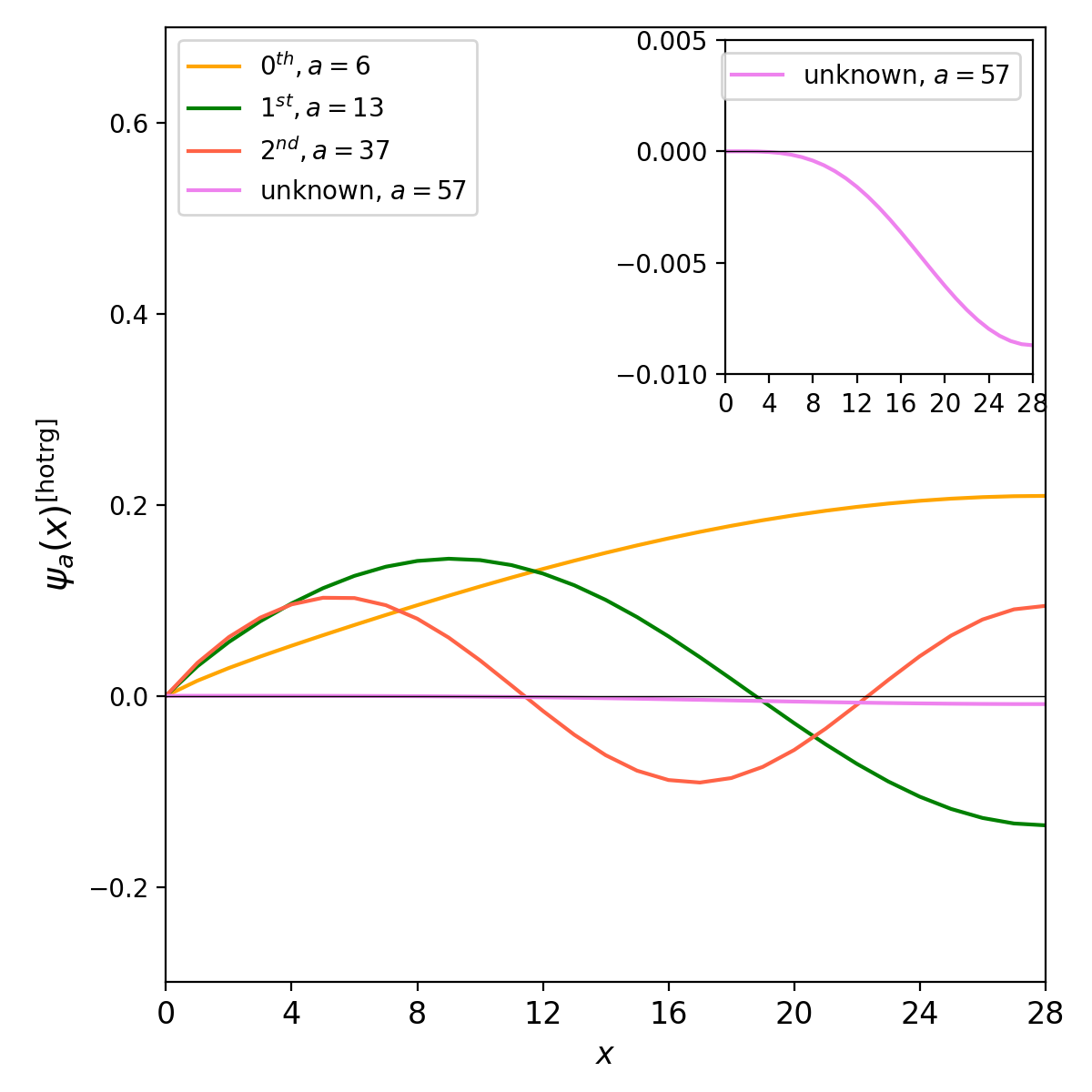}
\caption{}
\label{fig:wf_l56}
\end{subfigure}\hspace{10mm}
\begin{subfigure}[b]{0.4\textwidth}
\centering
\includegraphics[width=7cm,height=6.7cm]{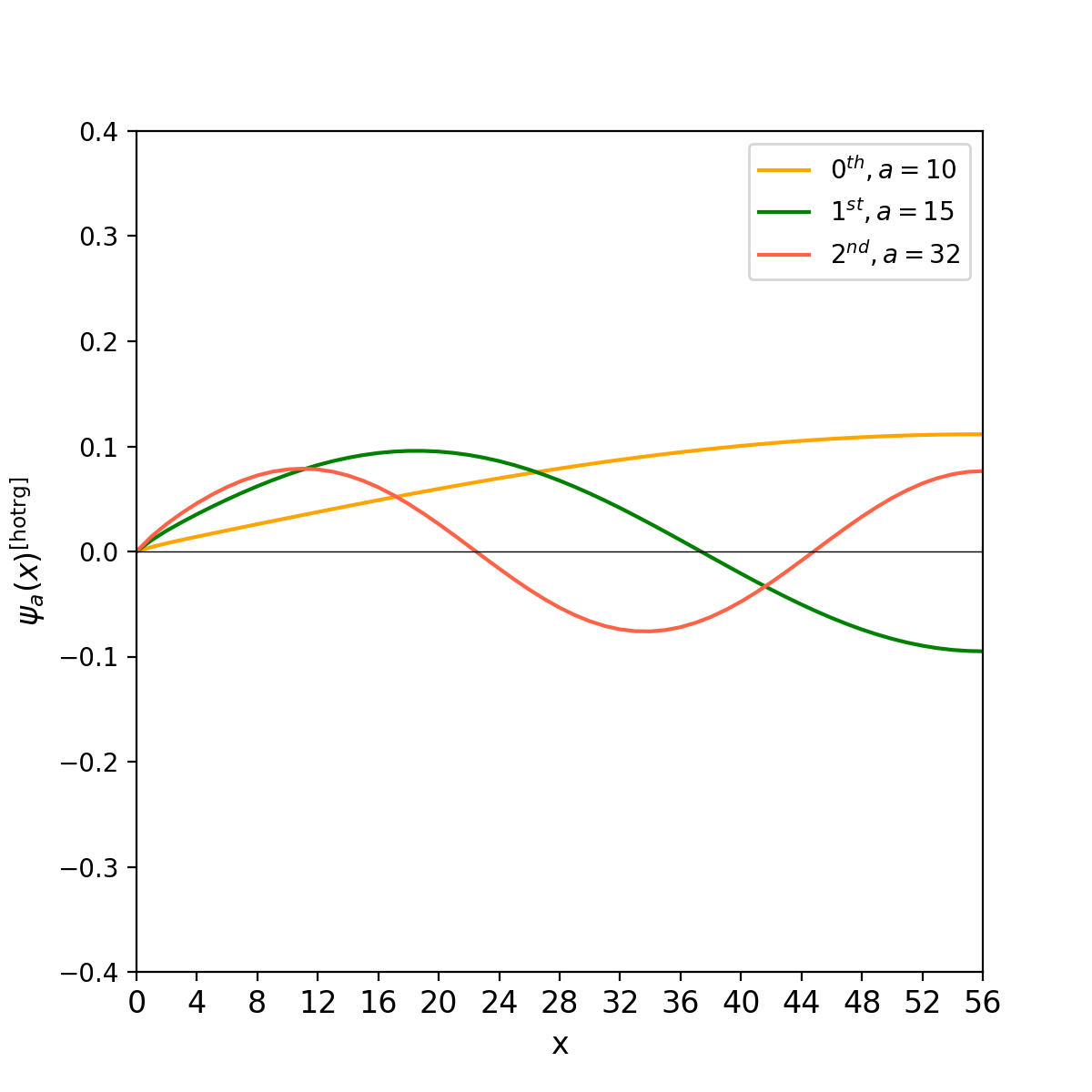}
\caption{}
\label{fig:wf_2p_d0}
\end{subfigure}
\caption{(a) The wave function $\psi_a(x)$ in $q=+1,P=0$ sector computed with $L_{\rm t}=8$ and $\chi=80$ for system size $L_{\rm s}=56$ corresponding to $a=6,13,37,57$.
The inset shows an enlarged view of the wave function for $a=57$.
(b) The wave function for $L_{\rm s}=112$ corresponding to eigenstates $a=10,15,32$.}
\label{fig:WFunction_2p}
\end{figure}
%%%%%%%%%

The wave function of the energy eigenstate $|a\rangle$ in the $q=+1, P=0$ sector can be computed by 
\be\label{eq:wf}
\psi_a(x)=\langle \Omega |\hat{\mathcal{O}}_2(x)|a\rangle.
\ee
The operator $\hat{\mathcal{O}}_2(x)$ is given by 
\be\label{eq:wf_operator}
\hat{\mathcal{O}}_2(x) = \frac{1}{L_{\rm s}}\sum_{x'=0}^{L_{\rm s}-1}\hat{s}_{x'}\hat{s}_{x'+x}
\ee
where $\hat{s}_{x'}$ is the single spin field operator and $x$ is the relative distance 
between the two operators \cite{BALOG2001315}.
From Eq.~(\ref{eq:wf}), we see that the wave function is obtained 
from matrix elements evaluated at each relative distance $x$.
In Fig.~\ref{fig:TN_wf}, we show the impurity tensor network 
corresponding to the operator $\hat{\mathcal{O}}_2(x)$
for $L_{\rm s}=4$ and $L_{\rm t}=4$.
The extension to larger spatial sizes $L_{\rm s}=8-112$ and $L_{\rm t}=4,8$ is straightforward.
The computational cost of evaluating the wave function using HOTRG algorithm scales as $O(L_{\rm s}^2\chi^7)$.
Owing to the symmetric property and PBC of the wave function,
\bea\label{eq:pbc_wf}
\psi_a(L_{\rm s}-x)&=&\psi_a(-L_{\rm_s}+x)\notag\\
&=&\psi_a(x)
\eea
which reduces the tensor network diagrams that need to be computed by half.
We coarse grain the impurity tensor network shown in Fig.~\ref{fig:TN_wf}
and subtitute the result into the Eq.~(\ref{eq:approx_Bba})
to obtain 
\be
\psi_a(x)^{[\rm hotrg]}\coloneqq\langle \Omega |\hat{\mathcal{O}}_2(x)|a\rangle^{[\rm hotrg]}\approx \langle \Omega |\hat{\mathcal{O}}_2(x)|a\rangle.
\ee

The numerical results \footnote{\label{fn:normalization}We apply the sign normalization for the wave function so that $\psi_a(1)^{[\rm hotrg]}>1$ for all $a$ and all $L_{\rm s}$  that is
\be
\psi_a(x)^{[\rm hotrg]}\Leftarrow\psi_a(x)^{[\rm hotrg]}\times \frac{\psi_a(1)^{[\rm hotrg]}}{|\psi_a(1)^{[\rm hotrg]}|}.
\ee} of the wave function\footnote{\label{fn:fwl56}
We show the wave function for $L_{\rm s}=56$, instead of $L_{\rm s}=64$, because $L_{\rm s}=56$ is the largest size where the unidentified state (possibly four-particle state) in $q=+1, P=0$ sector is obtained.} $\psi_a(x)^{[\rm hotrg]}$ computed with $L_{\rm t}=8$ and $L_{\rm s}=56$ coarse-grained with HOTRG using $\chi=80$ are shown in Fig.~\ref{fig:wf_l56}.
The wave functions, corresponding to $a=6,13,37$,
are the two-particle state wave function where
the values at $x=0$ are approximately zero, indicating that the two particles are never in contact.
Here, the $0^{\rm th},1^{\rm st},2^{\rm nd}$ wave functions are identified based on the number of nodes of $\psi_a(x)^{\rm [hotrg]}$ within the region $0<x\leq L_{\rm s}/2$, where the $0^{\rm th}$ wave function has no node, the $1^{\rm st}$ has one node, and so on.
In contrast, the wave function corresponding to eigenstate number $a=57$ exhibits qualitatively different behavior and has a relatively smaller amplitude. This behavior persists even at a larger bond-dimension, $\chi=90$. 
The similar wave function, as $a=57$ for $L_{\rm s}=56$, is also observed on other system sizes $L_{\rm s}=8$--$48$.

Furthermore, we show the wave function for $L_{\rm s}=112$
computed with $L_{\rm t}=8$ and $\chi=80$ in Fig.~\ref{fig:wf_2p_d0}.
The wave function from larger system size is preferable for the subsequent analysis, that is the extraction of the phase shift from the wave function approach which, will be discussed in Secs.~\ref{subsec:ps_2p_wf}--\ref{subsec:ps_2p_wf_inside}, 
and $L_{\rm s}=112$ is the largest size accessible with $\chi=80$, where at this size, two-particle state wave functions are obtained for the eigenstates $a=10,15,32$.

\subsection{Two-particle states analysis}
\label{subsec:2p_state}

%%%%%%%%%
\begin{figure}[t!]
\centering
\includegraphics[width=10cm,height=5.5cm]{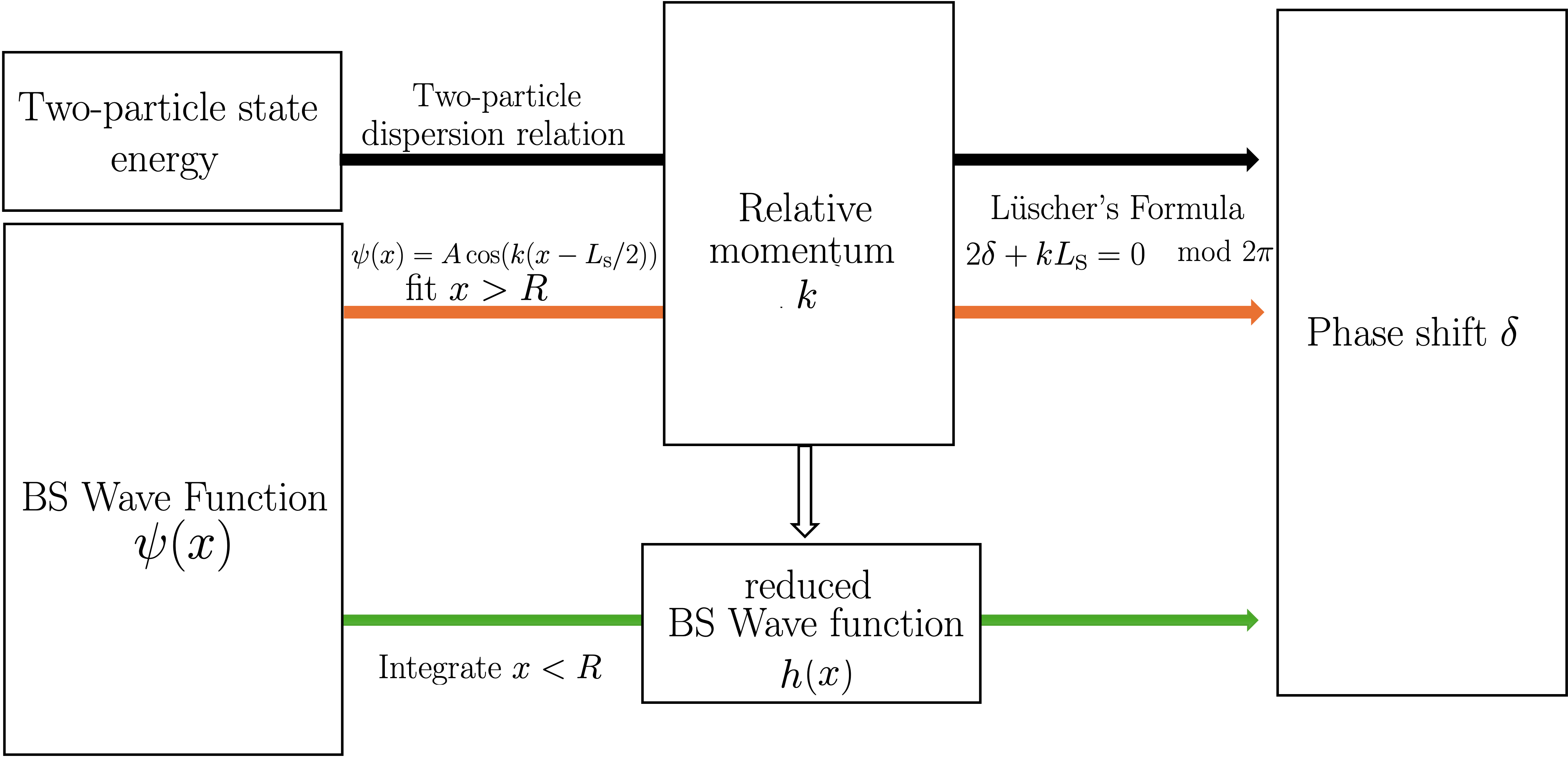}
\caption{A diagram of the two-particle state analysis performed in this work. The black arrow denotes the computation of the phase shift using only L\"uscher's formula, the orange arrow indicates the flow of analysis using the BS wave function outside the interaction range $R$ with the aid of L\"uscher's formula, and the green arrow represents the flow using the BS wave function inside $R$, where information on relative momentum is required.}
\label{fig:2p_analysis}
\end{figure}
%%%%%%%%%%%%%%%%%%%%

Hereafter, we focus on the two-particle states and their dynamics.
The scattering phase shift, as an important dynamical quantity,
can be extracted from the finite volume energy spectrum as well as from the wave function
of the two-particle states inside and outside interaction range, see Fig.~\ref{fig:2p_analysis}.
In the following, we show the procedure to determine the scattering phase shift using these three approaches.
\subsubsection{Phase shift from the energy spectrum}
\label{ssubec:ps_energy}
%%%%%%%%%
\begin{figure}[t!]
\centering
\begin{subfigure}[b]{1\textwidth}
\centering
\includegraphics[width=16cm,height=7cm]{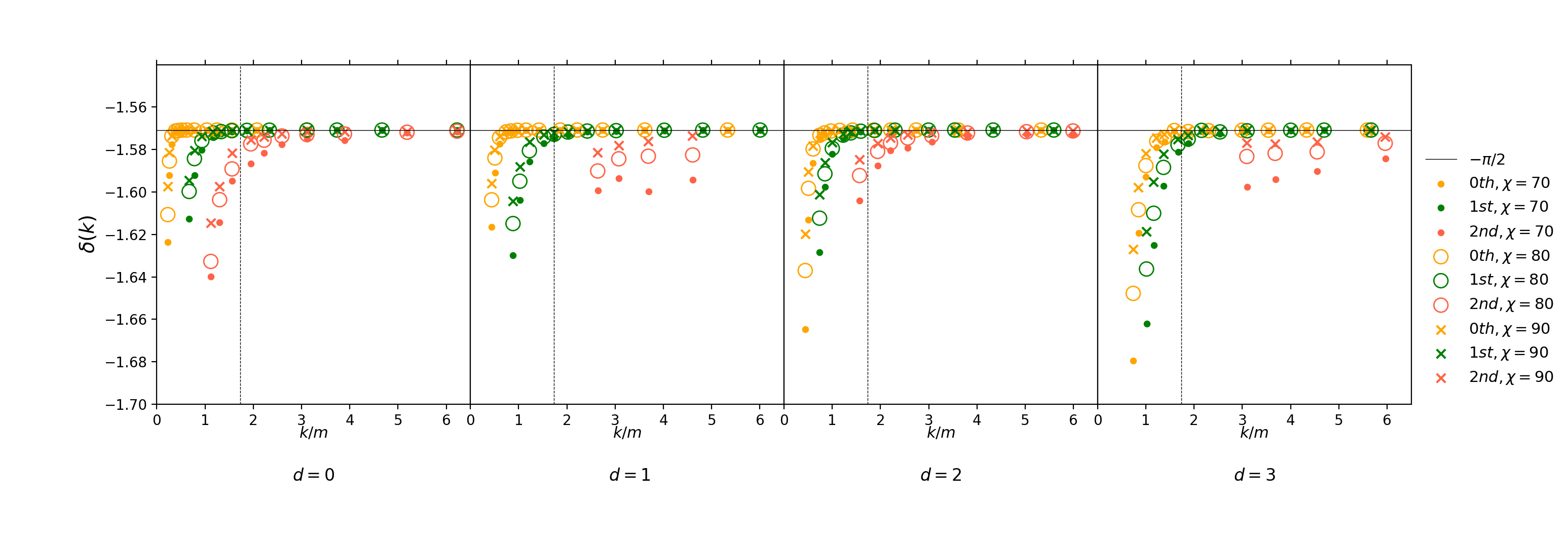}
\caption{}
\label{sfig:ps_lt4_d0}
\end{subfigure}\hspace{10mm}
\begin{subfigure}[b]{1\textwidth}
\centering
\includegraphics[width=16cm,height=7cm]{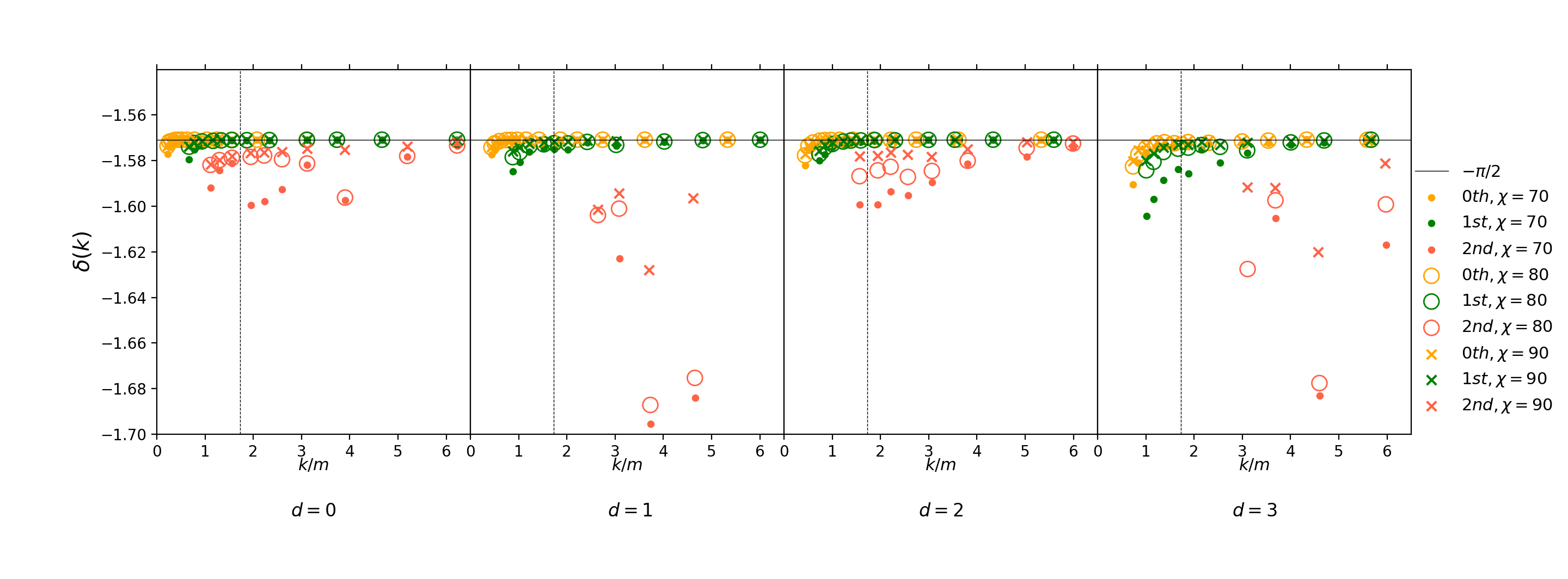}
\caption{}
\label{sfig:ps_lt8_d0}
\end{subfigure}
\caption{Scattering phase shift of two-particle state on CM and moving frame,
i.e. $d=0,1,2,3$ (from left to right)
computed using $\chi=80$ with (a) $L_{\rm t}=4$ 
and (b) $L_{\rm t}=8$. The vertical dashed line is the elastic limit $k/m=\sqrt{3}$.}
\label{fig:ps_d_dc80}
\end{figure}
%%%%%%%%%%%%%%%%%%%%
The scattering phase shift can be extracted
from the finite volume two-particle state energy spectrum in the CM frame ($P=0$) as well as in the moving frame ($P\neq 0$).
To this end, we compute the relative momentum $p$ using two-particle lattice dispersion relation \cite{PhysRevD.99.014501}
\be\label{eq:disrel_nonzeromom}
\omega_2^{(d)}=\cosh^{-1}\left(\cosh (m) +1 -\cos \left(\frac{\pi}{L_{\rm s}}d+p\right)\right) 
+\cosh^{-1} \left(\cosh (m) +1 -\cos \left(\frac{\pi}{L_{\rm s}}d-p\right)\right),
\ee
%%%%%%%%%
where the two-particle energy $\omega_a^{[\rm hotrg]}$ is used as the input $\omega_2^{(d)}$.
For $d\neq 0$, $\omega_2^{(d)}$ is taken to be the average of the degenerate energies.
As a reminder, $d=0,1,\ldots,L_{\rm s}-1$ is the integer related to the total momentum, as defined in Eq.~(\ref{eq:total_mom}).
Note that $p=\gamma k$, where $k$ is the relative momentum in the CM frame, 
with the Lorentz factor $\gamma$.
When $d=0$, Lorentz factor is $\gamma=1$, and the relative momentum satisfies $p=k$.
In this case, the relative momentum $p$ can be directly determined from Eq.~(\ref{eq:disrel_nonzeromom}).
However, for $d\neq 0$,
the equation must be solved numerically to obtain $p$, where in our analysis we employ the bisection method.
Using the resulting relative momentum $p$,
the phase shift can then be computed from L\"uscher's formula 
\cite{Luscher:1990ck,PhysRevD.99.014501,PhysRevD.88.014507} as follows
\be
\label{eq:ps_luscher}
2\delta^{(d)}+{pL_{\rm s}+\pi d} = 0\mod 2\pi.
\ee

In Fig.~\ref{fig:ps_d_dc80}, we show the scattering phase shift, \(\delta^{(d)}\), extracted from the two-particle state energy for \(d = 0\text{--}3\), computed using \(L_{\rm t} = 4, 8\) over $k/m$.
For \(d \neq 0\), the relative momentum $p$ is transformed to the CM momentum $k$ by first converting the moving frame energy to CM energy using \cite{PhysRevD.88.014507}
\be
\omega^{(\rm CM)} = \cosh^{-1}\left( \cosh\left(\omega^{(d)}\right) - 1 + \cos \frac{2\pi d}{L_{\rm s}} \right),
\ee
with $\omega^{(d)}$ as an input.
The resulting CM energy, $\omega^{(\rm CM)}$, is then substituted into Eq.~(\ref{eq:disrel_nonzeromom}) by setting \(d = 0\) to determine \(k\), which is used to compute \(k/m\) for the \(x\)-axis in Fig.~~\ref{fig:ps_d_dc80}.
The numerical results approximately agree with the theoretical Ising results \cite{Gattringer:1992np}
\be\label{eq:theoretical_phaseshift}
\delta_{[\rm ising]}=-\frac{\pi}{2}, 
\ee
up to some errors.
In general, the most accurate phase shift is obtained in the CM frame, i.e., for $d=0$. 
We observe that the phase shift at small $k/m$, which corresponds to larger system size $L_{\rm s}$, becomes less accurate due to the coarse-graining error.
We also see that increasing cut-off bond dimension $\chi$ 
reduces the error of the phase shift, as expected.

\subsubsection{Phase shift from the wave function outside interaction range}
\label{subsec:ps_2p_wf}

%%%%%%%%%
\begin{figure}[t!]
\centering
\begin{subfigure}[b]{0.4\textwidth}
\centering
\includegraphics[width=7cm,height=6cm]{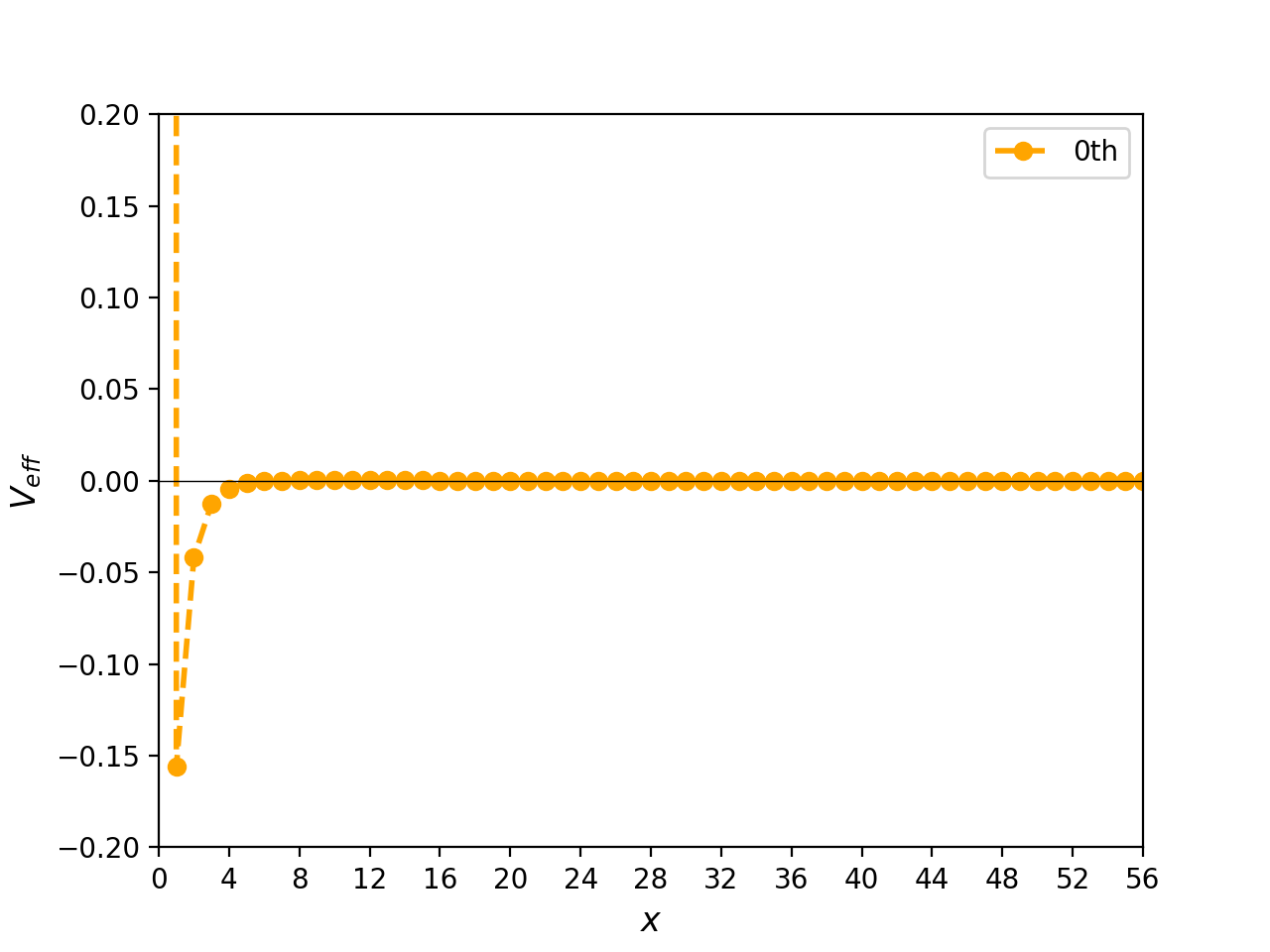}
\caption{}
\label{subfig:pot_l112_lt8}
\end{subfigure}\hspace{10mm}
\begin{subfigure}[b]{0.4\textwidth}
\centering
\includegraphics[width=7cm,height=6cm]{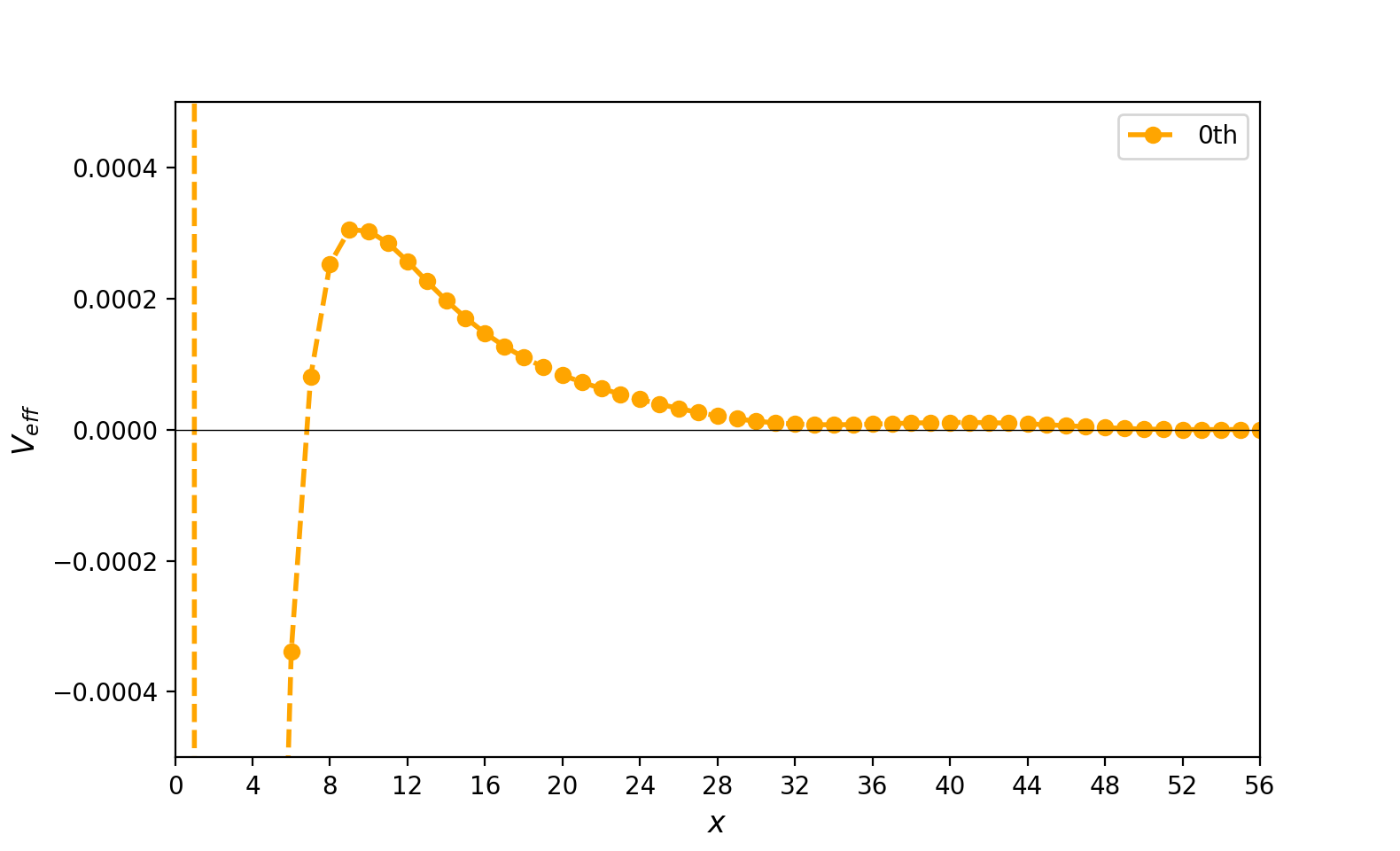}
\caption{}
\label{subfig:pot_l112_larger_lt8}
\label{subfig:pot_l112_lt8_b}
\end{subfigure}
\caption{(a) The effective potential from the two-particle ground state wave function for $L_{\rm s}=112$ 
and total momentum $P=0$ computed using $L_{\rm t}=8$ and $\chi=80$.
(b) The same potential in a smaller scale.}
\label{fig:pot_2p_d0}
\end{figure}
%%%%%%%%%
As previously mentioned,
the phase shift can be extracted directly from the two-particle state wave function in the $P=0$ sector, which is numerically computed by following the procedure described in Sec.~\ref{subsec:wf_2p}.
To extract the phase shift,
we first compute the effective potential $V_{\text{eff}}(x)$ of the scattering 
\be \label{eq:effective_pot}
V_{\rm eff} (x) =\frac{(\partial_x^{2})^{\rm lat}\psi(x)}{\psi(x)}+ k_{\rm lat}^2
\ee
where $(\partial_x^{2})^{\rm lat}\psi(x) =\psi(x+1)-2\psi(x)+\psi(x-1)$ with $\psi(x)^{[\rm hotrg]}$ as input.
Note that, the properties $\psi(x)=\psi(-x)$ is used to compute second derivative at $x=0$ and $x=L_{\rm s}/2$.
Here, 
\be\label{eq:klat}
k_{\rm lat}^2=2(1-\cos(k))
\ee
where $k$ is the relative momentum in CM, obtained from Eq.~(\ref{eq:disrel_nonzeromom}) with $d=0$.
The effective range $R$ is defined as the range where 
$V_{\rm eff}$ is zero, that is 
\be\label{eq:v=0}
V_{\rm eff}(x)=0\hspace{10mm}\text{for $x>R$.}
\ee
In Fig.~\ref{fig:pot_2p_d0}, we present $V_{\rm eff}(x)$ for the two-particle ground-state wave function of a system size $L_{\rm s}=112$, which was previously shown in Fig.~\ref{fig:wf_2p_d0}.
We observe that the effective potential $V_{\rm eff}(x)$ diverges to infinity as $x \to 0$, indicating a strongly repulsive interaction between the two particles at the origin 
\footnote{
Note that the numerical value of $\psi(0)$ is not exactly zero, but rather a very small number that may be either positive or negative.
In our normalization, as mentioned in footnote~\ref{fn:normalization}, the wave function for small $x$ can be expressed as 
$\psi(x)=c_1|x|+c_2x^2$
with $c_1>0$.
The second derivative of $\psi(x)$ is then given by
$\frac{d^2\psi(x)}{dx^2}=c_1\delta(x)+2c_2$ 
where $\delta(x)$ is the delta function.
Accordingly, the potential at $x=0$ is given by
\be
\lim_{x\to 0}V_{\rm eff}(x) =\lim_{x\to 0}\frac{c_1\delta(x)+2c_2}{c_1|x|+c_2x^2},
\ee
which ensures $V_{\rm eff}(x)\to +\infty$ for $x\to 0$.
}.

\begin{figure}[t!]
\centering
\begin{subfigure}[b]{0.4\textwidth}
\centering
\includegraphics[width=8cm,height=6cm]{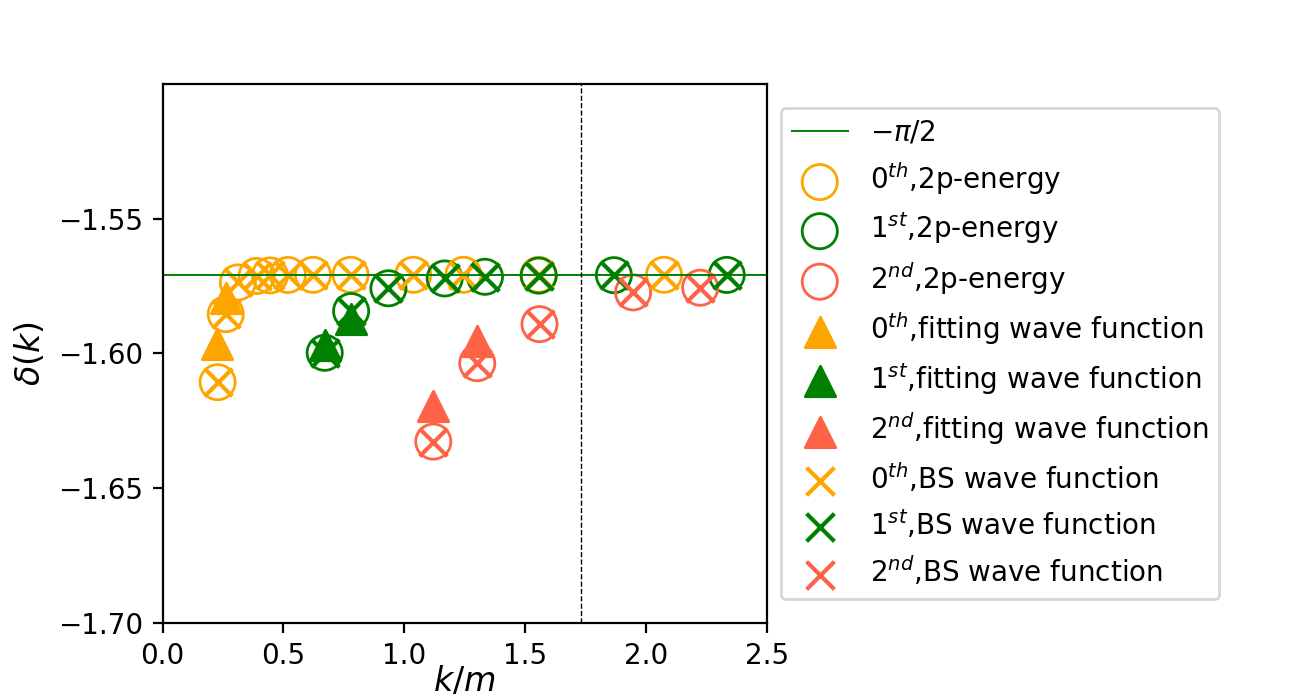}
\caption{}
\label{subfig:ps_energy_vs_wf_lt4}
\end{subfigure}\hspace{15mm}
\begin{subfigure}[b]{0.4\textwidth}
\centering
\includegraphics[width=8cm,height=6cm]{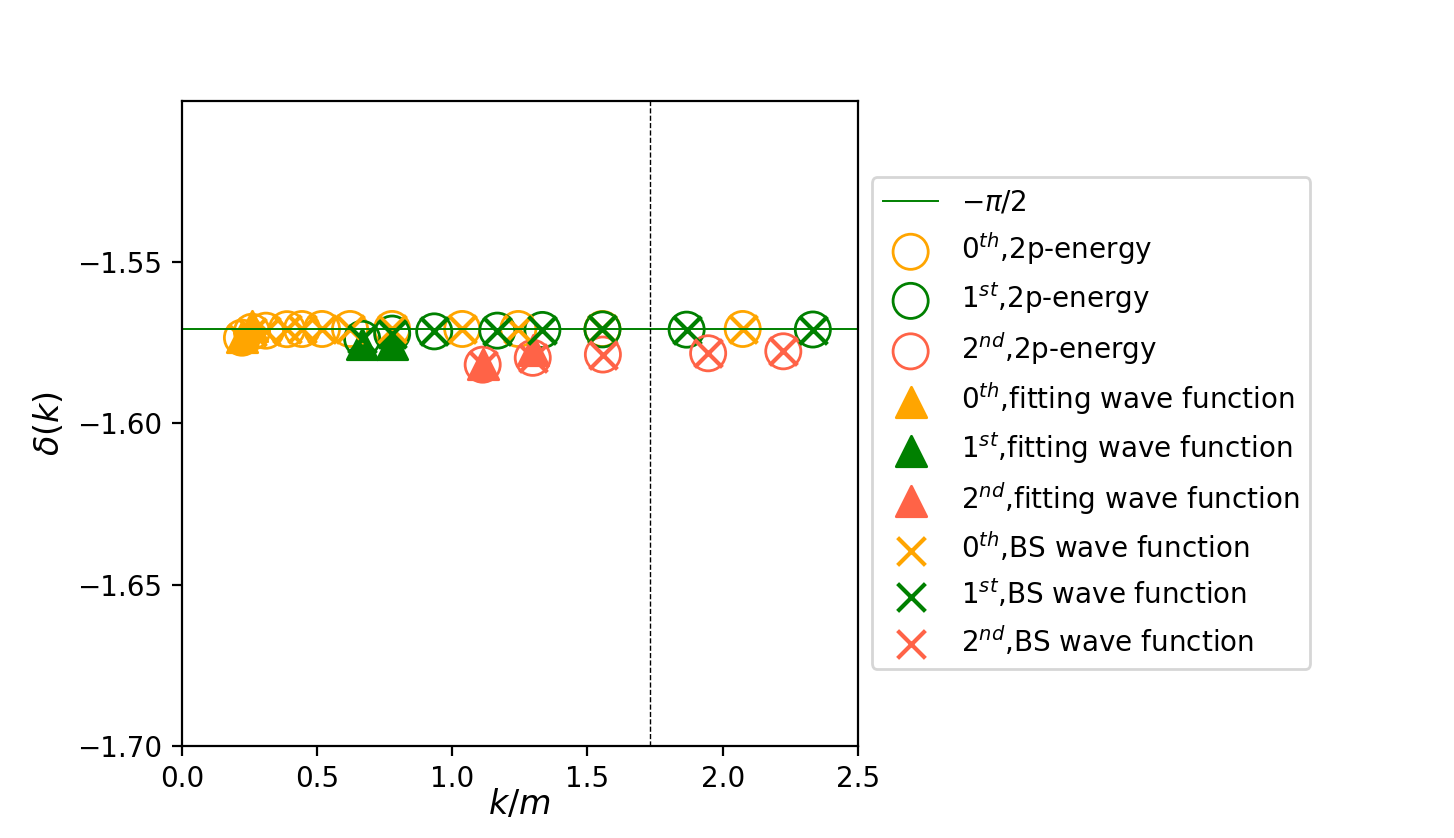}
\caption{}
\label{subfig:ps_energy_vs_wf_lt8}
\end{subfigure}
\caption{Comparison of the phase shift in CM frame, computed at $\chi=80$ using
the finite volume energy,
the fitting of the wave function outside the interaction range, and the Bethe-Salpeter wave function inside the interaction range for (a) $L_{\rm t}=4$ and (b) $L_{\rm t}=8$.}
\label{fig:ps_energy_vs_wf}
\end{figure}

For system at temperature $T=2.44$, we estimate the interaction range to be $R\approx 40$, 
since the effective potential $V_{\rm eff}(x)$ approaches zero for $x\geq40$, see Fig.~\ref{fig:pot_2p_d0}.
Outside interaction range $x>R$, that is in the free region, the wave function is described as \cite{BALOG2001315}
\bea\label{eq:wf_free}
\psi(x)^{[\rm free]}&=&
A \cos (kx+\delta)\notag\\
&=&A \cos (k(x-L_{\rm s}/2))
\eea
where the second line is obtained with the aid of L\"uscher's formula. Accordingly, $A$, a constant, and $k$, the relative momentum, are the fitting parameters.
We fit the numerical wave function $\psi(x)^{[\rm hotrg]}$, 
with the functional form $\psi(x)^{[\rm free]}$ in the free region, $R< x\leq L_{\rm s}/2$, extract $k$ and use the result to obtain the phase shift $\delta=-kL_{\rm s}/2$.
We apply this fitting procedure only for $L_{\rm s}=96,112$,
since there is no free region for $L_{\rm s}\leq 80$.
The scattering phase shift obtained using this approach is shown in Fig.~\ref{fig:ps_energy_vs_wf}.
For $L_{\rm t}=8$, the results agree with 
those from finite volume energy of order $O(10^{-2})$,
whereas for $L_{\rm t}=4$ larger discrepancies are observed,
particularly for the ground state and second excited state.
The reason is that, the wave functions from tensor network with large volumes $L_{\rm s}=96, 112$ and $L_{\rm t}=4$ 
are significantly affected by coarse-graining errors,
leading to a large error in the fitting results.
%%%%%%%%%
%%%%%%%%%
%%%%%%%%%%%%%%%%%%%%%
\subsubsection{Phase shift from the wave function inside interaction range}
\label{subsec:ps_2p_wf_inside}
%%%%%%%%%%%%%%%%%%%%%
In contrast to the fitting procedure, which is evaluated outside interaction range,
we show the extraction of phase shift from the inside of the interaction range
by employing the Bethe-Salpeter (BS) wave function method \cite{PhysRevD.98.011501,Namekawa:2019xiy,Yamazaki:2017gjl}.
For this purpose, first we recall scattering amplitude
for (1+1)d system in continuum space-time
\be
f(k)=\frac{e^{i\delta(k)}}{k}\sin \delta(k),
\ee
which corresponds to an amplitude $H(k)$, up to overall phase factor,
that is directly related to the BS wave function as follows
\bea\label{eq:H(k)}
H(k)= -\frac{1}{2k^2}\int_{-\infty}^{\infty} dx h(x) g(x).
\eea
Here, $g(x)$ is a function satisfying $\partial_x^2g(x)=-k^2g(x)$, namely $g(x)=\cos(kx)$ or $g(x)=\sin(k|x|)$.
Meanwhile, $h(x)$ is the reduced BS wave function
\be\label{eq:h_cont}
h(x)=\left(\partial_x^2+k^2\right)\psi(x)
\ee
where $\psi(x)$ is the BS wave function.
In the region $|x|>R$, $h(x)=0$ corresponding to $V_{\rm eff}(x)=0$ in Eq.~(\ref{eq:v=0}) and
$\psi(x)$ is merely a free wave function which may be written as
\be\label{eq:bs_wf}
\psi(x)=
\begin{cases}
A\cos(kx+\delta(k)) & x>R\\
A\cos(kx-\delta(k)) & x<-R\\
\end{cases}
\ee
Using integration by parts and Eq.~(\ref{eq:h_cont}), Eq.~(\ref{eq:H(k)}) can be written as
\be\label{eq:func_H_afterintegral}
2k^2H(k)=-\left[\left[\partial_x\psi(x)\right]g(x)-\psi(x)\left[\partial_xg(x)\right]\right]_{-R}^{R}.
\ee
For $g(x)=\cos(kx)$, the function $H(k)$ is denoted as $H_c(k)$. Substituting $g(x)=\cos(kx)$ and Eq.~(\ref{eq:bs_wf}) into Eq.~\ref{eq:func_H_afterintegral}, we obtain $H_c(k)=\sin(\delta(k))/k$.
Similarly, for $g(x)=\sin(k|x|)$, we denote $H(k)$ as $H_s(k)$, which is given by $H_s(k)=\cos (\delta(k))/k$.
Consequently, the ratio of $H_s/H_c$ determines the scattering phase shift
$\delta(k) =\cot^{-1}\left(\frac{H_s(k)}{H_c(k)}\right)$.

For the computation, the lattice counterparts of the amplitude, $H_s^{\rm lat}$ and $H_c^{\rm lat}$, are employed. For a given momentum, they are given by
\begin{align}
2k_{\rm lat}^2H_c^{\rm lat}(k;x)&=-\sum_{z=0}^{x}n(z)h^{\rm lat}(k;z)\cos(kz)\label{eq:H_lat1}\\
2k_{\rm lat}^2H_s^{\rm lat}(k;x)&=-\sum_{z=0}^{x}n(z)h^{\rm lat}(k;z)\sin(k|z|)\label{eq:H_lat2}
\end{align}
where $n(0)=n(L_{\rm s}/2)=1$, $n(z)=2$ for other $z$, and $x=1,2,\ldots,L_{\rm s}/2$.
Here, $h^{\rm lat}(k;z)$ is the lattice version of the reduced BS wave function in Eq.~(\ref{eq:h_cont}) that is given by
\be\label{eq:lattice_h(x)}
h^{\rm lat}(k;z)=\psi(z+1)-2\psi(z)+\psi(z-1)+k_{\rm lat}^2\psi(z).
\ee
Using Eqs.~(\ref{eq:H_lat1}-\ref{eq:lattice_h(x)}), with $\psi(x)^{[\rm hotrg]}$ as the input, the effective phase shifts as a function of $x$ is extracted from
\be\label{eq:phaseshift_bswf}
\delta (k;x) =\cot ^{-1}\left (\frac{H_s^{\rm lat}(k;x)}{H_c^{\rm lat}(k;x)}\right)\hspace{10mm}\text{for~~}0\leq x\leq L_{\rm s}/2.
\ee
%%%%%%%%%
\begin{figure}[t!]
\centering
\begin{subfigure}[b]{0.4\textwidth}
\centering
\includegraphics[width=7cm,height=6cm]{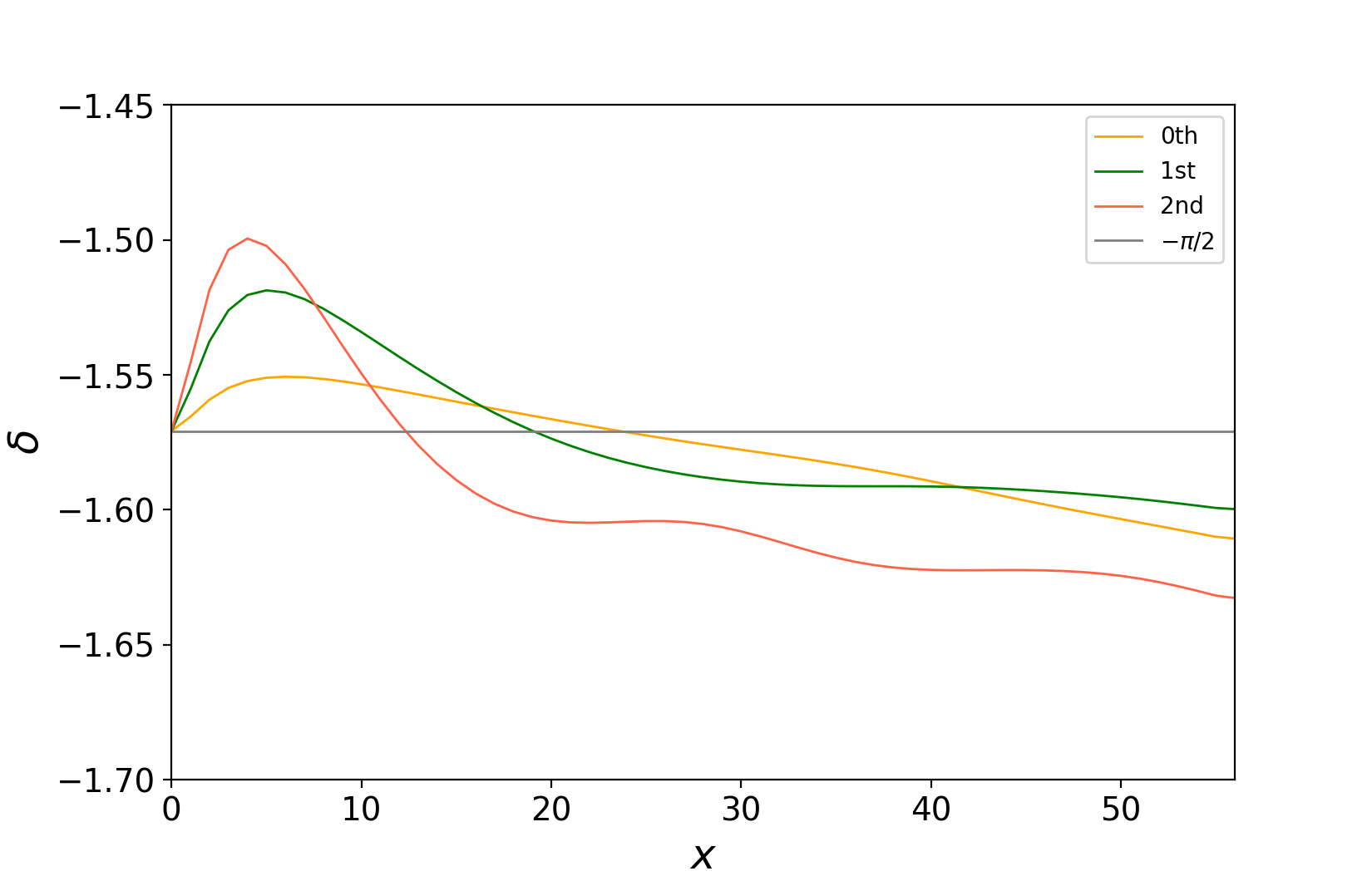}
\caption{}
\label{subfig:ps_BSwf_l112_lt4}
\end{subfigure}
\begin{subfigure}[b]{0.4\textwidth}
\centering
\includegraphics[width=7cm,height=6cm]{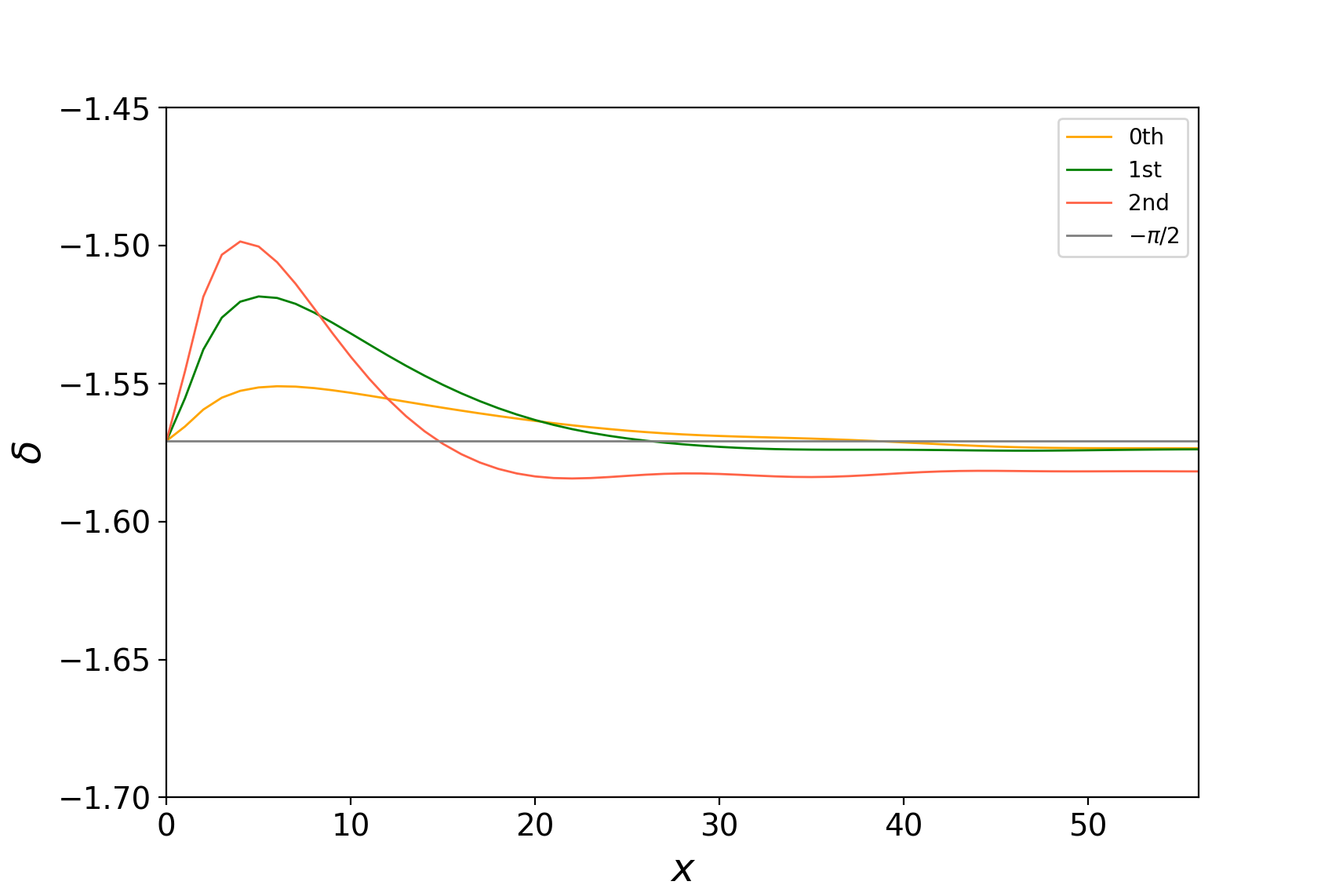}
\caption{}
\label{subfig:ps_BSwf_l112_lt8}
\end{subfigure}
\caption{Effective phase shift $\delta(k;x)$ computed by using Eq.~(\ref{eq:phaseshift_bswf})
over $0\leq x\leq L_{\rm s}/2$ at system size $L_{\rm s}=112$ computed with  $\chi=80$ and (a) $L_{\rm t}=4$, (b) $L_{\rm t}=8$.}
\label{fig:ps_BSwf_L112}
\end{figure}
%%%%%%%%%
In Fig.~\ref{fig:ps_BSwf_L112}, we show $\delta(k;x)$
from the wave function data for system size $L_{\rm s}=112$ and $L_{\rm t}=4,8$ with cut-off $\chi=80$.
This figure shows that the phase shift deviates from $-\pi/2$ for $x < R$ while it approaches $-\pi/2$ for $x>R$.
Comparing Figs.~\ref{subfig:ps_BSwf_l112_lt4} and \ref{subfig:ps_BSwf_l112_lt8},  we find that the data from $L_{\rm t}=8$ gives $\delta(k;x)$ closer to the theoretical value $-\pi/2$ for large $x$ compared to $L_{\rm t}=4$.

For every $L_{\rm s}=8-112$, we compute $\delta(k;x)$ using Eq.~(\ref{eq:phaseshift_bswf}).
We select the value at $x=L_{\rm s}/2$, that is $\delta(k;L_{\rm s}/2)$, and plot them in Fig.~\ref{fig:ps_energy_vs_wf}.
We observe that the result from L\"uscher's method 
and from the reduced BS wave function agree up to double precision
for all $L_{\rm s}$ and for both $L_{\rm t}=4,8$.
This agreement can be understood from the fact that Eq.~(\ref{eq:phaseshift_bswf}) turns into L\"uscher's formula for $x=L_{\rm s}/2$, as discussed in Appendix~\ref{sec:luschers_from_bswf}.
\subsubsection{Degeneracy of two-particle states}
\label{sec:two_p_states_degeneracy}
%%%%%%%%%
\begin{figure}[t!]
\centering
\includegraphics[width=16cm,height=7cm]{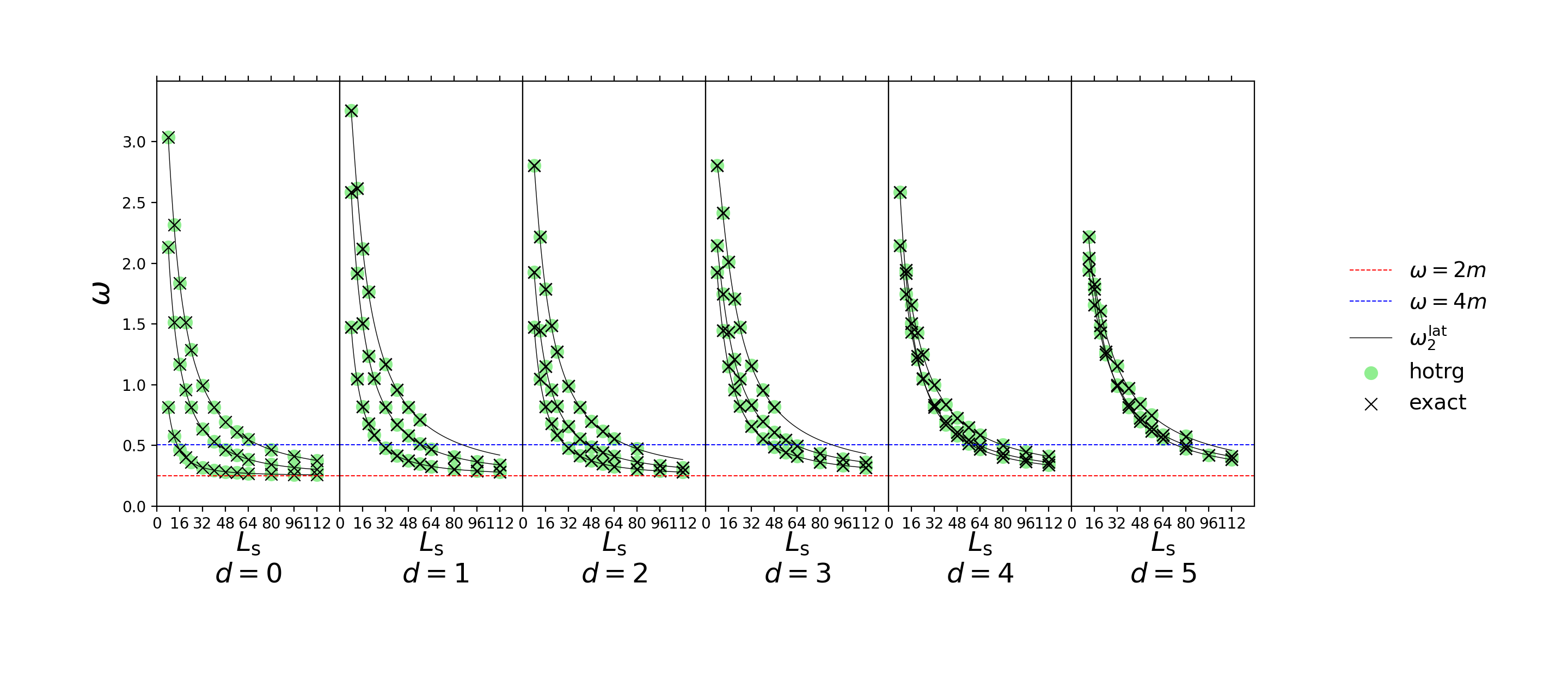}
\caption{Finite volume energy of two-particle states in the CM and moving frames, for $d=0$--$5$ computed with $L_{\rm t}=4$ and $\chi=80$, together with the two-particle states dispersion relation in Eq.~(\ref{eq:2p-disrel_0}). In the $d\neq 0$ sectors, the HOTRG data (green circles) represent the average of the degenerate energies.}
\label{fig:energy_2p_all_P_dc80}
\end{figure}
%%%%%%%%%%%%%%%%%%%%
From Secs.~\ref{subsec:2p_state}, \ref{subsec:ps_2p_wf}, and \ref{subsec:ps_2p_wf_inside}, we observe that the two-particle scattering phase shift of the (1+1)-dimensional Ising model is always $-\pi/2$ for any $k/m$, both inside and outside the elastic region $0 \leq k/m < \sqrt{3}$. This feature leads to four-fold degeneracies in the energy spectrum in the moving frame as shown in Sec.~\ref{subsec:numb_particles}, which can be understood as follows. First, inserting Eq.~(\ref{eq:theoretical_phaseshift}) into Eq.~(\ref{eq:ps_luscher}) makes the allowed relative momentum
\be\label{eq:relative_mom}
p=\frac{\pi}{L_{\rm s}}(1-d+2n),\hspace{10mm}n\in\mathbb{Z}.
\ee
Using Eq.~(\ref{eq:relative_momentum}) and $P=p_1+p_2$, the $p_1$ and $p_2$ are given by
\bea
\label{eq:p1}
p_1=\frac{(2n+1)\pi}{L_{\rm s}},
\hspace{5mm}
p_2=\frac{(2d-2n-1)\pi}{L_{\rm s}}.
\eea

Using Eq.~(\ref{eq:p1}) and assuming $p_1>p_2$ $(p>0)$, we find that there are four combinations of momenta $(p_1,p_2)$, as listed in Table~\ref{Tab:degeneracy} for $|d|=1$--$5$, that yield the same energy when substituted into the following two-particle dispersion relation:
\be\label{eq:2p-disrel_0}
\omega_2^{\rm lat}=\sum_{i=1}^{2}\cosh^{-1}(\cosh m +1 - \cos p_i).
\ee
Note that, this expression is the same as Eq.~(\ref{eq:disrel_nonzeromom}), with $p_1=\frac{\pi d}{L_{\rm s}}+p$ and $p_2=\frac{\pi d}{L_{\rm s}}-p$.
It can be verified that the information of the total momenta $P$ for four-fold degenerate energy in Table~\ref{Tab:degeneracy} are consistent with those previously listed in Table~\ref{Tab:spec_momB_2pstate} for $L_{\rm s}=64$.

%%%%%%%%%%%%%%%%%%%%%%%%%%%%%%%
\begin{table}[t!]
\begin{center}
\caption{The momentum of $p_1,p_2$ for two-particle state in $|d|=1$--$5$ sector for four-fold degenerate states.}
\label{Tab:degeneracy}
\begin{tabular}{|c|cccc|cccc|cccc|}
\hline
$P$&$\frac{2\pi}{L_{\rm s}}$&$\frac{-2\pi}{L_{\rm s}}$&$\frac{4\pi}{L_{\rm s}}$ &$\frac{-4\pi}{L_{\rm s}}$&$\frac{4\pi}{L_{\rm s}}$&$\frac{-4\pi}{L_{\rm s}}$&$\frac{6\pi}{L_{\rm s}}$&$\frac{-6\pi}{L_{\rm s}}$&$\frac{2\pi}{L_{\rm s}}$&$\frac{-2\pi}{L_{\rm s}}$&$\frac{8\pi}{L_{\rm s}}$&$\frac{-8\pi}{L_{\rm s}}$\\
\hline
$d$&$+1$&$-1$&+2&$-2$&$+2$&$-2$&$+3$&$-3$&$$+1&$-1$&$+4$&$-4$\\
\hline
$n$&$1$&$0$&$1$&$-1$&$2$&$0$&$2$&$-1$&$2$&$1$&$2$&$-2$\\
\hline
$p_1$&$\frac{3\pi} {L_{\rm s}}$&$\frac{\pi}{L_{\rm s}}$&$\frac{3\pi} {L_{\rm s}}$&$\frac{-\pi} {L_{\rm s}}$&$\frac{5\pi} {L_{\rm s}}$&$\frac{\pi} {L_{\rm s}}$&$\frac{5\pi} {L_{\rm s}}$&$\frac{-\pi} {L_{\rm s}}$&$\frac{5\pi} {L_{\rm s}}$&$\frac{3\pi} {L_{\rm s}}$&$\frac{5\pi} {L_{\rm s}}$&$\frac{-3\pi} {L_{\rm s}}$ \\
$p_2$&$\frac{-\pi} {L_{\rm s}}$&$\frac{-3\pi} {L_{\rm s}}$&$\frac{\pi} {L_{\rm s}}$&$\frac{-3\pi} {L_{\rm s}}$&$\frac{-\pi} {L_{\rm s}}$&$\frac{-5\pi} {L_{\rm s}}$&$\frac{\pi} {L_{\rm s}}$&$\frac{-5\pi} {L_{\rm s}}$&$\frac{-3\pi} {L_{\rm s}}$&$\frac{-5\pi} {L_{\rm s}}$&$\frac{3\pi} {L_{\rm s}}$&$\frac{-5\pi} {L_{\rm s}}$\\
\hline\hline
$P$&$\frac{6\pi}{L_{\rm s}}$&$\frac{-6\pi}{L_{\rm s}}$&$\frac{8\pi}{L_{\rm s}}$ &$\frac{-8\pi}{L_{\rm s}}$&$\frac{4\pi}{L_{\rm s}}$&$\frac{-4\pi}{L_{\rm s}}$&$\frac{10\pi}{L_{\rm s}}$&$\frac{-10\pi}{L_{\rm s}}$&$\frac{8\pi}{L_{\rm s}}$&$\frac{-8\pi}{L_{\rm s}}$&$\frac{10\pi}{L_{\rm s}}$&$\frac{-10\pi}{L_{\rm s}}$\\
\hline
$d$&$+3$&$-3$&$+4$&$-4$&$+2$&$-2$&$+5$&$-5$&$$+4&$-4$&$+5$&$-5$\\
\hline
$n$&$3$&$0$&$3$&$-1$&$3$&$1$&$3$&$-2$&$4$&$0$&$4$&$-1$\\
\hline
$p_1$&$\frac{7\pi} {L_{\rm s}}$&$\frac{\pi} {L_{\rm s}}$&$\frac{7\pi} {L_{\rm s}}$&$\frac{-\pi} {L_{\rm s}}$&$\frac{7\pi}{L_{\rm s}}$&$\frac{3\pi}{L_{\rm s}}$&$\frac{7\pi}{L_{\rm s}}$&$\frac{-3\pi}{L_{\rm s}}$&$\frac{9\pi}{L_{\rm s}}$&$\frac{\pi}{L_{\rm s}}$&$\frac{9\pi}{L_{\rm s}}$&$\frac{-\pi}{L_{\rm s}}$\\

$p_2$&$\frac{-\pi} {L_{\rm s}}$&$\frac{-7\pi} {L_{\rm s}}$&$\frac{\pi} {L_{\rm s}}$&$\frac{-7\pi} {L_{\rm s}}$&$\frac{-3\pi}{L_{\rm s}}$&$\frac{-7\pi}{L_{\rm s}}$&$\frac{3\pi}{L_{\rm s}}$&$\frac{-7\pi}{L_{\rm s}}$&$\frac{-\pi}{L_{\rm s}}$&$\frac{-9\pi}{L_{\rm s}}$&$\frac{\pi}{L_{\rm s}}$&$\frac{-9\pi}{L_{\rm s}}$\\
\hline
\end{tabular}
\end{center}
\end{table}
%%%%%%%%%%%%%%%%%%%%%%%%%%%%%%%

In addition, from Fig.~\ref{fig:energy_2p_all_P_dc80} we observe that the numerical two-particle state energy levels previously presented in Fig.~\ref{fig:2p_energy_oversize_nonzeromom} are consistent with those obtained from the dispersion relation given by Eq.~(\ref{eq:2p-disrel_0}). This provides further evidence for the identification of the total momentum and the number of particles for the eigenstates in the $q = +1$ sector, as discussed in Secs.~\ref{subsec:momentum} and \ref{subsec:numb_particles}.

\subsection{Three-particle states analysis}
\label{sec:3p-states}
\subsubsection{Three-particle states energy momentum relation}
\label{sec:three_p_states_disrel}

\begin{figure}[t!]
\centering
\includegraphics[width=18cm,height=7cm]{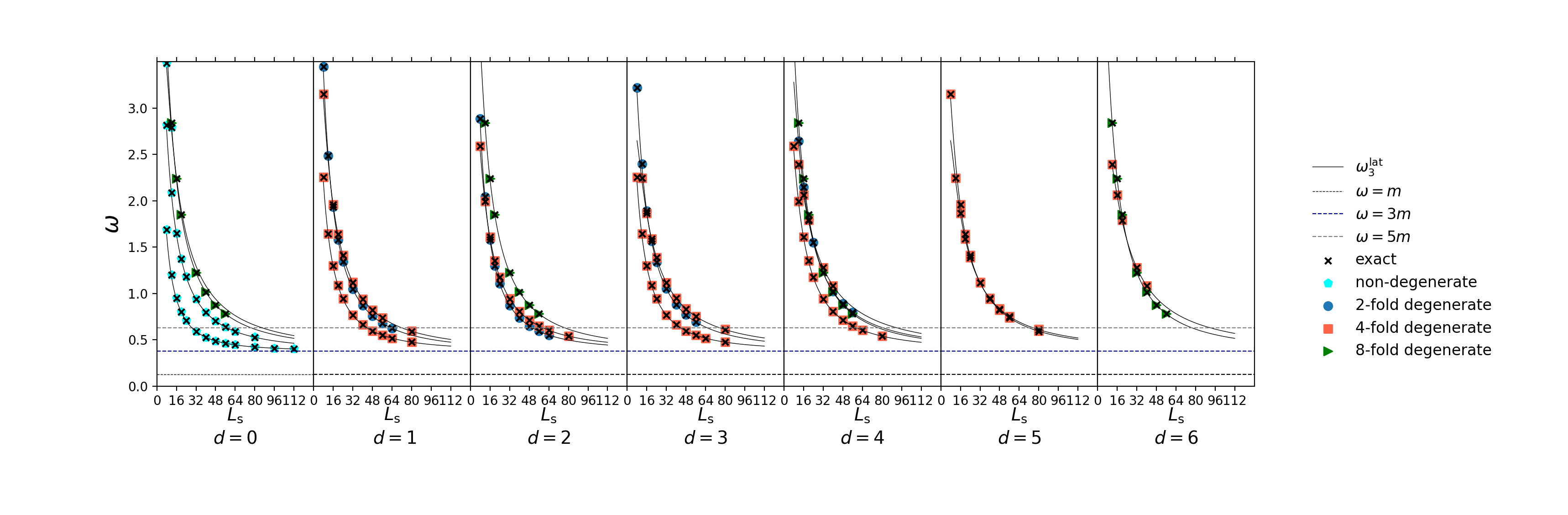}
\caption{Finite-volume energy of three-particle states for $d=0$--$6$ computed with $L_{\rm t}=4$ and $\chi=80$, together with the three-particle dispersion relation in Eq.~(\ref{eq:3p_disrel}). For degenerate states, the HOTRG data shown in the figure represent the average of the corresponding two-, four-, and eight-fold degenerate energy.}
\label{fig:energy_3p_all_P_dc80}
\end{figure}

In Sec.~\ref{subsec:momentum}, we mentioned that some of the eigenstates in the $q=-1$ sector for $L_{\rm s}=64$, namely $a=20, 48, 31$–$34, 36, 37, 49$–$52$, and $53, 54$, do not follow the one-particle dispersion relation. By applying the same procedure for other system sizes, such states can also be obtained.
For the $P = 0$ case, the eigenstates with $a = 20, 48$ for $L_{\rm s} = 64$ have been identified as three-particle states in Sec.~\ref{subsec:numb_particles}, as their energy approach $3m$ at large system sizes; see Fig.~\ref{fig:energy_oversize_qmin}.
The same argument can be applied for the $P\neq 0$ sector. However, some of the states cannot be clearly confirmed to belong to the three-particle sector solely on the energy dependence of system size, as they are only obtained up to relatively smaller sizes; see Fig.~\ref{fig:energy_3p_all_P_dc80}.

%%%%%%%%%
%%%%%%%%%%%%%%%%%%%%

Therefore, to verify the number of particles, we compute the three-particle lattice dispersion relation \cite{PhysRevD.99.014501}
\be\label{eq:3p_disrel}
\omega_3^{\rm lat}=\sum_{i=1}^{3}\cosh^{-1}(\cosh (m) + 1-\cos (p_i))
\ee
where $p_i$ for $i=1,2,3$ is the momentum.
The momenta $p_1,p_2$ can be computed from the quantization condition derived by assuming that only pairwise interactions occur, that is \cite{PhysRevD.99.014501}
\begin{align}
2(\delta(-q_{31})+\delta(q_{12})) + {PL_{\rm s}-p_1L_{\rm s}} &=  2\pi j_1,\label{eq:3p_quantization_condition}\\
2(\delta(-q_{23})+\delta(q_{12}))- {PL_{\rm s}+p_2L_{\rm s}}&= 2\pi j_2, \label{eq:3p_quantization_condition2}
\end{align}
where $j_1,j_2\in \mathbb{Z}$, while $p_3$ is constrained by 
$p_3=P-p_1-p_2$.
Here, $\delta$ is the two-particle state phase shift, which depends on the relative momentum of two interacting particles $q_{ij}=\frac{1}{2}(p_i-p_j)$, namely: $q_{12}=(p_1-p_2)/2$, $q_{23}=(p_2-p_3)/2$ and $q_{31}=(p_3-p_1)/2$. 
Inserting the theoretical prediction for the two-particle scattering phase shift in Eq.~(\ref{eq:theoretical_phaseshift}), we solve Eqs.~(\ref{eq:3p_quantization_condition}) and (\ref{eq:3p_quantization_condition2}) to obtain $p_1, p_2, p_3$ as follows
\be\label{eq:3p_resmomenta}
p_1=\frac{2\pi}{L_{\rm s}} (d+j_1),\hspace{5mm}
p_2= \frac{2\pi}{L_{\rm s}} (d-j_2),\hspace{5mm}
p_3=\frac{2\pi}{L_{\rm s}} (j_2-j_1-d),
\ee
where $d$ is the integer related to the total momentum $P=2\pi d/L_{\rm s}.$
For a given $d$, $(j_1,j_2)$ is chosen such that $q_{12}>0,$ and $q_{23}<0,q_{31}<0$ are satisfied.
By substituting the resulting values $p_1,p_2,p_3$ as listed in Table~\ref{Tab:degeneracy_3p_1}-\ref{Tab:degeneracy_3p_3}, into the dispersion relation in Eq.~(\ref{eq:3p_disrel}), we obtain predictions for the three-particle state energy.
In fact, the theoretical predictions agree with the numerical data well.
Following this procedure, we show in Fig.~\ref{fig:energy_3p_all_P_dc80} that the remaining unidentified eigenstates in $q=-1$ sector, obtained by our scheme for $d=0$--$6$, are indeed three-particle states. 
%%%%%%%%%%%%%%%%%%%%%%%%%%%%%
\subsubsection{Degeneracy of three-particle states}
\label{sec:three_p_states_degeneracy}
%%%%%%%%%%%%%%%%%%%%%%%%%%%%%
The degeneracy structure in three-particle sector of (1+1)d Ising model is more varied than one- or two-particle state.
For $P=0$, the three-particle states are non-degenerate.
In contrast, for $P\neq 0$, as a consequence of the value of the phase shift in Eq.~(\ref{eq:theoretical_phaseshift}) being unchanged for any kinematics region, the spectrum exhibits not only two-fold degeneracies, as in one-particle state case,
but also four-, and eight-fold degeneracies as shown in Fig.~\ref{fig:energy_3p_all_P_dc80}.
In Tables~\ref{Tab:degeneracy_3p_1}--\ref{Tab:degeneracy_3p_3}, we categorize all possible combinations of momenta $p_1, p_2, p_3$ that yield the same energy in Eq.~(\ref{eq:3p_disrel}) into a separate column.
In this case, Table~\ref{Tab:degeneracy_3p_1} shows the momenta for non- and two-fold degeneracies, Table~\ref{Tab:degeneracy_3p_2} shows the momenta for four-fold degeneracy, and 
Table~\ref{Tab:degeneracy_3p_3} shows the eight-fold degeneracy.

With this explanation in mind, one can revisit Table~\ref{Tab:spec_momB_1pstate} for $L_{\rm s}=64$, and identify non-degenerate, two-fold degenerate, and four-fold degenerate three-particle states. The non-degenerate states correspond to $a=20,48$, the two-fold degenerate states to $a=36,37,53,54$, and the four-fold degenerate states to $a=31$--$34$, $a=49$--$52$.
However, eight-fold degenerate states are not listed in the table, as for $L_{\rm s}=64$ they likely appear in eigenstates with $a>57$, which cannot be reliably extracted with $L_{\rm t}=4$ and $\chi=80$.
In our calculation, with cut-off $\chi=80$, the eight-fold degenerate states are only obtained for system sizes $L_{\rm s}=8$--$56$ as shown in Fig.~\ref{fig:energy_3p_all_P_dc80}.
%%%%%%%%%%%%%%%%%%%%%%%%%
%%%%%%%%%%%%%%%%%%%%%%%%%%%%%%%
\begin{table}[t!]

\begin{center}
\caption{The momentum of $p_1,p_2,p_3$ for three-particle state in $|d|=0$--$4$ sectors for non- and two-fold degenerate states.}
\label{Tab:degeneracy_3p_1}
\begin{tabular}{|c|c|c|c|cc|cc|cc|cc|cccc|cccc|cccc|cccc|cccc|}
\hline
$P$&$0$&$0$&$0$&$\frac{2\pi}{L_{\rm s}}$&$-\frac{2\pi}{L_{\rm s}}$&$\frac{4\pi}{L_{\rm s}}$&$-\frac{4\pi}{L_{\rm s}}$&$\frac{6\pi}{L_{\rm s}}$&$-\frac{6\pi}{L_{\rm s}}$&$\frac{8\pi}{L_{\rm s}}$&$-\frac{8\pi}{L_{\rm s}}$\\
\hline
$d$&$0$&$0$&$0$&$+1$&$-1$&$+2$&$-2$&$+3$&$-3$&$+4$&$-4$\\
\hline
$p_1$&$\frac{2\pi}{L_{\rm s}}$&$\frac{4\pi}{L_{\rm s}}$&$\frac{6\pi}{L_{\rm s}}$&$\frac{4\pi}{L_{\rm s}}$&$\frac{4\pi}{L_{\rm s}}$&$\frac{4\pi}{L_{\rm s}}$&$\frac{2\pi}{L_{\rm s}}$&$\frac{6\pi}{L_{\rm s}}$&$\frac{2\pi}{L_{\rm s}}$&$\frac{8\pi}{L_{\rm s}}$&$\frac{2\pi}{L_{\rm s}}$\\
$p_2$&$\frac{-2\pi}{L_{\rm s}}$&$\frac{-4\pi}{L_{\rm s}}$&$\frac{-6\pi}{L_{\rm s}}$&$\frac{-4\pi}{L_{\rm s}}$&$\frac{-4\pi}{L_{\rm s}}$&$\frac{-2\pi}{L_{\rm s}}$&$\frac{-4\pi}{L_{\rm s}}$&$\frac{-2\pi}{L_{\rm s}}$&$\frac{-6\pi}{L_{\rm s}}$&$\frac{-2\pi}{L_{\rm s}}$&$\frac{-8\pi}{L_{\rm s}}$\\
$p_3$&$0$&$0$&$0$&$\frac{2\pi}{L_{\rm s}}$&$\frac{-2\pi}{L_{\rm s}}$&$\frac{2\pi}{L_{\rm s}}$&$\frac{-2\pi}{L_{\rm s}}$&$\frac{2\pi}{L_{\rm s}}$&$\frac{-2\pi}{L_{\rm s}}$&$\frac{2\pi}{L_{\rm s}}$&$\frac{-2\pi}{L_{\rm s}}$\\
\hline
\end{tabular}
\end{center}
\end{table}
%%%%%%%%%%%%%%%%%%%%%%%%%%%%%%%
%%%%%%%%%%%%%%%%%%%%%%%%%%%%%%%
\begin{table}[t!]
\begin{center}
\caption{The momentum of $p_1,p_2,p_3$ for three-particle state in $|d|=1$--$6$ sectors for four-fold degenerate states.}
\label{Tab:degeneracy_3p_2}
\begin{tabular}{|c|cccc|cccc|cccc|cccc|cccc|}
\hline
$P$&$\frac{2\pi}{L_{\rm s}}$&$-\frac{2\pi}{L_{\rm s}}$&$\frac{6\pi}{L_{\rm s}}$&$-\frac{6\pi}{L_{\rm s}}$&$\frac{4\pi}{L_{\rm s}}$&$-\frac{4\pi}{L_{\rm s}}$&$\frac{8\pi}{L_{\rm s}}$&$-\frac{8\pi}{L_{\rm s}}$&$\frac{6\pi}{L_{\rm s}}$&$-\frac{6\pi}{L_{\rm s}}$&$\frac{10\pi}{L_{\rm s}}$&$-\frac{10\pi}{L_{\rm s}}$&$\frac{8\pi}{L_{\rm s}}$&$-\frac{8\pi}{L_{\rm s}}$&$\frac{12\pi}{L_{\rm s}}$&$-\frac{12\pi}{L_{\rm s}}$&$\frac{2\pi}{L_{\rm s}}$&$-\frac{2\pi}{L_{\rm s}}$&$\frac{10\pi}{L_{\rm s}}$&$-\frac{10\pi}{L_{\rm s}}$\\
\hline
$d$&$+1$&$-1$&$+3$&$-3$&$+2$&$-2$&$+4$&$-4$&$+3$&$-3$&$+5$&$-5$&$+4$&$-4$&$+6$&$-6$&$+1$&$-1$&$+5$&$-5$\\
\hline
$p_1$&$\frac{4\pi}{L_{\rm s}}$&$\frac{2\pi}{L_{\rm s}}$&$\frac{4\pi}{L_{\rm s}}$&$0$&$\frac{6\pi}{L_{\rm s}}$&$\frac{2\pi}{L_{\rm s}}$&$\frac{6\pi}{L_{\rm s}}$&$0$&$\frac{8\pi}{L_{\rm s}}$&$\frac{2\pi}{L_{\rm s}}$&$\frac{8\pi}{L_{\rm s}}$&$0$&$\frac{10\pi}{L_{\rm s}}$&$\frac{2\pi}{L_{\rm s}}$&$\frac{10\pi}{L_{\rm s}}$&$0$&$\frac{6\pi}{L_{\rm s}}$&$\frac{4\pi}{L_{\rm s}}$&$\frac{6\pi}{L_{\rm s}}$&$0$\\

$p_2$&$\frac{-2\pi}{L_{\rm s}}$&$\frac{-4\pi}{L_{\rm s}}$&$0$&$\frac{-4\pi}{L_{\rm s}}$&$\frac{-2\pi}{L_{\rm s}}$&$\frac{-6\pi}{L_{\rm s}}$&$0$&$\frac{-6\pi}{L_{\rm s}}$&$\frac{-2\pi}{L_{\rm s}}$&$\frac{-8\pi}{L_{\rm s}}$&$0$&$\frac{-8\pi}{L_{\rm s}}$&$\frac{-2\pi}{L_{\rm s}}$&$\frac{-10\pi}{L_{\rm s}}$&$0$&$\frac{-10\pi}{L_{\rm s}}$&$\frac{-4\pi}{L_{\rm s}}$&$\frac{-6\pi}{L_{\rm s}}$&$0$&$\frac{-6\pi}{L_{\rm s}}$\\

$p_3$&$0$&$0$&$\frac{2\pi}{L_{\rm s}}$&$\frac{-2\pi}{L_{\rm s}}$&$0$&$0$&$\frac{2\pi}{L_{\rm s}}$&$\frac{-2\pi}{L_{\rm s}}$&$0$&$0$&$\frac{2\pi}{L_{\rm s}}$&$\frac{-2\pi}{L_{\rm s}}$&$0$&$0$&$\frac{2\pi}{L_{\rm s}}$&$\frac{-2\pi}{L_{\rm s}}$&$0$&$0$&$\frac{4\pi}{L_{\rm s}}$&$\frac{-4\pi}{L_{\rm s}}$\\
\hline
\end{tabular}
\end{center}
\end{table}
%%%%%%%%%%%%%%%%%%%%%%%%%%%%%%%
%%%%%%%%%%%%%%%%%%%%%%%%%%%%%%%
\begin{table}[t!]
\begin{center}
\caption{The momentum of $p_1,p_2,p_3$ for three-particle state in $|d|=0,2,4,6$ sectors for eight-fold degenerate states.}
\label{Tab:degeneracy_3p_3}
\begin{tabular}{|c|cccccccc|}
\hline
$P$&$0$&$0$&$\frac{4\pi}{L_{\rm s}}$&$-\frac{4\pi}{L_{\rm s}}$&$\frac{8\pi}{L_{\rm s}}$&$-\frac{8\pi}{L_{\rm s}}$&$\frac{12\pi}{L_{\rm s}}$&$-\frac{12\pi}{L_{\rm s}}$\\
\hline
$d$&$0$&$0$&$+2$&$-2$&$+4$&$-4$&$+6$&$-6$\\
\hline
$p_1$&$\frac{4\pi}{L_{\rm s}}$&$\frac{6\pi}{L_{\rm s}}$&$\frac{6\pi}{L_{\rm s}}$&$\frac{4\pi}{L_{\rm s}}$&$\frac{6\pi}{L_{\rm s}}$&$\frac{2\pi}{L_{\rm s}}$&$\frac{6\pi}{L_{\rm s}}$&$\frac{-2\pi}{L_{\rm s}}$\\
$p_2$&$\frac{-6\pi}{L_{\rm s}}$&$\frac{-4\pi}{L_{\rm s}}$&$\frac{-4\pi}{L_{\rm s}}$&$\frac{-6\pi}{L_{\rm s}}$&$\frac{-2\pi}{L_{\rm s}}$&$\frac{-6\pi}{L_{\rm s}}$&$\frac{2\pi}{L_{\rm s}}$&$\frac{-6\pi}{L_{\rm s}}$\\
$p_3$&$\frac{2\pi}{L_{\rm s}}$&$\frac{-2\pi}{L_{\rm s}}$&$\frac{2\pi}{L_{\rm s}}$&$\frac{-2\pi}{L_{\rm s}}$&$\frac{4\pi}{L_{\rm s}}$&$\frac{-4\pi}{L_{\rm s}}$&$\frac{4\pi}{L_{\rm s}}$&$\frac{-4\pi}{L_{\rm s}}$\\
%\hline
%\hash~degeneracy&\multicolumn{8}{c|}{8}\\
\hline
\end{tabular}
\end{center}
\end{table}
%%%%%%%%%%%%%%%%%%%%%%%%%%%%%%%

\section{Summary}
In this paper, we have investigated the multi-particle states
by applying the spectroscopy scheme introduced in \cite{PhysRevD.110.034514}
with the updated coarse-graining strategy and demonstrated it to the (1+1)d Ising model. 
We find that tensor networks with $L_{\rm t}=4,8$
yield energy spectra with smaller errors than other choices of $L_{\rm t}$.
Consequently, the higher excited state energy than those reported in \cite{PhysRevD.110.034514}, as well as their corresponding quantum numbers $q=\pm 1$ can be reliably determined with the relative error in order $O(10^{-3})$.
We further confirmed that the errors for $L_{\rm t}=4,8$ decrease with increasing the bond dimension.

Next, we identified the total momentum of the eigenstates in the $q=-1$ and $q=+1$ sectors for $L_{\rm s}=8$–$112$ using the matrix elements of the appropriate operators.
From the $q=-1$ sector, we observed not only one-particle states but also three-particle states with $|P|=0$–$12\pi/L_{\rm s}$.
The classification of one- and three-particle states for each momentum is done by observing the behavior of the energy as a function of system size, as well as from their respective dispersion relations.
Similarly, in the $q=+1$ sector, the two-particle states with $|P|=0$–$10\pi/L_{\rm s}$ are also clearly identified.

In addition, we successfully computed the wave function of the two-particle state with $P=0$,
from the ground to the second excited state,
by employing the impurity tensor network method.
From the ground state wave function,
the effective potential is computed, and we use it to estimate the interaction range $R$, where for $T=2.44$ it is approximately $R\approx 40$.
Furthermore, we extracted the scattering phase shift of the two-particle state using three approaches:
the finite volume energy approach in both CM and moving frame using L\"uscher's formula, fitting the two-particle wave function outside the interaction range, and the BS wave function method from inside interaction range.
The results obtained from all three methods are consistent with each other as well as with the theoretical prediction up to the error.
From this calculation, we confirmed that the phase shift of (1+1)d Ising model is always given by $\delta=-\pi/2$ for any relative momentum. 
As a consequence, in the two-particle state sector, some states with different total momenta become four-fold degenerate. 

Lastly, in the three-particle sector, the numerical results agree with the theoretical prediction for the finite volume three-particle energy, which is computed under the assumption that no three-body interaction occur. In this sector, the energy are also degenerated where the degeneracies are more varied and include: two-fold degeneracy, as well as four-fold and eight-fold degeneracies.

For the future work, we will continue to the investigation of four-particle states, and the application of the scheme to other quantum field theories.

%%%%%%%%%%%%%%%%%%%%%%%%%%

\section*{ACKNOWLEDGEMENTS}
F.I.A.   is supported by JST SPRING, Grant No. JPMJSP2135.
S.T. is supported in part by JSPS KAKENHI Grants No.~21K03531, No.~22H05251, and No.~25K07280.
T.Y. is supported in part by JSPS KAKENHI Grant No.~23H01195 and No.~23K25891 , and MEXT as ``Program for Promoting Researchers on the Supercomputer Fugaku'' (Grant No. JPMXP1020230409).

%%%%%%%%%%%%%%%%%%%%%%%%%%
\appendix

%%%%%%%%%%%%%%%%%%%%%%%%%%
\section{L\"USCHER'S FORMULA FROM BETHE SALPETER WAVE FUNCTION}
\label{sec:luschers_from_bswf}
In this section, we will derive the L\"uscher's formula from the BS wave function. 
For this purpose, we compute $H_c^{\rm lat}(k;L_{\rm s}/2)$, and $H_s^{\rm lat}(k;L_{\rm s}/2)$ in Eqs.~(\ref{eq:H_lat1}-\ref{eq:H_lat2}) for $x=L_{\rm s}/2$, that is
\begin{align}
 2k^2_{\rm lat}H_{c}^{\rm lat}(k;L_{\rm s}/2)&=
-\sum_{z=0}^{L_{\rm s}/2} n(z) h(k;z)\cos(kz),\label{eq:hlat_subs1}\\
 2k^2_{\rm lat}H_{s}^{\rm lat}(k;L_{\rm s}/2)&=-\sum_{z=0}^{L_{\rm s}/2} n(z) h(k;z)\sin(k|z|)\label{eq:hlat_subs},
\end{align}
where $n(z)=1$ for $z=0,L_{\rm s}/2$ and $n(z)=2$ for else.
By substituting $h(k;z)=\psi(z+1)-2\psi(z)+\psi(z-1)+k_{\rm lat}^2\psi(z)$ with $k_{\rm lat}^2=2(1-\cos(k))$, Eq.~(\ref{eq:hlat_subs1}) can be simplified into
\bea\label{eq:Hc_derive}
2k^2_{\rm lat}H_{c}^{\rm lat}(k;L_{\rm s}/2)&=&
\left(-\psi(L_{\rm s}/2+1)+\psi(L_{\rm s}/2)\right)\cos(kL_{\rm s}/2)\notag\\
&&-\psi(L_{\rm s}/2)(2\sin (kL_{\rm s}/2)\sin (k))
+2(\psi(1)-\psi(-1)).
\eea
Thanks to the symmetric and periodic property of the wave function, it is easy to show that $\psi(-1)=\psi(1)$ and $\psi(L_{\rm s}/2-1)=\psi(L_{\rm s}/2+1)$. 
Using these properties, Eq.~(\ref{eq:Hc_derive}) can be simplified as
\be\label{eq:hc_bswf_n}
2k^2_{\rm lat}H_{c}^{\rm lat}(k;L_{\rm s}/2)=-\psi(L_{\rm s}/2)[2\sin(kL_{\rm s}/2)\sin(k)].
\ee
In a similar way, we can obtain
\be\label{eq:hs_bswf_n}
2k^2_{\rm lat}H_{s}^{\rm lat}(k;L_{\rm s}/2)=\psi(L_{\rm s}/2)[2\cos(kL_{\rm s}/2)\sin(k)].
\ee
Taking the ratio of Eq.~(\ref{eq:hc_bswf_n}) and (\ref{eq:hs_bswf_n}), we have
\be\label{eq:luscher_bswf_n}
\frac{H_{s}^{\rm lat}(k;L_{\rm s}/2)}{H_{c}^{\rm lat}(k;L_{\rm s}/2)}=\frac{-\cos(kL_{\rm s}/2)\sin(k)}{\sin(kL_{\rm s}/2)\sin(k)}=\cot(-kL_{\rm s}/2).
\ee
By comparing this result with Eq.~(\ref{eq:phaseshift_bswf}), one can see that Eq.~(\ref{eq:luscher_bswf_n}) gives L\"uscher's formula, namely $\delta = -kL_{\rm s}/2$.

\bibliography{biblio}
\end{document}